\documentclass[12pt]{article}

\usepackage{amssymb, amsmath, mathrsfs, amsfonts, graphicx, subfigure, natbib}
\usepackage{adjustbox}
\usepackage{array, epsfig, rotating, pdflscape, bm}

\numberwithin{equation}{section}

\newtheorem{them}{Theorem}[section]

\newcommand{\bmu} {\mbox{\boldmath $\mu$}}
\newcommand{\bnu} {\mbox{\boldmath $\nu$}}

\newcommand{\bbeta} {\mbox{\boldmath $\beta$}}

\newcommand{\bDelta} {\mbox{\boldmath $\Delta$}}

\newcommand{\be} {\mbox{\boldmath $e$}}

\newcommand{\bD} {\mathbf {D}}
\newcommand{\bF} {\mathbf {F}}
\newcommand{\bG} {\mathbf {G}}
\newcommand{\bH} {\mathbf {H}}
\newcommand{\bI} {\mathbf {I}}
\newcommand{\bK} {\mathbf {K}}
\newcommand{\bL} {\mathbf {L}}
\newcommand{\bM} {\mathbf {M}}

\newcommand{\bX} {\mathbf {X}}

\newcommand{\bZ} {\mathbf {Z}}

\newcommand{\bS} {\mathbf {S}}
\newcommand{\bT} {\mathbf {T}}
\newcommand{\bA} {\mathbf {A}}
\newcommand{\bB} {\mathbf {B}}
\newcommand{\bC} {\mathbf {C}}

\newcommand{\bR} {\mathbf {R}}

\newcommand{\bone} {\mbox{\boldmath $1$}}

\def\T{{ \mathrm{\scriptscriptstyle T} }}

\def\boldfacefake #1{%
    \hbox{%
        \mathsurround=0pt
        \hbox to 0.25pt{$#1$\hss}%
        \hbox to 0.25pt{$#1$\hss}%
        \hbox {$#1$}%
    }%
}

\renewcommand{\t}{{ \top }}

\newcommand{\diag}{\mbox{diag}}

\def\bmbeta{\bm{\beta}}

\def\bmmu{\bm{\mu}}

\def\blS{\mbox{$\boldfacefake S$}}
\def\blT{\mbox{$\boldfacefake T$}}

\def\cI{\mathcal{I}}
\def\cJ{\mathcal{J}}

\def\cL{\mathcal{L}}

\title{Local polynomial regression for pooled response data}
\date{}

\author{Dewei Wang \\ Department of Statistics, University of South Carolina\\
Xichen Mou \\ 
Division of Epidemiology, Biostatistics, and Environmental Health,\\University of Memphis\\
Xiang Li \\
JPMorgan Chase, Jersey City, New Jersey\\
\and Xianzheng Huang \\Department of Statistics, University of South Carolina}

\begin{document}
\maketitle

\begin{abstract}
We propose local polynomial estimators for the conditional mean of a continuous response when only pooled response data are collected under different pooling designs. Asymptotic properties of these estimators are investigated and compared. Extensive simulation studies are carried out to compare finite sample performance of the proposed estimators under various model settings and pooling strategies. We apply the proposed local polynomial regression methods to two real-life applications to illustrate practical implementation and performance of the estimators for the mean function. 

\end{abstract}

\section{Introduction}
Instead of measuring individual specimens to collect data for biomarkers or analytes of interest, collecting such data on pools of specimens has become increasingly common in epidemiological and environmental studies \citep{kendziorski2003efficiency, shih2004effects}. Collecting pooled data can reduce information loss when there is a detecting limit, and  offer a more timely manner to gather information, in addition to the obvious benefit of reducing cost of laboratory assays and preserving irreplaceable specimens. In some econometrics applications, pooled data are all that is available to researchers, such as data aggregated by family or by region. In these applications, data of other attributes at the individual level are often also recorded, and researchers are interested in associations between quantities at the individual level even though some data are collected at the pool level. Our study is motivated by these research questions that require methodologies for regression analysis based on pooled continuous response data and individual-level covariate data.

Traditional regression methodology applicable to individual response data cannot be directly used to analyze pooled response data, and there exist some research on regression analysis for pooled continuous responses. Under the parametric framework, \citet{malinovsky2012pooling} considered Gaussian random effects models for pooled repeated measures, and studied inference for variance components under different pooling strategies. \citet{mitchell2014regression} proposed a Monte Carlo expectation maximization algorithm to carry out regression analyses of pooled biomarker assessments assuming that the biomarker follows a log-normal distribution given covariates. \citet{mcmahan2016estimating} developed methods to infer receiver-operating characteristic curves using  pooled biomarker measurements. \citet{liu2017general} provided a general strategy based on Monte Carlo maximum likelihood for regression analysis of pooled data under generic parametric models assumed for the individual response given covariates. Under the semiparametric framework, \citet{mitchell2015semiparametric} proposed a semiparametric method for regression analysis of a right-skewed and positive response when data for the response are taken from pooled specimens. Without imposing parametric assumptions on the biomarker distribution, \citet{LinWang2018} developed a semiparametric approach for analyzing pooled biomarker measurements originating from a single-index model for the individual response. Under the nonparametric framework, \citet{linton2002nonparametric} proposed a kernel-based estimator for regression function for pooled data when covariate data are also aggregated, with both aggregated response data and covariate data subject to additive measurement error. 

Among the existing works on regression analysis of pooled response data, many consider various pooling designs. For example, \citet{ma2011cost} compared two pooling designs in the context of linear regression analysis for a pooled continuous response and aggregated covariates, one being random pooling where pools are randomly formed without taking into account covariate information, and the other termed as optimal pooling by the authors, where pools are formed by gathering specimens corresponding to similar covariate values. This latter strategy is better known as homogeneous pooling in the pool/group testing literature \citep{bilder2009bias}, and many researchers have shown efficiency gain in prediction and covariate effects estimation when homogeneous pooled data are used than when random pooled data are used \citep{vansteelandt2000regression, ma2011cost}. \citet{mitchell2014regression} developed a regression methodology for log-normal response data subject to a special form of homogeneous pooling where covariate values within a pool are identical. Like the regression analysis discussed in \citet{ma2011cost}, \citet{mitchell2014regression} also regressed the pooled continuous response on aggregated covariates to infer the association between the response and covariates at the individual level.   

In this article, we propose local polynomial estimators for the mean of a continuous response given covariates using pooled response data and individual-level covariate data. More specifically, the proposed estimators are for the mean function $m(x)=E(Y|X=x)$, where $Y$ is a continuous response of an experimental unit, $X$ is the covariate that can be vector-valued and relate to attributes of the experimental unit or individual. For ease of exposition, we consider a scalar covariate in this article. Observed data available for inferring $m(x)$ include pooled responses from $J$ groups of individuals, $\bZ = (Z_1, \ldots, \, Z_J)^\T$, where $Z_j=c_j^{-1}\sum_{k=1}^{c_j} Y_{jk}$, in which $c_j$ is the number of individuals in pool $j$, and $Y_{jk}$ is the unobserved response of individual $k$ in that pool, for $j=1, \ldots, J$, $k=1, \ldots, c_j$. Also observed are covariate data $\mathbb{X}=\{\tilde \bX_j, \, j=1, \ldots, J\}$, where $\tilde \bX_j=(X_{j1}, \ldots, X_{j,c_j})^\T$, with $X_{jk}$ being the covariate associated with individual $k$ in pool $j$, for $k=1, \ldots, c_j$ and $j=1, \ldots, J$. Three proposed local polynomial estimators for $m(x)$ based on data $(\bZ, \mathbb{X})$ are presented in Section~\ref{s:estimators1} next, where we assume that data arise from random pooling. Section~\ref{s:estimators2} presents local polynomial estimators based on homogeneous pooled data. Asymptotic properties of these estimators are investigated and compared in Section~\ref{s:asymptotics} under each of the two pooling designs. Section~\ref{s:bandwidth} describes bandwidth selection methods tailored for the proposed estimators. Section~\ref{s:empirical} presents a simulation study where we compare finite sample performance of the proposed estimators under different model settings and various pooling designs. We further illustrate the implementation and performance of the proposed methods in two real-life applications in Section~\ref{s:realdata}. Finally, in Section~\ref{s:discussion}, we summarize contributions of our study and discuss follow-up research directions. 

\section{Local polynomial estimators under random pooling}
\label{s:estimators1}
Local polynomial regression has been a well-received and widely applicable nonparametric strategy for estimating $m(x) $ when individual data are available \citep{Fan&Gijbels1996}. To estimate the regression function $m(x)$ based on individual data $\{(Y_{jk}, X_{jk}),k=1,\dots,c_j\}_{j=1}^J$, this strategy exploits the weighted least squares method to construct an objective function following a $p$-th order Taylor expansion of $m(s)$ around $x$, $m(s)\approx \sum_{\ell=0}^p \{m^{(\ell)}(x)/\ell !\} (s-x)^\ell$, with $m^{(\ell)}(x)$ equal to $(\partial^\ell/\partial s^\ell) m(s)$ evaluated at $s=x$. In particular, the objective function is given by 
\begin{align}
Q_0(\bbeta)=\sum_{j=1}^J\sum_{k=1}^{c_j} \left\{Y_{jk}-\sum_{\ell=0}^p \beta_\ell (X_{jk}-x)^\ell\right\}^2 K_h(X_{jk}-x),
\label{eq:Q0}
\end{align}
where $K_h(t)=K(t/h)/h$, $K(t)$ is a symmetric kernel, $h$ is a bandwidth, $\beta_\ell=m^{(\ell)}(x)/\ell!$, for $\ell=0, 1, \ldots, p$, and $\bbeta=(\beta_0, \beta_1, \ldots, \beta_p)^\T$. Minimizing $Q_0(\bbeta)$ with respect to $\bbeta$ yields an estimate of $m(x)(=\beta_0)$, along with estimates of $m^{(\ell)}(x)(=\ell! \beta_\ell)$, for $\ell=1, \ldots, p$. Denote by $\hat m_0(x)$ the so-obtained estimator for $m(x)$. 

In what follows, we revise $Q_0(\bbeta)$ to construct new objective functions to adapt the local polynomial regression strategy to pooled response data from random pooling. 

\subsection{The average-weighted estimator}
\label{s:firstest}
Now that individual responses $\{Y_{jk},k=1,\dots, c_j\}_{j=1}^J$ in (\ref{eq:Q0}) are unobserved but pooled responses $\{Z_j\}_{j=1}^J$ are instead, it is natural to switch  attention from $E(Y_i|X_i)$ to $E(Z_j|\tilde \bX_j)=c_j^{-1}\sum_{k=1}^{c_j} m(X_{jk})$, as if one were regressing $Z$ on the accompanying covariates in a pool collectively. This motivates the following weighted least squares objective function,
\begin{equation}
Q_1(\bbeta) = \sum_{j=1}^J \left\{Z_j-\sum_{\ell=0}^p \beta_\ell c_j^{-1}\sum_{k=1}^{c_j}(X_{jk}-x)^\ell \right\}^2 \left\{c_j^{-1}\sum_{k=1}^{c_j}K_h(X_{jk}-x)\right\}. \label{eq:Q1}
\end{equation}
In (\ref{eq:Q0}), the weight function $K_h(X_i-x)$ quantifies the proximity of the $i$th covariate data point to $x$, producing a larger weight for an individual whose covariate value is closer to $x$. In (\ref{eq:Q1}), the average of such proximity measures associated with $c_j$ covariate data points in pool $j$ is used to assess the overall closeness of this collection of covariate values to $x$. 

Minimizing $Q_1(\bbeta)$ with respect to $\bbeta$ and extracting the first element of the resultant minimizer gives a $p$-th order local polynomial estimator for $m(x)$. This estimator can be explicitly expressed as $\hat m_1(x)  = \be_1^\T \bS_1^{-1}(x)\bT_1(x)$,
where $\be_1^\T=(1, 0, \ldots, 0)_{1\times (p+1)}$, $\bS_1(x) = \bD_1(x)^\T \bK_1(x) \bD_1(x)$, and $\bT_1(x) = \bD_1(x)^\T \bK_1(x) \bZ$, in which, $\bD_1(x)$ is a $J\times(p+1)$ matrix with $\bD_1(x)[j,\ell+1]=c_j^{-1}\sum_{k=1}^{c_j}(X_{jk}-x)^\ell$, for $j=1, \ldots, J$, $\ell=0, 1, \ldots, p$, and $\bK_1(x) =\textrm{diag}\{c_1^{-1}\sum_{k=1}^{c_1}K_h(X_{1k}-x), \, \ldots, \, c_J^{-1}\sum_{k=1}^{c_J} K_h(X_{Jk}-x)\}$. Elaborated expressions of entries in $\bS_1(x)$ and $\bT_1(x)$ are given in Appendix A. To highlight the weight function construction in (\ref{eq:Q1}), $\hat m_1(x)$ is referred to as the average-weighted estimator in this article.

\subsection{The product-weighted estimator}
\label{s:secondest}
Instead of averaging individual-level weights to construct a weight function as in $Q_1(\bbeta)$, one may view $\tilde \bX_j$ as a multivariate covariate resulting from stacking the $c_j$ individual-level covariates in pool $j$ on top of each other, and an alternative weight function can be formulated to measure the nearness of this multivariate covariate to $x\bone_{c_j}$, where $\bone_{c_j}$ denotes the $c_j \times 1$ vector of one's. Mimicking the product kernel used in multivariate kernel density estimation, we propose the following weighted least squares objective function with a different weight function, 
\begin{equation}
Q_2(\bbeta) = \sum_{j=1}^J \left\{Z_j-\sum_{\ell=0}^p \beta_\ell c_j^{-1}\sum_{k=1}^{c_j}(X_{jk}-x)^\ell \right\}^2 \left\{\prod_{k=1}^{c_j}K_h(X_{jk}-x)\right\}.\label{eq:Q2}
\end{equation}

More succinctly, the estimator for $m(x)$ resulting from minimizing $Q_2(\bbeta)$ is given by $\hat m_2(x)  = \be_1^\T \bS_2^{-1}(x)\bT_2(x)$, where $\bS_2(x)= \bD_1(x)^\T \bK_2(x) \bD_1(x)$ and $\bT_2(x) = \bD_1(x)^\T \bK_2(x) \bZ$, in which the weight matrix is given by $\bK_2(x)=\textrm{diag}\{\prod_{k=1}^{c_1}K_h(X_{1k}-x), \ldots, \prod_{k=1}^{c_J}K_h(X_{Jk}-x)\}$. Detailed expressions of entries in $\bS_2(x)$ and $\bT_2(x)$ are provided in Appendix B. Due to the construction of the weight function in (\ref{eq:Q2}), we call $\hat m_2(x)$ the product-weighted estimator in the sequel. 

\subsection{The marginal-integration estimator}
\label{s:thirdest}
The first two estimators are motivated by the mean of $Z_j$ given all covariate data in pool $j$. The third estimator is inspired by the mean of $c_j Z_j$ given one arbitrary individual's covariate in pool $j$ derived next under the assumption that $Y_{jk'} \perp X_{jk}$ for $k' \ne k$ and the pools are formed randomly independent of covariate information. By the definition of $Z_j$, we have 
\begin{align*}
E(c_j Z_j|X_{jk}=x) & = \sum_{k'=1, k' \ne k}^{c_j} E(Y_{j k'}|X_{jk}=x)+ E(Y_{jk}|X_{jk}=x) \\
& =\sum_{k'=1, k' \ne k}^{c_j} E(Y_{j k'})+ m(x) = (c_j-1) \mu+m(x), 
\end{align*}
where $\mu=E(Y_{j k'})$ for $k'=1, \ldots, c_j$ and $j=1, \ldots, J$.
Hence, 
\begin{align}
E\{c_j Z_j-(c_j-1) \mu|X_{jk}=x\} & = m(x).
\label{eq:mean1}
\end{align}
If one views $c_j Z_j-(c_j-1) \mu$ as a pseudo response, (\ref{eq:mean1}) is reminiscent of the conditional mean model for individual-level data, $E(Y_i|X_i=x)=m(x)$, except for the dependence of the pseudo response on the unknown parameter $\mu$. Since $\mu$ is the marginal mean of $Y$, one may use the overall sample mean response, $\hat \mu=N^{-1}\sum_{j=1}^J c_j Z_j$, to estimate $\mu$, where $N=\sum_{j=1}^J c_j$. This yields a surrogate of the pseudo response defined by $R_j=c_j Z_j-(c_j-1)\hat \mu$, for $j=1, \ldots, J$. Heuristically, $R_j$ can be viewed as an ``estimate" for $Y_{jk}$, writing it as $\hat Y_{jk}$ for the meantime as a reminder that one tries to return to the mean model one would use had individual responses been available, $E(Y_{jk}|X_{jk}=x)=m(x)$. Certainly, $E(\hat Y_{jk}|X_{jk}=x)\ne m(x)$ due to the estimation of $\mu$ in $\hat Y_{jk}$. In fact, one can show that $E(\hat Y_{jk}|X_{jk}=x)= m(x)+\{\mu-m(x)\}(c_j-1)/N$. 

Using the surrogate of the pseudo response and (\ref{eq:mean1}), we formulate the following weighted least squares objective function,  
\begin{equation}
Q_3(\bbeta) = \sum_{j=1}^J \sum_{k=1}^{c_j}\left\{R_j-\sum_{\ell=0}^p \beta_\ell (X_{jk}-x)^\ell \right\}^2 K_h(X_{jk}-x). \label{eq:Q3}
\end{equation}
Minimizing $Q_3(\bbeta)$ with respect to $\bbeta$ yields a $p$-th order local polynomial estimator for $m(x)$ yields our third proposed estimator for $m(x)$, denoted by $\hat m_3(x)$. As one can see from the elaborated expression of it given in Appendix C that $\hat m_3(x)$ is simply $\hat m_0(x)$ with $Y_{jk}$ replaced by $R_j$, for $j=1, \ldots, J$, $k=1, \ldots, c_j$. The construction of $\hat m_3(x)$ follows the same strategy of marginal integration used in \citet{LinWang2018}. For this reason, we refer to $\hat m_3(x)$ as the marginal-integration estimator henceforth. 

All three estimators reduce to $\hat m_0(x)$ when $c_j=1$ for $j=1, \ldots, J$, but are otherwise typically very different from each other. In-depth comparisons between the three estimators that go beyond their formulations demand more systematic investigation on their theoretical properties. This is the content of Section~\ref{s:asymptotics}, where we look into the asymptotic bias and variance of these estimators under each of the two considered pooling designs.  

\section{Local polynomial estimators under homogeneous pooling}
\label{s:estimators2}
When pooled data result from homogeneous pooling, it is no longer sensible to consider the mean of $c_j Z_j$ given one ``arbitrary" covariate data point in pool $j$ as we just did to construct $\hat m_3(x)$, since individuals' covariates within a pool are not that ``arbitrary" now after all, and $E(Y_{jk'}|X_{jk}=x)$ is typically not equal to $E(Y_{jk'})$ for $k' \ne k$. But it is still meaningful to consider the mean of $Z_j$ given all covariate data in pool $j$ as we did under random pooling that leads to $\hat m_1(x)$ and $\hat m_2(x)$. 

To be more concrete, consider the homogeneous pooling design following which pools of individuals are created according to the sorted covariate data in $\mathbb{X}$. This yields covariate data associated with pool $j$, for $j=1, \ldots, J$, given by $\tilde \bX_{(j)}=(X_{(j1)}, \ldots, X_{(jc_j)})^\T$, where $X_{(11)} \le X_{(12)} \le \ldots \le X_{(1c_1)} \le X_{(21)} \le \ldots \le X_{(2c_2)} \le \ldots \le X_{(J1)} \le \ldots \le X_{(Jc_J)}$. Even though the response data are not sorted, we use $Z_{(j)}=c_j^{-1}\sum_{k=1}^{c_j} Y_{(jk)}$ to denote the corresponding pooled response, where $Y_{(jk)}$ is the response of the individual whose covariate value is $X_{(jk)}$, for $k=1, \ldots, c_j$, and $j=1, \ldots, J$. Evaluating the objective functions in (\ref{eq:Q1}) and (\ref{eq:Q2}) at $\{(Z_{(j)}, \tilde \bX_{(j)})\}_{j=1}^J$ give the following objective functions one maximizes with respect to $\bbeta$ in order to obtain the average-weighted estimator, $\hat m_1(x)$, and the product-weighted estimator, $\hat m_2(x)$, respectively, under homogeneous pooling, 
\begin{align*}
Q_1(\bbeta) & = \sum_{j=1}^J \left\{Z_{(j)}-\sum_{\ell=0}^p \beta_\ell c_j^{-1}\sum_{k=1}^{c_j}(X_{(jk)}-x)^\ell \right\}^2 \left\{c_j^{-1}\sum_{k=1}^{c_j}K_h(X_{(jk)}-x)\right\}, \\
Q_2(\bbeta) & = \sum_{j=1}^J \left\{Z_{(j)}-\sum_{\ell=0}^p \beta_\ell c_j^{-1}\sum_{k=1}^{c_j}(X_{(jk)}-x)^\ell \right\}^2 \left\{\prod_{k=1}^{c_j}K_h(X_{(jk)}-x)\right\}.
\end{align*}

\section{Comparisons between different estimators}
\label{s:asymptotics}

\subsection{Asymptotic bias and variance}
Under certain regularity conditions listed in Appendix A, we derive asymptotic means and variances of the proposed estimators for $\bbeta$ as $J \to \infty$ with $\max_{1\le j \le J} c_j$ bounded. Conditions listed there relate to $m(x)$, the variance function $\sigma^2(x)=\textrm{Var}(Y|X=x)$, the density function of $X$, $f_{\hbox {\tiny $X$}}(x)$, and the kernel $K(t)$, which are mostly common conditions seen in the context of local polynomial regression using individual-level data. In what follows, we summarize findings from these derivations (with details provided in the appendices) in two theorems that highlight some interesting contrasts between different estimators for $m(x)$ when pools are of equal size with $c_j=c$, for $j=1, \ldots, J$, with additional conditions imposed in each theorem when needed. Several quantities appearing in these theorems are defined next for ease of reference: 
\begin{align*}
\bmu^*_\ell & =(\mu_\ell, \mu_{\ell+1}, \ldots, \mu_{\ell+p})^\T, \quad \tilde \bmu_\ell=[\mu_{\ell_1+\ell_2+\ell}]_{\ell_1, \ell_2=0, 1, \ldots, p}, \\
 \tilde \bnu_0 & =[\nu_{\ell_1+\ell_2}]_{\ell_1, \ell_2=0, 1, \ldots, p}, \quad \bR^*_p=(R_{0,p}(x), R_{1,p}(x), \ldots, R_{p,p}(x))^\T, \\
\bDelta^*_0(x) & = (1, \delta_1(x), \ldots, \delta_p(x))^\T, \quad \tilde \bDelta_0(x)= [\delta_{\ell_1+\ell_2}(x)]_{\ell_1, \ell_2=0, 1, \ldots, p},
\end{align*}
where $R_{\ell,p}(x)= E[(X-x)^\ell \{m(X)-\sum_{\ell=0}^p \beta_\ell (X-x)^\ell\}]$ and $\delta_\ell(x)=E\{(X-x)^\ell\}$, for $\ell=0, 1, \ldots, 2p$.

The first theorem concerns the three estimators under random pooling. Appendices A, B, and C provide the proof for the three parts of this theorem that allow unequal pool sizes.  
\begin{them}
\label{th:randomp}
As $J\to \infty$ and $h\to 0$, one has the following results regarding the difference between an estimator for $m(x)$ and $m(x)$. 
\begin{itemize}
\item[(i)] If the $\ell$-th moment of $X$ exists, for $\ell=1, \ldots, 2p$, then
\begin{align} 
\hat m_1(x)-m(x) = &\ \be_1^\T \bM_0^{-1}(x) \left\{ \bL_0(x)-h f^{-1}_{\hbox {\tiny $X$}}(x) f'_{\hbox {\tiny $X$}}(x) \bM_1(x) \bM_0^{-1}(x) \bL_0(x)\right.\nonumber\\
&\left. +O\left(h^2\right) \right\} +\sqrt{c}\times O_{\hbox {\tiny $P$}}\left(\frac{1}{\sqrt{Nh}}\right), \label{eq:m1asy}
\end{align}
where 
\begin{align*}
\bL_0(x) =&\ \frac{c-1}{c^2}\left\{R_{0,p}(x) \be_1+\bR_p^*(x)\right\}+\frac{(c-1)(c-2)_+}{c^2}R_{0,p}(x)\bDelta_0^*(x), \\
\bM_0(x) =&\ \frac{\tilde \bmu_0}{c^2}+\frac{c-1}{c^2}\left\{\tilde \bDelta_0(x)+\bDelta^*_0(x) \bmu_0^{*\T}+\bmu_0^*\bDelta_0^{*\T}(x)\right\} \\
& + \frac{(c-1)(c-2)_+}{c^2}\bDelta_0^*(x) \bDelta_0^{*\T}(x), \\
\bM_1(x) =&\ \frac{\tilde \bmu_1}{c^2}+\frac{c-1}{c^2}\left\{\bDelta^*_0(x) \bmu_1^{*\T}+\bmu_1^*\bDelta_0^{*\T}(x)\right\}, 
\end{align*}
in which $(t)_+=\max(t, 0)$. 
\item[(ii)] If $m(x)$ is $(p+3)$th-order continuously differentiable, then 
\begin{align}
&\ \hat m_2(x)-m(x) \nonumber \\
= &\ \be_1^\T h^{p+1} \left\{ \beta_{p+1} \left\{ \tilde \bmu_0+(c-1) \bmu_0^*\bmu_0^{*\T} \right\}^{-1} \left\{\bmu^*_{p+1}+(c-1) \mu_{p+1}\bmu^*_0 \right\} \right.\nonumber \\
& + hf^{-1}_{\hbox {\tiny $X$}}(x)\left[ \left\{\beta_{p+2}f_{\hbox {\tiny $X$}}(x)+\beta_{p+1}f'_{\hbox {\tiny $X$}}(x)\right\}\left\{\tilde \bmu_0+(c-1)\bmu_0^* \bmu_0^{*\T}  \right\}^{-1}\right. \nonumber \\
&\times \left\{\bmu^*_{p+2} +(c-1) \mu_{p+2}\bmu^*_0  \right\} -\beta_{p+1} f'_{\hbox {\tiny $X$}}(x)
\left\{\tilde \bmu_0+(c-1)\bmu_0^* \bmu_0^{*\T} \right\}^{-1}  \nonumber \\
& \times \left\{ \tilde \bmu_1 +(c-1) \left(\bmu_0^*\bmu_1^{*\T}+\bmu_1^*\bmu_0^{*\T} \right)\right\}\left\{\tilde \bmu_0+(c-1)\bmu_0^* \bmu_0^{*\T}  \right\}^{-1} \nonumber\\
& \times \left.\left.\left\{\bmu^*_{p+1}+(c-1)\mu_{p+1}\bmu_0^*\right\}\right]+O(h^2)\right\}+\sqrt{c}\times O_{\hbox {\tiny $P$}}\left(\frac{1}{\sqrt{Nh^c}}\right). \nonumber 
\end{align}
\item[(iii)] Let $\bar \sigma^2=E\{\sigma^2(X)\}$. If $\textrm{Var}(Y)$ exists, then 
\begin{align}
\hat m_3(x)-m(x) = &\ \be_1^\T h^{p+1}\left\{\beta_{p+1} \tilde\bmu_0^{-1}\bmu_{p+1}^*+hf^{-1}_{\hbox {\tiny $X$}}(x)\left[\left\{\beta_{p+2}f_{\hbox {\tiny $X$}}(x) \right.\right.\right.\nonumber \\
&\left.\left.\left.+\beta_{p+1}f'_{\hbox {\tiny $X$}}(x)\right\}\tilde\bmu_0^{-1}\bmu_{p+2}^*-\beta_{p+1}f'_{\hbox {\tiny $X$}}(x)\tilde\bmu_0^{-1}\tilde\bmu_1\tilde\bmu_0^{-1}\bmu_{p+1}^*\right]\right. \nonumber \\
&\left. +O(h^2)\right\}+\sqrt{\sigma^2(x)+(c-1)\bar\sigma^2}\times O_{\hbox {\tiny $P$}}\left(\frac{1}{\sqrt{Nh}}\right). \label{eq:m3asy}
\end{align}
\end{itemize}
\end{them}

Theorem~\ref{th:randomp}-(i) indicates that $\hat m_1(x)$ is an inconsistent estimator for $m(x)$, with the dominating bias given by $\be_1^\T \bM^{-1}_0(x) \bL_0(x)$ that does not depend on $h$, and thus does not diminish as $h\to 0$, but it does vanish when $c=1$. Considering a local constant estimator by setting $p=0$ in (\ref{eq:m1asy}), we show in Appendix A that 
\begin{align*}
&\ \hat m_1(x)-m(x) \\
= &\ \frac{c-1}{c}E\{m(X)-m(x)\} +\frac{h^2 \mu_2 }{c} \left\{\beta_1 \frac{f'_{\hbox {\tiny $X$}}(x)}{f_{\hbox {\tiny $X$}}(x)}+\beta_2\right\}+O(h^4)+O_{\hbox {\tiny $P$}}\left(\frac{1}{\sqrt{Jh}}\right), 
\end{align*}
of which the second term (of order $h^2$) is $c^{-1}$ times the dominating bias of the Nadaraya-Watson estimator based on individual-level data. 

Theorem~\ref{th:randomp}-(ii) suggests that $\hat m_2(x)$ is a consistent estimator for $m(x)$ with the asymptotic variance of order $O\{1/(Jh^c)\}$, which inflates quickly as $c$ increases. Comparing (ii) and (iii) of Theorem~\ref{th:randomp} reveals that $\hat m_2(x)$ and $\hat m_3(x)$ typically do not share the same dominating bias except when $c=1$, and $\hat m_3(x)$ exhibits the same asymptotic bias as that of $\hat m_0(x)$ regardless of the pool size. The variability of $\hat m_3(x)$ is understandably higher than that of $\hat m_0(x)$, but it only grows linearly in $c$ and thus is much less inflated than the variance of $\hat m_2(x)$. More specifically, (\ref{eq:m3asy}) implies that the amount of variance inflation of $\hat m_3(x)$ depends linearly on the pool size and $\bar \sigma^2$. 

Summarizing these implications of Theorem~\ref{th:randomp}, we conclude that the marginal-integration estimator $\hat m_3(x)$ is the preferred estimator among the three proposed under random pooling. It outperforms the average-weighted estimator $\hat m_1(x)$ for its consistency, and it surpasses the product-weighted estimator $\hat m_2(x)$ for its much less inflated variance when compared with $\hat m_0(x)$. However, $\hat m_3(x)$ is no longer well justified under homogeneous pooling as pointed out in Section~\ref{s:estimators2}. The following theorem is regarding the average-weighted estimator and the product-weighted estimator applied to data from the homogeneous pooling design. Appendix D provides the proof for this theorem.
\begin{them} 
\label{th:homop}
Assume that $x$ is an interior point of a compact and nondegenerate interval $\cI$, the pdf of X, $f_{\hbox {\tiny $X$}}(\cdot)$, is bounded away from zero on an interval $\cJ$, where $\cI\subset \cJ$, and $K(|t|)=0$ for $|t|>1$, with $K'(t)$ bounded. Then, as $J\to \infty$, $h\to 0$, and $Jh^4\to \infty$, 
\begin{align}
&\ \hat m_1(x)-m(x) \nonumber \\
= &\ \be_1^\T h^{p+1}\left\{\beta_{p+1} \tilde\bmu_0^{-1}\bmu_{p+1}^*+hf^{-1}_{\hbox {\tiny $X$}}(x)\left[\left\{\beta_{p+2}f_{\hbox {\tiny $X$}}(x)+\beta_{p+1}f'_{\hbox {\tiny $X$}}(x)\right\}\tilde\bmu_0^{-1}\bmu_{p+2}^*\right. \right. \nonumber\\
& \left.\left.-\beta_{p+1}f'_{\hbox {\tiny $X$}}(x)\tilde\bmu_0^{-1}\tilde\bmu_1\tilde\bmu_0^{-1}\bmu_{p+1}^*\right]+O(h^2)\right\} +O_{\hbox {\tiny $P$}}\left(\frac{1}{\sqrt{Nh}}\right), \label{eq:m1asyh}
\end{align}
and
\begin{equation}
\textrm{Var}\left\{\hat m_1(x)|\mathbb{X}\right\} = \frac{\sigma^2(x)}{Nhf_{\hbox {\tiny $X$}}(x)}\be_1^\T\tilde \bmu_0^{-1}\tilde \bnu_0\tilde \bmu_0\left\{1+o_{\hbox {\tiny $P$}}(1)  \right\}.\label{eq:m1vh}
\end{equation}
If Condition (C5) is satisfied for the kernel defined by $K^\dagger(t)=K^c(t)$, then 
\begin{align}
&\ \hat m_2(x)-m(x) \nonumber \\
= &\ \be_1^\T h^{p+1}\left\{\beta_{p+1} \tilde\bmu_{\dagger,0}^{-1}\bmu_{\dagger, p+1}^*+hf^{-1}_{\hbox {\tiny $X$}}(x)\left[\left\{\beta_{p+2}f_{\hbox {\tiny $X$}}(x)+\beta_{p+1}f'_{\hbox {\tiny $X$}}(x)\right\}\tilde\bmu_{\dagger,0}^{-1}\bmu_{\dagger,p+2}^*\right. \right. \nonumber\\
& \left.\left.-\beta_{p+1}f'_{\hbox {\tiny $X$}}(x)\tilde\bmu_{\dagger,0}^{-1}\tilde\bmu_{\dagger,1}\tilde\bmu_{\dagger,0}^{-1}\bmu_{\dagger,p+1}^*\right]+O(h^2)\right\} +O_{\hbox {\tiny $P$}}\left(\frac{1}{\sqrt{Nh}}\right), \label{eq:m2asyh} 
\end{align}
and
\begin{equation}
\textrm{Var}\left\{\hat m_2(x)|\mathbb{X}\right\} = \frac{\sigma^2(x)}{Nhf_{\hbox {\tiny $X$}}(x)}\be_1^\T \tilde \bmu_{\dagger,0}^{-1}\tilde \bnu_{\dagger,0}\tilde \bmu_{\dagger,0}\left\{1+o_{\hbox {\tiny $P$}}(1)  \right\},\label{eq:m2vh}
\end{equation}
where $\bmu^*_{\dagger,\ell}$, $\tilde \bmu_{\dagger, \ell}$, and $\tilde\bnu_{\dagger,0}$ are the counterparts of $\bmu^*_\ell$, $\tilde \bmu_\ell$, and $\tilde\bnu_0$, respectively, with $K(t)$ replaced by $K^\dagger(t)$. 
\end{them}

Among the additional assumptions imposed in Theorem~\ref{th:homop}, the one on $x$ and the assumption on $K(t)$ are similar to Conditions (T1) and (T5) in \citet{delaigle2012nonparametric}, respectively. Theorem~\ref{th:homop} indicates that both $\hat m_1(x)$ and $\hat m_2(x)$ are consistent estimators for $m(x)$ under homogeneous pooling, with the former sharing the same dominating bias as that of $\hat m_0(x)$, and the latter exhibiting the same form of dominating bias with a re-defined kernel that depends on $c$. Moreover, the asymptotic variances of both estimators are of the same order as that of $\hat m_0(x)$ despite the pool size. The practical implication of Theorem~\ref{th:homop} is that, if one uses homogeneous pooled data to infer $m(x)$ via either one of the two proposed local polynomial estimators, one only needs $J$ assays without losing accuracy or efficiency asymptotically compared with when un-pooled data are used that require $N=cJ$ assays. 

\subsection{Further remarks}
\label{s:remarks}
We are now in the position to reflect on the findings in Theorems~\ref{th:randomp} and \ref{th:homop} to gain a deeper understanding of the three proposed estimators for $m(x)$ using pooled data. 

The stark contrast between properties of the average-weighted estimator under the two pooling designs may seem peculiar at first glance. As natural as it initially appears to be, the use of average weights is the root cause for the persistent bias of $\hat m_1(x)$ under random pooling. For ease of exposition, assume for the time being $c_j=2$, for $j=1, \ldots, J$. The objective function $Q_1(\bbeta)$ in (\ref{eq:Q1}) associated with $\hat m_1(x)$ is essentially constructed for estimating $m^*(x_1, x_2)\triangleq\{m(x_1)+m(x_2)\}/2$ evaluated at $(x_1, x_2)=x\bone_2$. The same weight, $\{K_h(X_{j1}-x)+K_h(X_{j2}-x)\}/2$, is assigned to both individuals in pool $j$ whose covariate values are $\tilde \bX_j=(X_{j1}, X_{j2})^\T$. This can yield misleading weight when, for example, $X_{j1}$ is close to $x$ but $X_{j2}$ is far away from $x$, which can often happen under random pooling. In contrast, the product weight in $Q_2(\bbeta)$ in (\ref{eq:Q2}) associated with $\hat m_2(x)$ avoids such misleading weighting scheme because $K_h(X_{j1}-x)K_h(X_{j2}-x)$ is small if either one of the two individual weights is small, and thus $\tilde \bX_j$ will only contribute more in estimating $m^*(x, x)=m(x)$ when both $X_{j1}$ and $X_{j2}$ are closer to $x$. In particular, when $K(t)$ is the Gaussian kernel, the product weight function amounts to evaluating the bivariate Gaussian density function at the Euclidean distance between $\tilde \bX_j$ and $x\bone_2$, whereas the average weight function lacks such connection with a meaningful distance measure between the two points in $\mathbb{R}^2$. 

\begin{figure} 
	\centering
	\setlength{\linewidth}{1\textwidth}
	\includegraphics[width=\linewidth]{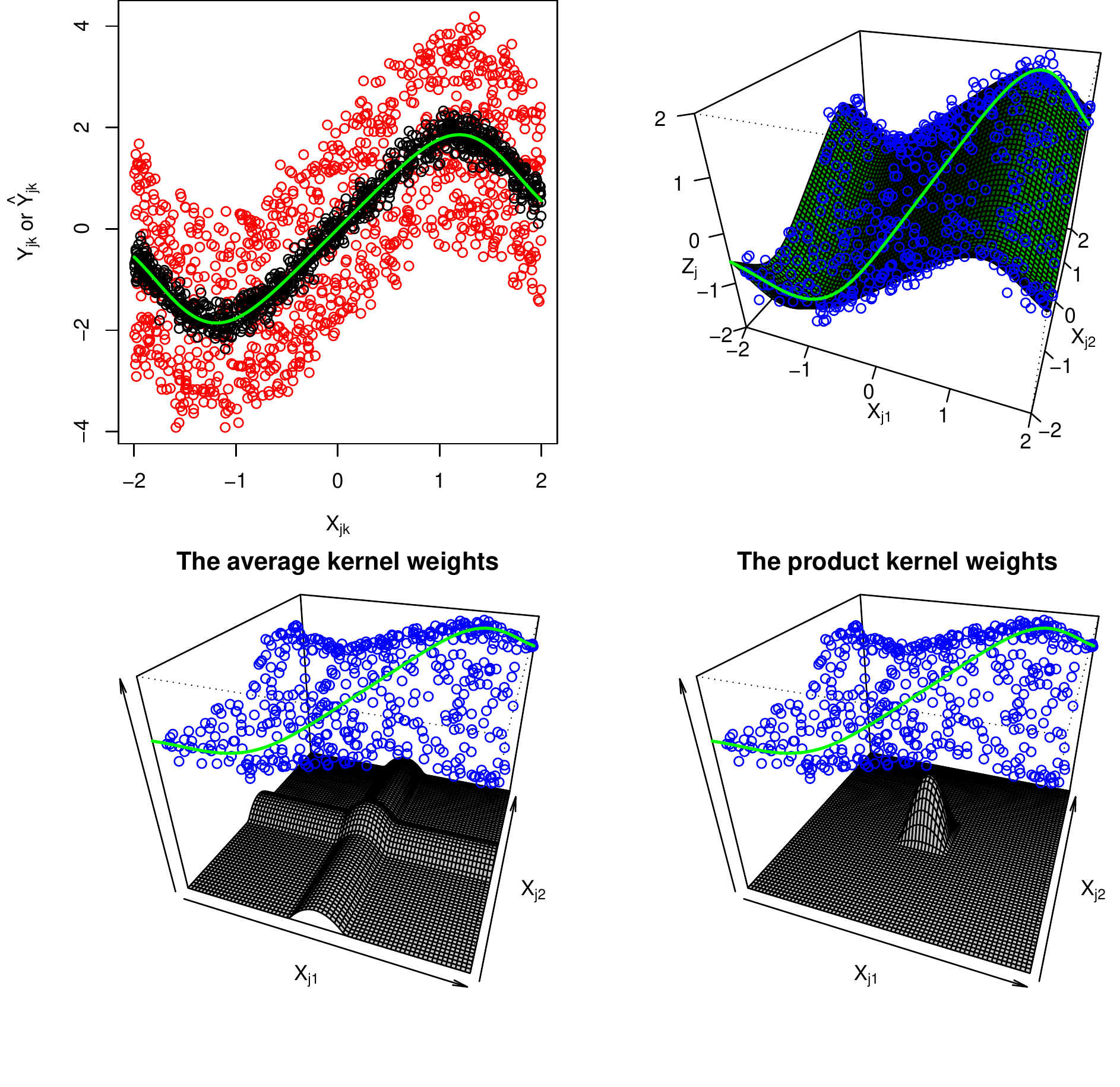} 
	\caption{\linespread{1.3}\selectfont{}Plots under random pooling. Top-left panel: Individual-level data $\{(Y_{jk}, X_{jk}), k=1, 2\}_{j=1}^J$ as black circles, pseudo individual-level responses and covariate data $\{(\hat Y_{jk}, X_{jk}), k=1, 2\}_{j=1}^J$ as red circles, overlaid with the true $m(x)$ as the green curve. Top-right panel: the bivariate function $m^*(x_1, x_2)=\{m(x_1)+m(x_2)\}/2$ as the curved light green surface, with its value evaluated at $(x, x)$, i.e., $m^*(x, x)=m(x)$, highlighted as the green curve, overlaid with the pool-level data $\{(X_{j1}, X_{j2}, Z_j)\}_{j=1}^J$ as blue circles. Bottom-left panel: the shape of the average kernel weights $\{[K(\{X_{j1}-x\}/h)+K(\{X_{j2}-x\}/h)]/2\}_{j=1}^J$ when $x=0$, along with $m^*(x,x)$ and the pool-level data. Bottom-right panel: the shape of the product kernel weights $\{[K(\{X_{j1}-x\}/h)K(\{X_{j2}-x\}/h)]/2\}_{j=1}^J$ when $x=0$, along with $m^*(x,x)$ and the pool-level data. \label{f:random4}}
\end{figure}

Even though $\hat m_2(x)$ exploits a more sensible weight function when comparing with $\hat m_1(x)$ under random pooling, downplaying $X_{j1}$ even when it is close to $x$ simply because the covariate value of the other individual in the same pool is far away from $x$ is not an efficient use of data. And such waste of data information is more severe when the pool size is bigger, which is essentially the curse-of-dimensionality when one estimates the multivariate function $m^*(x\bone_c)$ based on a response along with a $c$-dimensional covariate. It is such inefficient use of data that causes the much inflated variance concluded in Theorem~\ref{th:randomp} for $\hat m_2(x)$. Figure~\ref{f:random4} illustrates the average weight function and the product weight function (in bottom panels) under random pooling when $c=2$ and $K(t)$ is the Epanechnikov kernel. Also shown in Figure~\ref{f:random4} (see the top-left panel) are individual-level data generated according to the model specified in (D1) described in Section~\ref{s:empirical}, overlaid with the pseudo response data from random pooling, which are used for the construction of $\hat m_3(x)$. From there one can see that the pseudo data, $\{(\hat Y_{jk}, X_{jk}), k=1, 2\}_{j=1}^J$, are much more variable than the original data used to obtain $\hat m_0(x)$, and thus the increased variance of $\hat m_3(x)$ is expected when compared with $\hat m_0(x)$. Despite the higher variability, the pseudo data cloud does preserve the overall pattern of the original data cloud, which explains the common dominating bias shared between $\hat m_3(x)$ and $\hat m_0(x)$. Unlike $Q_1(\bbeta)$ and $Q_2(\bbeta)$, the construction of $Q_3(\bbeta)$ in (\ref{eq:Q3}) is directly designed for estimating the univariate function $m(x)$ instead of $m^*(x\bone_c)$, and thus $\hat m_3(x)$ overcomes the pitfall of misleading weight assignment in $\hat m_1(x)$, as well as the curse-of-dimensionality that $\hat m_2(x)$ suffers. 

\begin{figure} 
	\centering
	\setlength{\linewidth}{1\textwidth}
	\includegraphics[width=\linewidth]{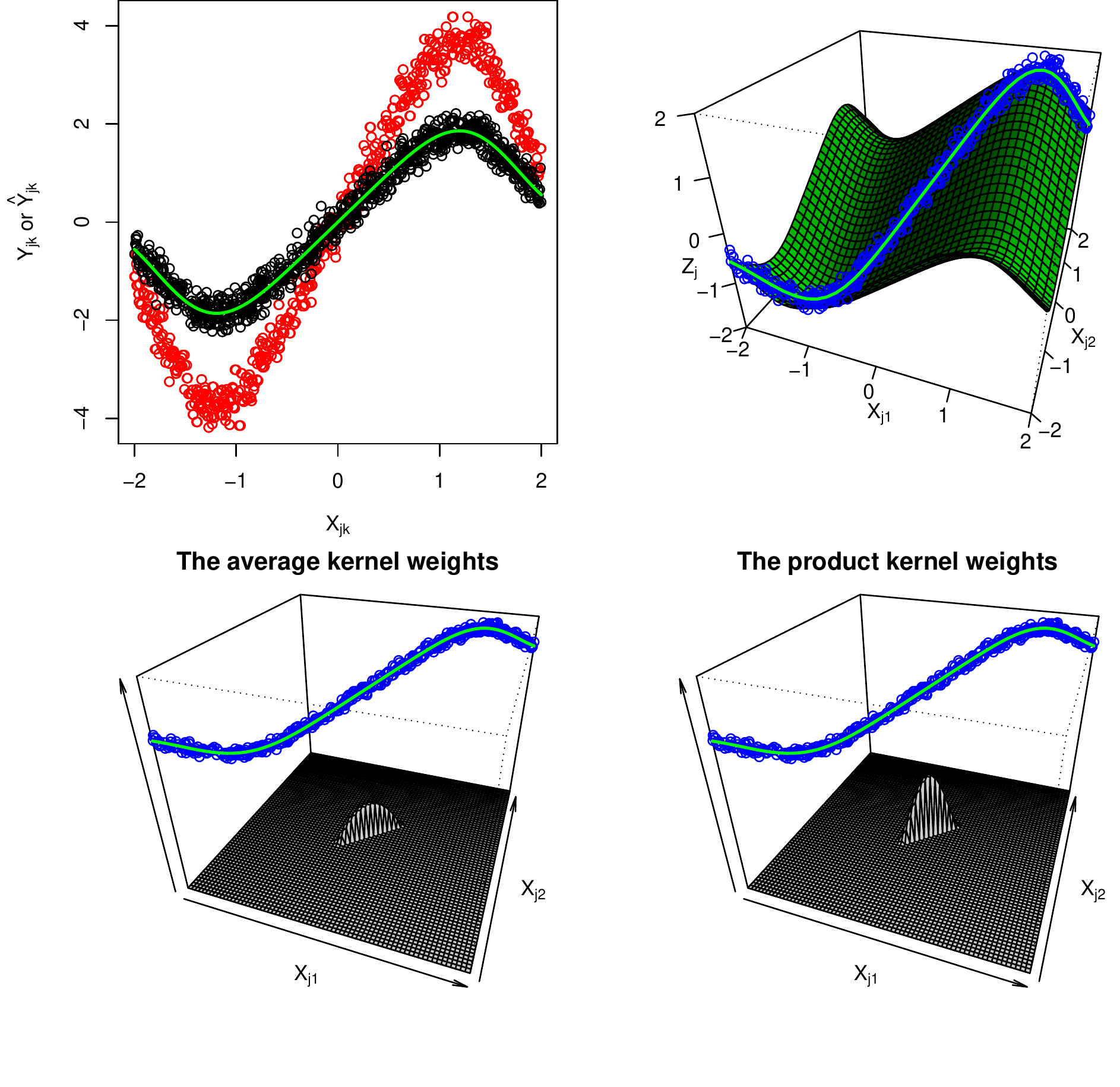} 
	\caption{\linespread{1.3}\selectfont{}Plots under homogeneous pooling. Top-left panel: Individual-level data $\{(Y_{(jk)}, X_{(jk)}), k=1, 2\}_{j=1}^J$ as black circles, pseudo individual-level responses and covariate data $\{(\hat Y_{(jk)}, X_{(jk)}), k=1, 2\}_{j=1}^J$ as red circles, overlaid with the true $m(x)$ as the green curve. Top-right panel: the bivariate function $m^*(x_1, x_2)=\{m(x_1)+m(x_2)\}/2$ as the curved light green surface, with its value evaluated at $(x, x)$, i.e., $m^*(x, x)=m(x)$, highlighted as the green curve, overlaid with pool-level data $\{(X_{j1}, X_{j2}, Z_j)\}_{j=1}^J$ as blue circles. Bottom-left panel: the shape of the average kernel weights $\{[K(\{X_{j1}-x\}/h)+K(\{X_{j2}-x\}/h)]/2\}_{j=1}^J$ when $x=0$, along with $m^*(x,x)$ and the pool-level data. Bottom-right panel: the shape of the product kernel weights $\{[K(\{X_{j1}-x\}/h)K(\{X_{j2}-x\}/h)]/2\}_{j=1}^J$ when $x=0$, along with $m^*(x,x)$ and the pool-level data. \label{f:homo4}}
\end{figure}

Figure~\ref{f:homo4} is the counterpart of Figure~\ref{f:random4} under homogeneous pooling. One can see (in the top-left panel) in Figure~\ref{f:homo4} that the pseudo data, $\{(\hat Y_{(jk)}, X_{(jk)}), k=1, 2\}_{j=1}^N$, clearly distort the original data pattern, and thus are inappropriate for estimating $m(x)$. With individuals sharing similar covariates values gathering in the same pool, the concern relating to $\hat m_1(x)$ of assigning inadequate weight no longer exists, neither does the concern relating to $\hat m_2(x)$ of inefficient use of data. The bottom panels of Figure~\ref{f:homo4} depict the average weight function and the product weight function, both are reminiscent of some symmetric kernel function. 

\section{Bandwidth selection}
\label{s:bandwidth}
The choice of bandwidths in local polynomial estimators plays a key role in the performance of these estimators. Besides the usual challenges encountered in bandwidth selection in local polynomial regression, a unique complication we face here is the lack of individual-level response data, which makes loss functions used for bandwidth selection that are based on individual-level residuals (or prediction errors) inapplicable in our context. Next we develop leave-one-out cross-validation (CV) procedures to choose bandwidths in three proposed local polynomial estimators for $m(x)$ using random pooled data. 
  
For the average-weighted estimator, $\hat m_1(x)$, we choose the bandwidth $h$ that minimizes the following pool-level residual sum of squares, 
\begin{align}
\textrm{RSS}_1(h) & = \sum_{j=1}^J \sum_{k=1}^{c_j} \left\{Z_j-c_j^{-1}\sum_{k=1}^{c_j} \hat m_{1, h}(X_{jk}) \right\}^2, \label{eq:RSS1}
\end{align}	
where $\hat m_{1,h}(X_{jk})$ is the realization of $\hat m_1(X_{jk})$ based on the observed data $(\bZ, \mathbb{X})$ excluding data from pool $j$, $(Z_j, \tilde \bX_j)$, with the bandwidth set at $h$.
The bandwidth in the product-weighted estimator, $\hat m_2(x)$, is chosen by minimizing a CV criterion similarly defined as (\ref{eq:RSS1}),	
\begin{align}
\textrm{RSS}_2(h) & = \sum_{j=1}^J \sum_{k=1}^{c_j} \left\{Z_j-c_j^{-1}\sum_{k=1}^{c_j} \hat m_{2, h}(X_{jk}) \right\}^2. \label{eq:RSS2}
\end{align}	
Admittedly, CV criteria or loss functions constructed based on prediction errors at the pool level may not be sensitive to the influence of $h$ on prediction power at the individual level, and thus may not serve as effective model criteria for the purpose for choosing bandwidths.

Given (\ref{eq:RSS1}) and (\ref{eq:RSS2}), one can easily envision a similar CV criterion, denoted by $\textrm{RSS}_3(h)$, defined for choosing $h$ in $\hat m_3(x)$. We however take into account the close tie between $\hat m_3(x)$ and local polynomial estimators designed for individual-level data, and propose a new and more effective CV criterion. This new criterion tailored for $\hat m_3(x)$ is mostly thanks to the pseudo individual-level observations, $\{(\hat Y_{jk}, X_{jk}), \, k=1,\dots, c\}_{j=1}^J$, used in $\hat m_3(x)$. In particular, we choose $h$ used in $\hat m_3(x)$ that minimizes the following pseudo (individual-level) residual sum of squares, 
\begin{align}
\textrm{PRSS}_3(h) & ={\sum_{j=1}^J \sum_{k=1}^{c_j}} \{\hat Y_{jk}-\hat m_{3,h}(X_{jk})\}^2, \label{eq:PRSS3}
\end{align}
where $\hat m_{3,h}(X_{jk})$ is the realization of $\hat m_3(X_{jk})$ based on the pseudo individual-level data excluding one pseudo data point, $(\hat Y_{jk}, X_{jk})$, with the bandwidth set at $h$. Empirical evidence suggest that $\textrm{PRSS}_3(h)$ is a more effective CV criterion for bandwidth selection than $\textrm{RSS}_3(h)$.  

When one is concerned about undesirable boundary effects on the prediction assessment of a fitted model, and in turn on the choice of $h$, one may exclude data near the boundary in the pseudo individual-level prediction in (\ref{eq:PRSS3}). For instance, one may only consider computing $\hat m_{3,h}(X_{jk})$ when $X_{jk}\in [a, b]$, where $a$ and $b$ are the $2.5$th and $97.5$th sample quantiles of observed covariate data, respectively. Similar data exclusion strategy can be employed in computing (\ref{eq:RSS1}) and (\ref{eq:RSS2}). Lastly, when homogeneous pooled data are used, we employ the same CV criteria defined above, although evaluated at homogeneous pooled data, to choose bandwidths. 

\section{Simulation study}
\label{s:empirical}
\subsection{Design of simulation experiments}
To compare different estimators of $m(x)$ in regard to their finite sample performance, and to explore other factors that may influence the estimation, we carry out an empirical study using synthetic data. More specifically, we adopt the following data generating processes reported in \citet{delaigle2009design} to generate individual-level response data: 
\begin{itemize}
\item[(D1)] $m(x)=x^3 \exp(x^4/1000)\cos x$, $\epsilon \sim N(0, 0.6^2)$, $X\sim 0.8 X_1+0.2 X_2$, where $X_1$ follows a distribution with pdf given by $0.1875 x^2 I(-2\le x \le 2)$ and $X_2\sim \textrm{uniform}(-1, 1)$; 
\item[(D2)] $m(x)=2x \exp(-10 x^4/81)$, $\epsilon \sim (0, 0.2^2)$, $X\sim 0.8 X_1+0.2 X_2$, where the distributions of $X_1$ and $X_2$ are as those specified in (D1); 
\item[(D3)] $m(x)=x^3$, $\epsilon \sim N(0, 1.2^2)$, $X\sim N(0, 1)$; 
\item[(D4)] $m(x)=x^4$, $\epsilon \sim N(0, 4^2)$, $X\sim N(0, 1)$. 
\end{itemize}

Under each data generating process, we generate individual-level data of size $N=600$, $\{(Y_i, X_i)\}_{i=1}^N$. Given an individual-level data set, we create pooled data, first using random pooling and then using homogeneous pooling, with a common pool size $c=2, 3, 4, 5, 6$ across all $J$ pools. Given each pooled data set, we obtain three local linear estimates for the mean function, $\hat m_1(x)$, $\hat m_2(x)$, and $\hat m_3(x)$. In addition, we also compute the local linear estimate using individual-level data, $\hat m_0(x)$, as a benchmark estimate. In all four estimators, we set $K(t)$ as the Epanechnikov kernel. The empirical integrated squared error (ISE) is the metric we use to assess the quality of an estimated mean function, defined by $\textrm{ISE}=\sum_{j=1}^J\sum_{k=1}^c \{Y_{jk}-\hat m(X_{jk})\}^2$ for an estimator $\hat m(\cdot)$.

\subsection{Simulation results}
\begin{landscape}
\begin{figure} 
	\centering
	\setlength{\linewidth}{1.4\textwidth}
	\includegraphics[width=\linewidth]{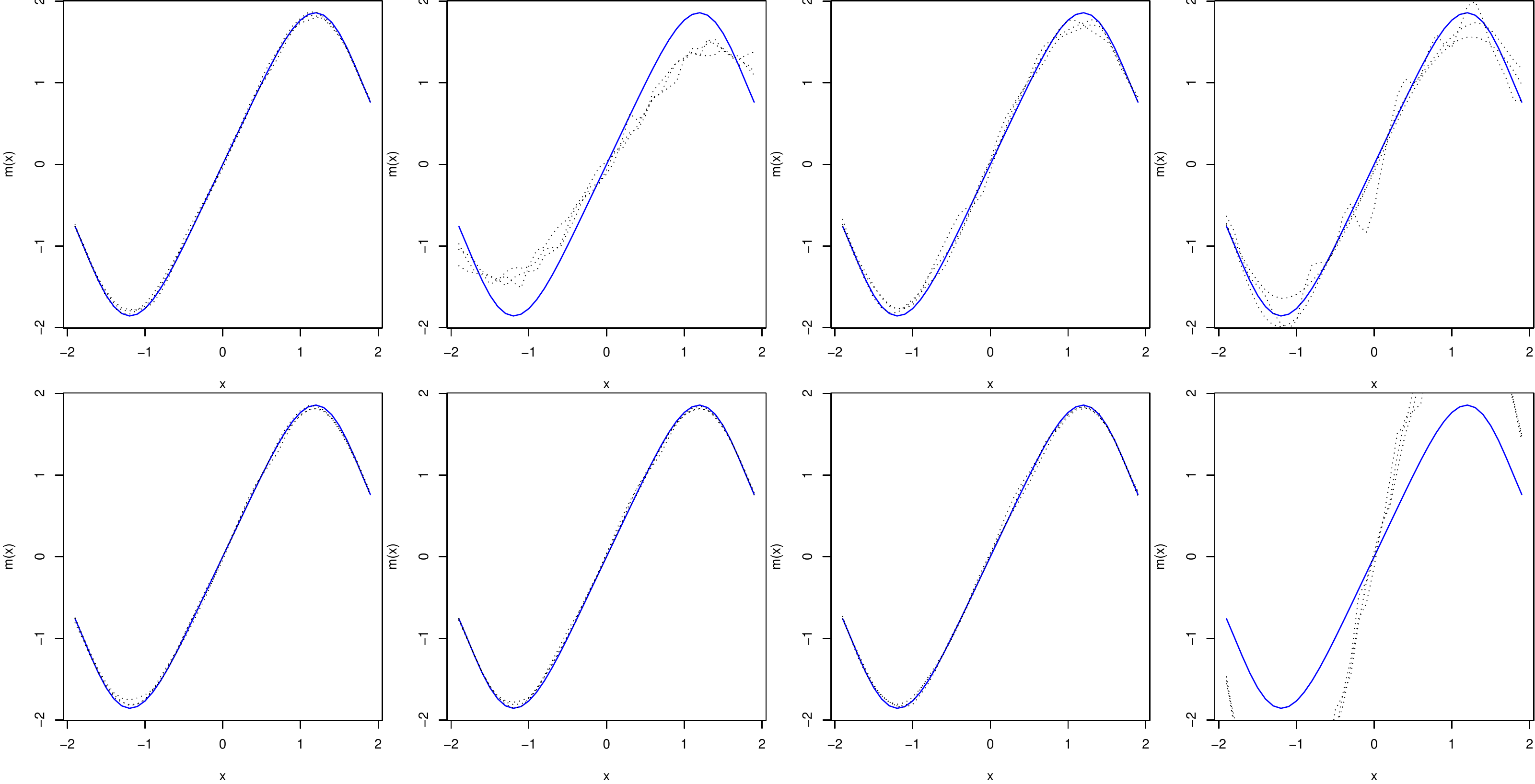} 
	\caption{\linespread{1.3}\selectfont{}Four estimates for $m(x)$ under (D2): the local linear estimate based on individual-level data (the first column), $\hat m_0(x)$, the average-weighted estimate (the second column), $\hat m_1(x)$, the product-weighted estimate (the third column), $\hat m_2(x)$, and the marginal-integration estimate (the fourth column), $\hat m_3(x)$. The latter three estimates are based on random pooled data in the upper panels, and are based on homogeneous pooled data in the lower panels. Within each panel, the blue curve is the true function $m(x)$, and the three dotted lines are three realizations of the estimator depicted in that panel whose ISE's are equal to the three quartiles among the 500 realizations of ISE's associated with the estimator.}
	\label{f:D2}
\end{figure}
\end{landscape}

Figure \ref{f:D2} depicts the three proposed estimators when $c=2$ along with the benchmark estimate $\hat m_0(x)$ using data generated according to (D2). Appendix E provides parallel results under the other three designs of data generating processes and those when $c=6$. 

Under random pooling (see upper panels of Figure \ref{f:D2}), the average-weighted estimator $\hat m_1(x)$ is unable to capture the shape of $m(x)$, and it fails more miserably around regions with more curvature. The product-weighted estimator $\hat m_2(x)$ is able to recover the overall shape of $m(x)$, although exhibiting a higher variability than $\hat m_0(x)$, especially around the inflection points of $m(x)$. With $c=2$, the marginal-integration estimator $\hat m_3(x)$ performs similarly as $\hat m_2(x)$. When one increases $c$ (see Appendix E), one can see that $\hat m_3(x)$ shows a much more stable performance in estimating $m(x)$ than $\hat m_2(x)$ does. This is in line with the implication of Theorem~\ref{th:randomp} that the variance of $\hat m_2(x)$ inflates faster as the pool size increases than the variance of $\hat m_3(x)$ does. 

Under homogeneous pooling (see lower panels of Figure \ref{f:D2}), the marginal-integration estimator $\hat m_3(x)$ distorts the functional form of $m(x)$, whereas both $\hat m_1(x)$ and $\hat m_2(x)$ perform similarly as $\hat m_0(x)$, in regard to both accuracy and precision.

\section{Real-life applications}
\label{s:realdata}
In this section, we analyze data from two real-life applications to illustrate the proposed local linear estimators for a conditional mean function. The individual-level observations are available in both applications, making it feasible to compute the local linear estimate based on individual-level data, $\hat m_0(x)$, which we compare our proposed estimates based on pooled data with. In all considered estimators, we set $K(t)$ as the Epanechnikov kernel. 

\noindent 
{\it Example 1 (Perfluorinated chemicals):} The first data set is from the National Health and Nutrition Examination Survey, relating to a study of the bioaccumulation of perfluorinated chemicals (PFCs) in human bodies. PFCs are widely used in the coating of industrial products, such as food packaging foams and non-stick cookware surfaces, many of which are toxic and accumulate in human bodies. \citet{karrman2006levels} studied the relationship between the concentration levels of PFCs in an individual's blood and one's age, gender, and geographic region using pooled serum samples of individuals in Australia. The particular data we entertain here include concentration levels of multiple PFCs in the serum samples of 1,904 residents in the United States between 2011 and 2012, along with their demographic information. The goal of our analysis is to infer the relationship between the concentration level of one particular type of PFCs, perfluorohexane sulfonic acid (PFHxS, $Y$), in an individual's blood and his/her age ($X$). 

To assess the uncertainty of each estimation procedure, we generate 500 bootstrap samples from the raw individual-level data. Based on each bootstrap version of the individual-level data, we compute the local linear estimate, $\hat m_0(x)$, for the mean concentration level of PFHxS given one's age. Additionally, using the original data, we randomly create 952 pools, each of size two, producing a set of random pooled data; and we also create 952 pools of equal size based on the sorted data for age, producing a set of homogeneous pooled data. With the pool composition under each pooling design fixed, 500 bootstrap versions of random pooled data, and 500 bootstrap versions of homogeneous pooled data are generated by resampling pools with replacement. Using each pooled data set, we compute $\hat m_1(x)$, $\hat m_2(x)$, and $\hat m_3(x)$, resulting in 500 realizations of each estimator. 

\begin{figure} 
	\centering
	\includegraphics[width=\linewidth]{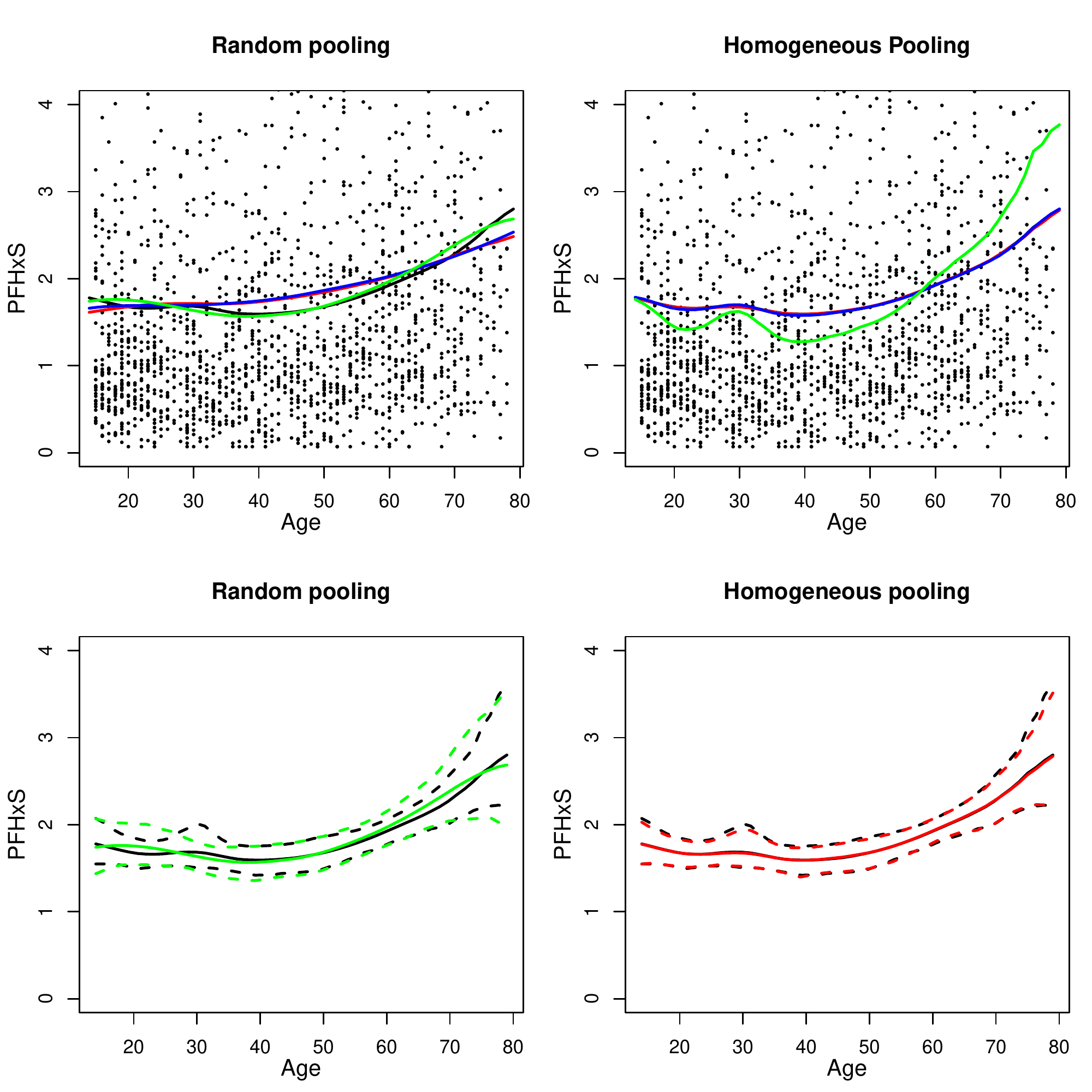} 
	\caption{\linespread{1.3}\selectfont{}Results from Example 1 (Perfluorinated chemicals). Top panels depict the average of each considered estimate across 500 bootstraps. The black dots are individual observations, with observations far larger than 4 omitted. Within each panel, the solid black line corresponds to the local linear estimate based on individual-level data, $\hat m_0(x)$; the solid red, blue, and green lines correspond to the average-weight estimate $\hat m_1(x)$, the product-weighted estimate $\hat m_2(x)$, and the marginal-integration estimate $\hat m_3(x)$, respectively. Bottom panels show two quantiles of the estimates across 500 bootstraps. The dashed black, red, and green lines are $5\%$ and $95\%$ quantiles of $\hat m_0(x)$, $\hat m_1(x)$, and $\hat m_3(x)$, respectively.}
	\label{f:nhanes-combined-revised}
\end{figure}

Figure~\ref{f:nhanes-combined-revised} depicts the average of each estimate across 500 bootstrap samples and two quantiles of selected estimates. When random pooled data are used, the marginal-integration estimate $\hat m_3(x)$ matches closely with the benchmark estimate based on individual-level data, $\hat m_0(x)$, both indicating a relatively stable level of PFHxS with a slight decrease as one approaches age 40, and then a steep increase of the concentration level once one passes around age 50. This pattern can be explained by the fact that PFHxS can be partly eliminated from the human body via, for instance, gastrointestinal activities, menstrual bleeding, and breast feeding \citep{genuis2013gastrointestinal}, but many of these pathways of PFCs elimination become less proactive or are completely lost (such as due to menopause) after one reaches certain age. In contrast, the average-weighted estimate, $\hat m_1(x)$, and the product-weighted estimate, $\hat m_2(x)$, suggest a much slower and nearly a constant increase in the concentration level as one gets older across the entire observed age range. We believe that this is one case where $\hat m_1(x)$ fails to capture the underlying pattern of $m(x)$ due to its inherent inconsistency in estimation, and $\hat m_2(x)$ also misses this pattern due to its high uncertainty in estimation. In conclusion, when only random pooled data are available, $\hat m_3(x)$ provides a more reliable estimate for the underlying relationship between one's PFHxS level in blood and age than the other two proposed estimates, although its variability is slightly higher than that of $\hat m_0(x)$ according to the bootstrap quantiles of the two estimates.

When homogeneous pooled data are used (see the top-right panel of Figure~\ref{f:nhanes-combined-revised}), $\hat m_3(x)$ appears to exaggerate the curvature of the conditional mean function, resulting in a much faster increase in the concentration level once one passes age 50, compared to the rate of increase indicated by the same estimate under random pooling. Despite the use of pooled data, $\hat m_1(x)$ and $\hat m_2(x)$ are nearly indistinguishable from $\hat m_0(x)$, and these three estimates mostly preserve the earlier estimated pattern of $m(x)$ that can be justified on scientific grounds. Moreover, the variability of $\hat m_1(x)$ is comparable with that of $\hat m_0(x)$ according to the comparison of the bootstrap quantiles associated with these two estimates. In conclusion, the marginal-integration estimate $\hat m_3(x)$ based on homogeneous pooled data leads to misleading inference for the underlying truth, whereas the other two estimates based on pooled data provide inference similar to those from the estimate based on individual-level data without noticeable efficiency loss.

\noindent 
{\it Example 2 (Chemokines):} The second data set we use to illustrate local linear estimation using different types of data is from the Collaborative Perinatal Project (CPP), which is a long-standing, collaborative project on maternal and child health in the United States. More specifically, this data  include chemokine levels collected from 388 pregnant females recruited in CPP, with measurements taken at the individual level as well as the pool level, with 194 non-overlapping pools of size two randomly formed. Chemokines are a family of small proteins related to the homeostatic and inflammatory process in the human body. Medical researchers have studied extensively the role that chemokines play in the immune system. For example, regarding to two particular chemokines, MCP-3 and GRO-$\alpha$, \citet{dhawan2002role} investigated the role of the former in tumorigenesis, and \citet{tsou2007critical} studied the latter in monocyte mobilization.

\begin{figure} 
	\centering
	\includegraphics[width=.9\textwidth]{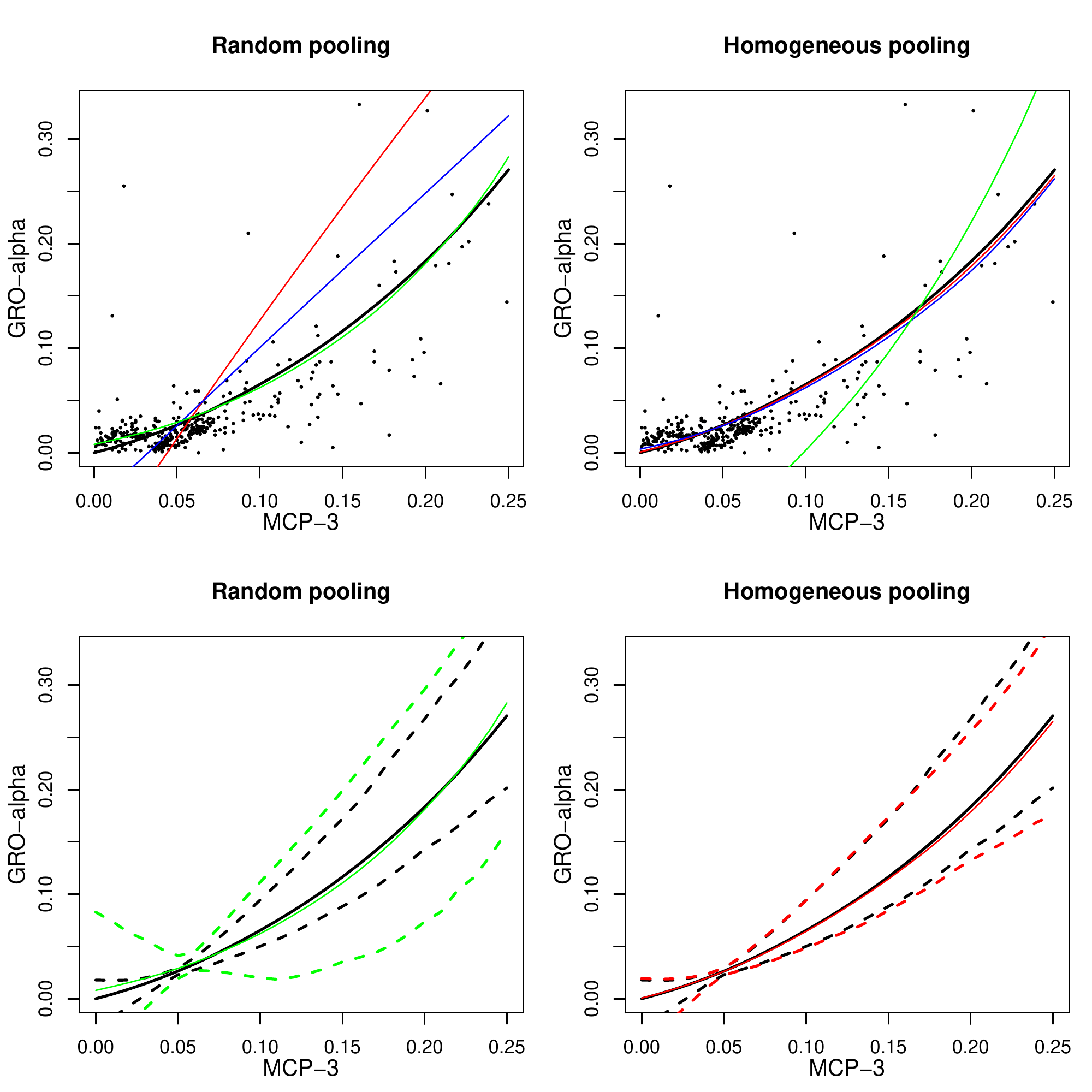} 
	\caption{\linespread{1.3}\selectfont{}Results from Example 2 (Chemokines). Top panels depict the average of each considered estimate across 500 bootstraps. The black dots are individual observations. Within each panel, the solid black line corresponds to the local linear estimate based on individual-level data, $\hat m_0(x)$; the solid red, blue, and green lines correspond to the average-weight estimate $\hat m_1(x)$, the product-weighted estimate $\hat m_2(x)$, and the marginal-integration estimate $\hat m_3(x)$, respectively. Bottom panels show two quantiles of the estimates across 500 bootstraps. The dashed black, red, and green lines are $5\%$ and $95\%$ quantiles of $\hat m_0(x)$, $\hat m_1(x)$, and $\hat m_3(x)$, respectively.}
	\label{f:CPP-combined-revised}
\end{figure}

Based on the observed individual-level data and the random pooled data available in CPP, we infer the conditional mean concentration of GRO-$\alpha$ ($Y$) given MCP-3 ($X$). For illustration purpose, we generate another pooled data set, with a common pool size of two, following the homogeneous pooling design based on sorted MCP-3 levels. To assess the uncertainty of each estimation method, we generate 500 bootstrap samples for each of the three data types, individual-level data, random pooled data, and homogeneous pooled data, following the same resampling process described in the first example. Figure~\ref{f:CPP-combined-revised} shows the average of each considered estimate across 500 bootstrap samples and two quantiles of selected estimates. 

Similar to the phenomena in the first example, the marginal-integration estimate $\hat m_3(x)$ yields a similar estimate for the mean concentration level of GRO-$\alpha$ given the level of MCP-3 as that of $\hat m_0(x)$ when random pooled data are used; but it grossly deviates from this benchmark estimate when homogeneous pooled data are used. In contrast, the other two proposed local linear estimates based on random pooled data go through an obviously uninteresting region of the observed data, yet both estimates applied to homogeneous pooled data follow closely the benchmark estimate $\hat m_0(x)$, and they only show slight discrepancy from it around the region where data are relatively scarce. 

\section{Discussion}
\label{s:discussion}
We present in this article methods for estimating the mean of a continuous response given covariates via local polynomial regression when only pooled response data are observed along with individual-level covariates. Two commonly adopted pooling designs in practice are considered when formulating the local polynomial estimators, and properties of the proposed estimators are compared under each of the pooling designs. We use two real-life applications to illustrate the implementation and performance of the proposed estimators in comparison with their counterpart estimator when individual response data are available. Findings from the two applications are in line with observations on their finite sample performance using synthetic data from the simulation study, which agree with the theoretical implications of the large-sample properties derived for the proposed estimators. 

In summary, the marginal-integration estimator $\hat m_3(x)$ is the winner among the three proposed when pooled data are from a random pooling design, but it fails when pools are not formed randomly; the average-weighted estimator $\hat m_1(x)$ performs the best when homogeneous pooled data are used, but it is an inconsistent estimator for the mean function when pools are formed randomly; the product-weighted estimator $\hat m_2(x)$ is a consistent estimator under both pooling designs, but is subject to high variability under random pooling. Based on our discussions in Section~\ref{s:remarks}, we believe that there is still room for improvement by more carefully/selectively incorporating individual covariate information within a pool to relate to the pooled response of that pool, as opposed to either using all covariate information (as in $\hat m_1(x)$ and $\hat m_2(x)$) or using one individual's covariate information (as in $\hat m_3(x)$). Following this more selective incorporation of covariate information for each pool, an alternative construction of the weight function in the objective function may be needed accordingly to exploit a more sensible measure of distance between selected individuals' covariate information and $x$, the value at which the mean function is of interest. We are hopeful that this more refined strategy for constructing the objective function can lead to a local polynomial estimator that outperforms all three estimators proposed in the current study despite the pooling design. 

Another follow-up research is motivated by the fact that, in many applications, covariates of interest cannot be measured precisely or observed directly. It is of interest then to carry out local polynomial regression to infer $m(x)$ using pooled response data and individual-level covariate data that are prone to measurement error.
 
\setcounter{equation}{0} 
\def\theequation{A.\arabic{equation}}
\renewcommand\thesubsection{A.\arabic{subsection}}
\section*{Appendix A: Proof of Theorem~1-(i)}
\subsection{Regularity conditions}
Define the variance function as $\sigma^2(x)=\textrm{Var}(Y|X=x)$, and denote by $f_{\hbox {\tiny $X$}}(x)$ the probability density function (pdf) of $X$. The following conditions imposed on $m(x)$, $\sigma^2(x)$, $f_{\hbox {\tiny $X$}}(x)$, and $K(t)$ are assumed throughout the appendices. 
\begin{itemize}
\item[(C1)] The mean function $m(x)$ is $(p+3)$-th order continuously differentiable.
\item[(C2)] The variance function $\sigma^2(x)$ is second order continuously differentiable.
\item[(C3)] The density function $f_{\hbox {\tiny $X$}}(x)$ is second order continuously differentiable, and $f(x)>0$. 
\item[(C4)] The kernel function $K(t)$ is an even function. 
\item[(C5)] $\mu_\ell=\int t^\ell K(t)dt$ and $\nu_\ell=\int t^\ell K^2(t) dt$ are well-defined for $\ell=0, 1, \ldots, 2p$.
\end{itemize}
All integrals are over the entire real line $\mathbb{R}$ in this article. 

Conditions (C1)--(C3) relate to certain degree of smoothness of the mean function, the variance function, and the density function of $X$ that are typically required for establishing consistency of local polynomial estimators when individual-level data are available. Conditions (C4) and (C5) are also commonly imposed assumptions for the consistency of kernel density estimators besides local polynomial estimators for a smooth function. Additional conditions associated with a particular estimator under a specific pooling design are stated in the relevant theorem when needed.  

The average-weighted $p$-th order local polynomial estimator of $m(x)$, denoted by $\hat m_1(x)$, results from minimizing the following objective function, 
\begin{equation*}
Q_1(\bbeta) = \sum_{j=1}^J \left\{Z_j-\sum_{\ell=0}^p \beta_\ell c_j^{-1}\sum_{k=1}^{c_j}(X_{jk}-x)^\ell \right\}^2 \left\{c_j^{-1}\sum_{k=1}^{c_j}K_h(X_{jk}-x)\right\},
\end{equation*}
where $\beta_\ell=m^{(\ell)}(x)/\ell!$, for $\ell=0, 1, \ldots, p$. Define $\bbeta=(\beta_0, \beta_1, \ldots, \beta_p)^\T$, with its dependence on $x$ suppressed. Equivalently, one has $\hat m_1(x)  = \be_1^\T \bS_1^{-1}(x)\bT_1(x)$,
where $\bS_1(x)=\bD_1(x)^\T \bK_1(x) \bD_1(x)=\left[S_{1,\ell_1,\ell_2}(x)\right]_{\ell_1, \ell_2=0, 1, \ldots, p}$, and $\bT_1(x)=\bD_1(x)^\T \bK_1(x) \bZ = (T_{1,0}(x),\,  T_{1,1}(x), \, \ldots, \, T_{1, p}(x))^\T$, in which $
\bZ=(Z_1, \ldots, \, Z_J)^\T$, 
\begin{equation}
\label{eq:form1}
\begin{aligned}
\bD_1(x) & = 
\begin{bmatrix}
1      & \bar X_1-x &  c_1^{-1}\sum_{k=1}^{c_1}(X_{1k}-x)^2  & \ldots & c_1^{-1}\sum_{k=1}^{c_1}(X_{1k}-x)^p \\
\vdots & \vdots   & \vdots         & \ldots & \vdots \\
1      & \bar X_J-x &  c_J^{-1}\sum_{k=1}^{c_J}(X_{Jk}-x)^2  & \ldots & c_J^{-1}\sum_{k=1}^{c_J}(X_{Jk}-x)^p 
\end{bmatrix},\\
\bK_1(x) & =\textrm{diag}\left(c_1^{-1}\sum_{k=1}^{c_1}K_h(X_{jk}-x), \, \ldots, \, c_J^{-1}\sum_{k=1}^{c_J} K_h(X_{Jk}-x)\right),
\end{aligned}
\end{equation}
and $\bar X_j=c_j^{-1}\sum_{k=1}^{c_j} X_{jk}$, for $j=1, \ldots, J$. The entry on the $(\ell_1+1)$-th row and $(\ell_2+1)$-th column of $\bS_1(x)$ is, for $\ell_1, \ell_2=0, 1, \ldots, p$,
\begin{align}
&\ S_{1,\ell_1,\ell_2}(x)\nonumber \\
= &\ \sum_{j=1}^J \left\{c_j^{-1}\sum_{k=1}^{c_j} (X_{jk}-x)^{\ell_1}\right\} \left\{c_j^{-1}\sum_{k=1}^{c_j} (X_{jk}-x)^{\ell_2}\right\} \left\{c_j^{-1}\sum_{k=1}^{c_j}K_h(X_{jk}-x)\right\};\label{eq:S1entry} 
\end{align}
and the $(\ell+1)$-th entry of $\bT_1(x)$ is, for $\ell=0, 1, \ldots, p$,
\begin{equation} 
T_{1,\ell}(x) =\sum_{j=1}^J Z_j \left\{c_j^{-1}\sum_{k=1}^{c_j} (X_{jk}-x)^\ell\right\} \left\{c_j^{-1}\sum_{k=1}^{c_j}K_h(X_{jk}-x)\right\}.\label{eq:T1entry}
\end{equation}

In what follows, we study the mean and variance of $J^{-1}\bS_1(x)$ and $\bC^{(1)}_p(x)=J^{-1}\{\bT_1(x)-\bS_1(x)\bbeta\}$ in order to reveal dominating terms of $\hat \bbeta_1-\bbeta$, where $\hat \bbeta_1=\bS_1^{-1}(x)\bT_1(x)$. 

\subsection{Frequently used notations and results}
We first defined some frequently used notations and list some useful results to be referenced later. Throughout this document, $\ell$ and $p$ are non-negative integers. For a kernel $K(t)$, define 
\begin{align*}
\mu_\ell & =\int t^\ell K(t)dt, \quad \nu_\ell  =\int t^\ell K^2(t)dt,\\
\bmu^*_\ell & =(\mu_\ell, \mu_{\ell+1}, \ldots, \mu_{\ell+p})^\T, \quad \tilde\bmu_\ell = [\mu_{\ell_1+\ell_2+\ell}]_{\ell_1, \ell_2=0, 1, \ldots, p}, \\
\bnu^*_\ell & =(\nu_\ell, \nu_{\ell+1}, \ldots, \nu_{\ell+p})^\T, \quad 
\tilde\bnu_\ell = [\nu_{\ell_1+\ell_2+\ell}]_{\ell_1, \ell_2=0, 1, \ldots, p}. 
\end{align*}
All integrals in this document are over the entire real line unless otherwise stated. 
By definition, $\bmu^*_\ell$ is a $(p+1)\times 1$ vector, and $\tilde \bmu_\ell$ is a $(p+1)\times (p+1)$ matrix of which the entry on the $(\ell_1+1)$th row and $(\ell_2+1)$th column is $\mu_{\ell_1+\ell_2+\ell}$, for $\ell_1, \ell_2=0, 1, \ldots, p$. The structures of $\bnu^*_\ell$ and $\tilde\bnu_\ell$ are similar to those of $\bmu^*_\ell$ and $\tilde \bmu_\ell$, respectively. Moreover, because $K(t)$ is a kernel, $\mu_0=1$; and, with $K(t)$ being a symmetric kernel as we assume throughout the study, $\mu_\ell=\nu_\ell=0$ for an odd integer $\ell$.

Given the bandwidth $h$, define the following $(p+1)\times (p+1)$ matrices, $\bH =\diag(1, h, \ldots, h^p)$, and $\bH^* = [h^{2(\ell_1+\ell_2)}]_{\ell_1, \ell_2=0, 1, \ldots, p}$. Given the pool composition, define the following constants that are of order $O(1)$ as $J\to \infty$, 
\begin{align*}
t_{k0} & = \frac{1}{J}\sum_{j=1}^J \frac{1}{c^k_j}, \textrm{ for $k \in \{0, 1, 2\}$,}\\
t_{k_1 k_2} & =\frac{1}{J}\sum_{j=1}^J \frac{[\prod_{k=1}^{k_2} (c_j-k)]_+}{c_j^{k_1}}, \textrm{ for $k_1, k_2\in \{1, 2, 3\}$}, 
\end{align*}
where $[s]_+=\max(0, s)$ for a constant $s$. Clearly, when $c_j=1$ for all $j=1, \ldots, J$, $t_{k0}=1$ and $t_{k_1 k_2}=0$ for all $k, k_1, k_2$. 

For a real number $x$ in the support of $X$, define $\delta_\ell(x) =E\{(X-x)^\ell\}$ and $R_{\ell,p} = E\{(X-x)^\ell r_p(X, x)\}$, assuming that the expectations exist, where $r_p(X, x) = m(X)-\sum_{\ell=0}^p \beta_\ell (X-x)^\ell$. Then define the following vectors and matrices, 
\begin{align*}
\bDelta^*_{\ell}(x) & = (\delta_{\ell}(x), \delta_{\ell+1}(x), \ldots, \delta_{\ell+p}(x))^\T, \quad \tilde \bDelta_\ell(x)= [\delta_{\ell_1+\ell_2+\ell}(x)]_{\ell_1, \ell_2=0, 1, \ldots, p}, \\
\bR^*_p(x)& =(R_{0,p}(x), \ldots, R_{p,p}(x))^\T.
\end{align*}

Lastly, define the following expectations assuming they exist, for $\ell=0, 1, \ldots, 2p$, 
\begin{equation}
\label{eq:kappas} 
\left\{
\begin{aligned}
\kappa_{1,\ell}(h, x)& =  E\{(X-x)^\ell K_h(X-x)\}, \\
\kappa_{2,\ell}(h, x)& =  E\{(X-x)^\ell K^2_h(X-x)\},\\
\kappa^*_{1,\ell}(g, h, x) & = E\{g(X)(X-x)^\ell K_h(X-x)\},\\
\kappa^*_{2,\ell}(g, h, x) & = E\{g(X)(X-x)^\ell K^2_h(X-x)\},
\end{aligned}
\right.
\end{equation}
where $g(x)$ is a generic function that is fourth-order continuously differentiable. When it causes no confusion given the context, we suppress the dependence of these quantities on $(h, x)$ in some derivations later. For example, we sometimes write $\kappa^*_{1,\ell}(g)$ in place of $\kappa^*_{1, \ell}(g, h, x)$. 
Assuming that the density of $X$, $f_{\hbox {\tiny $X$}}(x)$, is fourth-order continuously differentiable, one can show the following regarding (\ref{eq:kappas}),
\begin{align}
\kappa_{1,\ell}(h,x) & = \sum_{q=0}^3 h^{\ell+q}\mu_{\ell+q}f^{(q)}_{\hbox {\tiny $X$}}(x)/q!+O\left(h^{\ell+4}\right) \label{eq:EXK} \\
\kappa_{2,\ell}(h,x) & = \sum_{q=0}^3\ h^{\ell+q-1} \nu_{\ell+q}f^{(q)}_{\hbox {\tiny $X$}}(x)/q!+ O\left(h^{\ell+3}  \right). \label{eq:EXK2} 
\end{align}
With $\ell=0$ , one has, for a positive integer $c$,
\begin{align}
\kappa^c_{1,0}(h,x) & = f^{c-1}_{\hbox {\tiny$X$}}(x)\left\{f_{\hbox {\tiny $X$}}(x)+c h^2 \mu_2 f_{\hbox {\tiny $X$}}''(x)/2  \right\}+O(h^4), \label{eq:EKc}\\
\kappa^c_{2,0}(h,x) & = h^{-c}\nu_0f^c_{\hbox {\tiny$X$}}(x)+O(h^{2-c}). \label{eq:EK2c}
\end{align}
Moreover, 
\begin{align}
\kappa^*_{1,\ell}(g, h, x) & = \ h^\ell \mu_\ell g(x) f_{\hbox {\tiny $X$}}(x)+h^{\ell+1}\mu_{\ell+1}\left\{g(x) f'_{\hbox {\tiny $X$}}(x)+g'(x) f_{\hbox {\tiny $X$}}(x)\right\}+\nonumber \\
& \quad\, h^{\ell+2}\mu_{\ell+2}\left\{g(x)f''_{\hbox {\tiny $X$}}(x)/2+g'(x) f'_{\hbox {\tiny $X$}}(x)+g''(x)f_{\hbox {\tiny $X$}}(x)/2 \right\}+\nonumber \\
& \quad\,O\left(h^{\ell+3} \right), \label{eq:EgXK}\\
\kappa^*_{2,\ell}(g, h, x) & = h^{\ell-1} \nu_\ell g(x) f_{\hbox {\tiny $X$}}(x)+h^\ell\nu_{\ell+1}\left\{g(x) f'_{\hbox {\tiny $X$}}(x)+g'(x) f_{\hbox {\tiny $X$}}(x)\right\}+\nonumber \\
& \quad\, h^{\ell+1}\nu_{\ell+2}\left\{g(x)f''_{\hbox {\tiny $X$}}(x)/2+g'(x) f'_{\hbox {\tiny $X$}}(x)+g''(x)f_{\hbox {\tiny $X$}}(x)/2 \right\}+\nonumber \\
& \quad\, O\left(h^{\ell+1} \right). \label{eq:EgXK2}
\end{align}

\subsection{About $J^{-1}\bS_1(x)$}
By (\ref{eq:S1entry}), one has 
\vspace{-0.25in}
\begin{align*}
E\{J^{-1}S_{1,\ell_1, \ell_2}(x)\} 
& = J^{-1} \sum_{j=1}^J E\left[\left\{c_j^{-1}\sum_{k=1}^{c_j} (X_{jk}-x)^{\ell_1}\right\} \left\{c_j^{-1}\sum_{k=1}^{c_j} (X_{jk}-x)^{\ell_2}\right\} \right.\\
& \quad \left.\times \left\{c_j^{-1}\sum_{k=1}^{c_j}K_h(X_{jk}-x)\right\}\right]\\
& = J^{-1}\sum_{j=1}^J c_j^{-3}E\left[\sum_{k=1}^{c_j} (X_{jk}-x)^{\ell_1+\ell_2} K_h(X_{jk}-x)\right. \\ 
& \quad + \sum_{k_1\ne k_2}\left\{(X_{jk_1}-x)^{\ell_1+\ell_2}K_h(X_{jk_2}-x)\right. \\
& \quad + (X_{jk_1}-x)^{\ell_1}(X_{jk_2}-x)^{\ell_2}K_h(X_{jk_2}-x)\\
& \quad \left. +(X_{jk_1}-x)^{\ell_1}(X_{jk_2}-x)^{\ell_2}K_h(X_{jk_1}-x)\right\} \\
& \quad \left. +\sum_{k_1\ne k_2 \ne k_3}(X_{jk_1}-x)^{\ell_1}(X_{jk_2}-x)^{\ell_2}K_h(X_{jk_3}-x)\right].
\end{align*}
Thus, 
\begin{align*}
E\{J^{-1}S_{1,\ell_1, \ell_2}(x)\} 
& = J^{-1}\sum_{j=1}^J c_j^{-3}\left[c_j \kappa_{1, \ell_1+\ell_2}(h, x)+c_j(c_j-1)\left\{\delta_{\ell_1+\ell_2}(x) \kappa_{1,0}(h, x) \right.\right. \\ 
& \quad \left. +\delta_{\ell_1}(x)\kappa_{1, \ell_2}(h, x)+\delta_{\ell_2}(x)\kappa_{1, \ell_1}(h, x)\right\}\\
& \quad \left. +c_j(c_j-1)(c_j-2) \delta_{\ell_1}(x)\delta_{\ell_2}(x)\kappa_{1,0}(h,x)\right].
\end{align*}
Using (\ref{eq:EXK}) in the last expression gives 
\begin{align}
& E\{J^{-1}S_{1,\ell_1, \ell_2}(x)\}\nonumber  \\
= &\ t_{20}\left\{ h^{\ell_1+\ell_2}\mu_{\ell_1+\ell_2}f_{\hbox {\tiny $X$}}(x)+h^{\ell_1+\ell_2+1}\mu_{\ell_1+\ell_2+1}f'_{\hbox {\tiny $X$}}(x)+0.5h^{\ell_1+\ell_2+2}\mu_{\ell_1+\ell_2+2}f''_{\hbox {\tiny $X$}}(x)\right.\nonumber\\
& \left. +O\left( h^{\ell_1+\ell_2+3} \right)\right\}+ t_{21}\left[\delta_{\ell_1+\ell_2}(x) \left\{\mu_0 f_{\hbox {\tiny $X$}}(x)+0.5 h^2 \mu_2 f''_{\hbox {\tiny $X$}}(x)+O(h^3)\right\} \right. \nonumber\\
&\ +\delta_{\ell_1}(x) \left\{h^{\ell_2}\mu_{\ell_2}f_{\hbox {\tiny $X$}}(x)+h^{\ell_2+1}\mu_{\ell_2+1}f'_{\hbox {\tiny $X$}}(x) + 0.5 h^{\ell_2+2}\mu_{\ell_2+2}f''_{\hbox {\tiny $X$}}(x)+O\left(h^{\ell_2+3}\right)\right\}\nonumber\\
& \left. +\delta_{\ell_2}(x) \left\{h^{\ell_1}\mu_{\ell_1}f_{\hbox {\tiny $X$}}(x)+h^{\ell_1+1}\mu_{\ell_1+1}f'_{\hbox {\tiny $X$}}(x) + 0.5 h^{\ell_1+2}\mu_{\ell_1+2}f''_{\hbox {\tiny $X$}}(x)+O\left(h^{\ell_1+3}\right)\right\}\right]\nonumber\\
&\ +t_{22}\delta_{\ell_1}(x)\delta_{\ell_2}(x)\left\{\mu_0 f_{\hbox {\tiny $X$}}(x)+0.5 h^2 \mu_2 f''_{\hbox {\tiny $X$}}(x)+O(h^3)\right\}\nonumber\\
= &\ \{t_{21}\delta_{\ell_1+\ell_2}(x) +t_{22}\delta_{\ell_1}(x)\delta_{\ell_2}(x)\}\left\{\mu_0 f_{\hbox {\tiny $X$}}(x)+0.5 h^2 \mu_2 f''_{\hbox {\tiny $X$}}(x)+O(h^3)\right\}\nonumber\\
&\ +t_{21}\left[\delta_{\ell_1}(x) \left\{h^{\ell_2}\mu_{\ell_2}f_{\hbox {\tiny $X$}}(x)+h^{\ell_2+1}\mu_{\ell_2+1}f'_{\hbox {\tiny $X$}}(x) + 0.5 h^{\ell_2+2}\mu_{\ell_2+2}f''_{\hbox {\tiny $X$}}(x)+O\left(h^{\ell_2+3}\right)\right\} \right. \nonumber\\
& \left.+ \delta_{\ell_2}(x) \left\{h^{\ell_1}\mu_{\ell_1}f_{\hbox {\tiny $X$}}(x)+h^{\ell_1+1}\mu_{\ell_1+1}f'_{\hbox {\tiny $X$}}(x) + 0.5 h^{\ell_1+2}\mu_{\ell_1+2}f''_{\hbox {\tiny $X$}}(x)+O\left(h^{\ell_1+3}\right)\right\}\right]\nonumber\\
&\ + t_{20}\left\{ h^{\ell_1+\ell_2}\mu_{\ell_1+\ell_2}f_{\hbox {\tiny $X$}}(x)+h^{\ell_1+\ell_2+1}\mu_{\ell_1+\ell_2+1}f'_{\hbox {\tiny $X$}}(x)+0.5h^{\ell_1+\ell_2+2}\mu_{\ell_1+\ell_2+2}f''_{\hbox {\tiny $X$}}(x)\right.\nonumber\\
& \left.+O\left( h^{\ell_1+\ell_2+3} \right)\right\}. \label{eq:ES1entry}
\end{align}
Hence, 
\begin{align}
& E\left\{J^{-1}\bS_1(x)\right\} \nonumber \\
= &\ \left\{\mu_0 f_{\hbox {\tiny $X$}}(x) +0.5 h^2 \mu_2 f''_{\hbox {\tiny $X$}}(x)+O(h^4)\right\}\left\{t_{21}\tilde \bDelta_0(x)+t_{22}\bDelta^*_0(x)\bDelta^{*\T}_0(x)\right\}+ \nonumber \\
&\ t_{20} \bH\left\{f_{\hbox {\tiny $X$}}(x) \tilde \bmu_0+hf'_{\hbox {\tiny $X$}}(x) \tilde \bmu_1+0.5 h^2  f''_{\hbox {\tiny $X$}}(x)\tilde\bmu_2+O(h^4)\right\}\bH + \nonumber \\
& \ t_{21}\left(\bDelta^*_0(x)\left\{f_{\hbox {\tiny $X$}}(x) \bmu^{*\T}_0+hf'_{\hbox {\tiny $X$}}(x)\bmu^{*\T}_1+0.5 h^2  f''_{\hbox {\tiny $X$}}(x)\bmu^{*\T}_2\right\}\bH+ \right. \nonumber \\
&\ \bH\left\{f_{\hbox {\tiny $X$}}(x) \bmu^*_0+hf'_{\hbox {\tiny $X$}}(x)\bmu^*_1+0.5 h^2  f''_{\hbox {\tiny $X$}}(x)\bmu^*_2\right\}\bDelta^{*\T}_0(x)+ \nonumber \\
& \left. +O\left[h^3\left\{\bDelta_0^*(x) \bmu_3^{*\T}\bH +\bH\bmu^*_3\bDelta_0^{*\T}(x)\right\}\right]\right). \label{eq:ES1}
\end{align}
If $c_j=1$ for $j=1, \ldots, J$, then $t_{20}=1$ and $t_{21}=t_{22}=0$, and thus (\ref{eq:ES1}) reduces to 
$$\bH\left\{f_{\hbox {\tiny $X$}}(x) \tilde \bmu_0+hf'_{\hbox {\tiny $X$}}(x) \tilde \bmu_1+0.5 h^2  f''_{\hbox {\tiny $X$}}(x)\tilde\bmu_2+O(h^4)\right\}\bH,$$
which is exactly the counterpart result in local polynomial regression using individual-level data. If $c_j>1$ for some pools, the first term (before $t_{20}$) in (\ref{eq:ES1}) remains there despite the order of the polynomial, i.e., the value of $p$. In fact, the dominating term in each entry of $E\{J^{-1}\bS_1(x)\}$, that is, (\ref{eq:ES1entry}), depend on $(\ell_1, \ell_2)$ only via their dependence on $\delta_{\ell}(x)$, for $\ell=\ell_1, \ell_2, \ell_1+\ell_2$, which are constant functions free of $h$. This suggests that orders of the dominating term in (\ref{eq:ES1entry}) remain the same for all $(\ell_1, \ell_2)$. This stands in stark contrast with the other two proposed estimators, $\hat m_2(x)$ and $\hat m_3(x)$, as to become clear in Appendices B and C. 

The order of $\textrm{Var}\{J^{-1}S_{1,\ell_1, \ell_2}(x)\}$ is determined the order of 
\begin{align}
& \ J^{-2} \sum_{j=1}^J E\left[\left\{c_j^{-1}\sum_{k=1}^{c_j} (X_{jk}-x)^{\ell_1}\right\}^2 \left\{c_j^{-1}\sum_{k=1}^{c_j} (X_{jk}-x)^{\ell_2}\right\}^2 \times\right. \nonumber \\
& \left.\left\{c_j^{-1}\sum_{k=1}^{c_j}K_h(X_{jk}-x)\right\}^2\right].
\label{eq:Eprod2m1}
\end{align}
According to (\ref{eq:EXK}) and (\ref{eq:EXK2}), the expectation as the summand of (\ref{eq:Eprod2m1}) is dominated by $E\{K_h^2(X-x)\}$ for all $(\ell_1, \ell_2)$, which is of order $O(h^{-1})$. It follows that (\ref{eq:Eprod2m1}) is of order $O\{1/(Jh)\}$, and so is $\textrm{Var}\{J^{-1}S_{1,\ell_1,\ell_2}(x)\}$. Along with (\ref{eq:ES1}), we now have   
\begin{align}
& J^{-1}\bS_1(x) \nonumber \\
= & \ \left\{\mu_0 f_{\hbox {\tiny $X$}}(x) +0.5 h^2 \mu_2 f''_{\hbox {\tiny $X$}}(x)+O(h^4)\right\}\left\{t_{21}\tilde \bDelta_0(x)+t_{22}\bDelta^*_0(x)\bDelta^{*\T}_0(x)\right\}+ \nonumber \\
& \ t_{20} \bH\left\{f_{\hbox {\tiny $X$}}(x) \tilde \bmu_0+hf'_{\hbox {\tiny $X$}}(x) \tilde \bmu_1+0.5 h^2  f''_{\hbox {\tiny $X$}}(x)\tilde\bmu_2+O(h^4)\right\}\bH + \nonumber \\
& \ t_{21}\left(\bDelta^*_0(x)\left\{f_{\hbox {\tiny $X$}}(x) \bmu^{*\T}_0+hf'_{\hbox {\tiny $X$}}(x)\bmu^{*\T}_1+0.5 h^2  f''_{\hbox {\tiny $X$}}(x)\bmu^{*\T}_2\right\}\bH+ \right. \nonumber \\
& \ \bH\left\{f_{\hbox {\tiny $X$}}(x) \bmu^*_0+hf'_{\hbox {\tiny $X$}}(x)\bmu^*_1+0.5 h^2  f''_{\hbox {\tiny $X$}}(x)\bmu^*_2\right\}\bDelta^{*\T}_0(x)+ \nonumber \\
& \left. +O\left[h^3\left\{\bDelta_0^*(x) \bmu_3^{*\T}\bH +\bH\bmu^*_3\bDelta_0^{*\T}(x)\right\}\right]\right)+O_{\hbox {\tiny $P$}}\left(\frac{1}{\sqrt{Jh}}\bI_{p+1} \right).
\label{eq:Sdecomm1}
\end{align}

It follows that 
\begin{align}
& \left\{J^{-1}\bS_1(x)\right\}^{-1} \nonumber \\
=& \ \bH^{-1}(\bA+h\bB+h^2\bC)^{-1}\bH^{-1} +O\left(h^4\bH^{-2}\right)+O_{\hbox {\tiny $P$}}\left(\frac{1}{\sqrt{Jh}} \right) \nonumber \\
= & \ \bH^{-1} \bA^{-1}\left\{\bA-h\bB+h^2\left(\bB\bA^{-1}\bB+\bC\right)-h^3\left(\bB \bA^{-1}\bB\bA^{-1}\bB+\bC\bA^{-1}\bB+\right. \right.\nonumber \\
& \left. \left. \bB\bA^{-1}\bC\right)\right\}\bA^{-1}\bH^{-1}+O\left(h^4\bH^{-2}\right)+O_{\hbox {\tiny $P$}}\left(\frac{1}{\sqrt{Jh}} \bI_{p+1} \right) ,
\label{eq:Sinvdecomm1}
\end{align}
\begin{align*}
\bA = &  \ f_{\hbox {\tiny $X$}}(x) \left[t_{20}\tilde \bmu_0 +\bH^{-1}\left\{t_{21}\tilde \bDelta_0(x)+t_{22}\bDelta_0^*(x)\bDelta_0^{*\T}(x)\right\}\bH^{-1}+  \right.  \\
& \left. t_{21}\left\{\bH^{-1}\bDelta_0^* (x) \bmu_0^{*\T} +\bmu^*_0 \bDelta_0^{*\T}(x) \bH^{-1} \right\}\right],\\
\bB = & \ f'_{\hbox {\tiny $X$}}(x) \left[ t_{20}\tilde \bmu_1 +  t_{21}\left\{\bH^{-1}\bDelta_0^* (x) \bmu_1^{T} +\bmu^*_1 \bDelta_0^{*\T}(x) \bH^{-1} \right\}\right],  \\
\bC = & \ 0.5f''_{\hbox {\tiny $X$}}(x) \left[ t_{20}\tilde \bmu_2 +\mu_2\bH^{-1}\left\{t_{21}\tilde \bDelta_0(x)+t_{22}\bDelta_0^*(x)\bDelta_0^{*\T}(x)\right\}\bH^{-1}+  \right. \\
& \left. t_{21}\left\{\bH^{-1}\bDelta_0^* (x) \bmu_2^{*\T} +\bmu^*_2 \bDelta_0^{*\T}(x) \bH^{-1} \right\}\right].
\end{align*}

\subsection{About $\bC^{(1)}_p(x)$}
\label{s:ECm1}
Write $\bC^{(1)}_p(x)$ as $(C^{(1)}_{p,0}(x), C^{(1)}_{p,1}(x), \ldots, C^{(1)}_{p, p}(x))^\T$. By (\ref{eq:S1entry}) and (\ref{eq:T1entry}), and using iterated expectations, one has, for $\ell=0, 1, \ldots, p$, 
\begin{align*}
& E\left\{C^{(1)}_{p,\ell}(x)\right\}\\
= &\ J^{-1}E\left[  \sum_{j=1}^J Z_j \left\{c_j^{-1} \sum_{k=1}^{c_j} (X_{jk}-x)^\ell\right\} \left\{c_j^{-1} \sum_{k=1}^{c_j} K_h(X_{jk}-x)\right\}-\right.\\
& \left. \sum_{\ell_2=0}^p \beta_{\ell_2}\sum_{j=1}^J \left\{c_j^{-1} \sum_{k=1}^{c_j} (X_{jk}-x)^\ell\right\}\left\{c_j^{-1} \sum_{k=1}^{c_j} (X_{jk}-x)^{\ell_2}\right\} \left\{c_j^{-1} \sum_{k=1}^{c_j} K_h(X_{jk}-x)\right\}\right]\\
= &\ J^{-1}E\left[  \sum_{j=1}^J \left\{c_j^{-1}\sum_{k=1}^{c_j} m(X_{jk}) \right\} \left\{c_j^{-1} \sum_{k=1}^{c_j} (X_{jk}-x)^\ell\right\} \left\{c_j^{-1} \sum_{k=1}^{c_j} K_h(X_{jk}-x)\right\}-\right.\\
& \left. \sum_{j=1}^J \left\{c_j^{-1} \sum_{k=1}^{c_j} (X_{jk}-x)^\ell\right\}  \left\{c_j^{-1} \sum_{k=1}^{c_j} K_h(X_{jk}-x)\right\}\sum_{\ell_2=0}^p \beta_{\ell_2}\left\{c_j^{-1} \sum_{k=1}^{c_j} (X_{jk}-x)^{\ell_2}\right\}\right]\\
= &\ J^{-1}\sum_{j=1}^J E\left[\left\{c_j^{-1} \sum_{k=1}^{c_j} (X_{jk}-x)^\ell\right\}  \left\{c_j^{-1} \sum_{k=1}^{c_j} K_h(X_{jk}-x)\right\} \left\{c_j^{-1} \sum_{k=1}^{c_j}r_p(X_{jk}, x)\right\} \right] \\
= &\ J^{-1}\sum_{j=1}^J c_j^{-3} \left[  c_j E\left\{(X-x)^\ell K_h(X-x)  r_p(X, x) \right\}+  \right. \\
  &\ c_j (c_j-1)E\left\{(X-x)^\ell K_h(X-x)\right\} E\left\{  r_p(X, x)\right\}+ \nonumber\\
  &\ c_j (c_j-1)E\left\{(X-x)^\ell\right\}E\left\{ K_h(X-x) r_p(X, x) \right\}+ \nonumber\\
  &\ c_j (c_j-1)E\left\{(X-x)^\ell  r_p(X, x) \right\}E\left\{K_h(X-x)\right\}+ \nonumber\\
  &\ \left. c_j (c_j-1)(c_j-2)E\left\{(X-x)^\ell\right\}E\left\{ r_p(X, x) \right\}E\left\{ K_h(X-x)\right\}\right].
\end{align*}
That is, 
\begin{align*}
& E\left\{C^{(1)}_{p,\ell}(x)\right\}\\
= &\ t_{20} E\left\{(X-x)^\ell  r_p(X, x) K_h(X-x)\right\} +
t_{21} R_{0,p}(x) E\left\{(X-x)^\ell K_h(X-x)\right\}+ \\
 &\ t_{21} \delta_\ell(x)E\left\{r_p(X,x) K_h(X-x)\right\}+ 
t_{21} E\left\{(X-x)^\ell r_p(X,x) \right\} E\left\{K_h(X-x)\right\}+ \\
 &\ t_{22}\delta_\ell(x) R_{0,p}(x) E\left\{K_h(X-x)\right\}\\
= &\ t_{20} \left\{\kappa^*_{1,\ell}(m, h, x)-\sum_{\ell_1=0}^p \beta_{\ell_1}\kappa_{1,\ell+\ell_1}(h,x)\right\} +t_{21} R_{0,p}(x) \kappa_{1,\ell}(h,x) +\\
& \ t_{21}\delta_\ell (x) \left\{\kappa^*_{1,0}(m, h, x)-\sum_{\ell_1=0}^p \beta_{\ell_1}\kappa_{1,\ell_1}(h,x)\right\}+t_{21} R_{\ell, p}(x)\kappa_{1,0}(h,x)+\\
& \ t_{22} \delta_\ell (x) R_{0,p}(x) \kappa_{1,0}(h,x). 
\end{align*}

Using (\ref{eq:EXK}) and (\ref{eq:EgXK}) in the above expression gives
\begin{align}
E\left\{C^{(1)}_{p,0}(x) \right\} = &\ t_{11} R_{0,p}(x) \left\{f_{\hbox {\tiny $X$}}(x)+0.5 h^2\mu_2 f^{(2)}_{\hbox {\tiny $X$}}(x) \right\} +\nonumber \\
&\ h^2 \mu_2 t_{10}\left\{\beta_1 f'_{\hbox {\tiny $X$}}(x)I(p=0)+\beta_2 f_{\hbox {\tiny $X$}}(x)I(p\le 1)\right\} +O(h^4), \label{eq:EC0pm1}
\end{align}
and, for $\ell>0$, 
\begin{align}
E\left\{C^{(1)}_{p,\ell}(x)\right\} = &\ \left\{t_{21} R_{\ell,p}(x) + t_{22} \delta_\ell(x) R_{0,p}(x)  \right\}\left\{f_{\hbox {\tiny $X$}}(x) +0.5 h^2 \mu_2 f''_{\hbox {\tiny $X$}}(x)  \right\}+\nonumber \\
 &\ h^\ell t_{21} R_{0,p}(x) \left\{\mu_\ell f_{\hbox {\tiny $X$}}(x)+h \mu_{\ell+1}f'_{\hbox {\tiny $X$}}(x)\right\}+O(h^4). \label{eq:EC1pm1}
\end{align}

Putting results in (\ref{eq:EC0pm1}) and (\ref{eq:EC1pm1}) in a vector, one has 
\begin{align}
E\{\bC^{(1)}_p(x)\} = &\ \left\{f_{\hbox {\tiny $X$}}(x)+0.5 h^2 \mu_2f''_{\hbox {\tiny $X$}}(x)  \right\}
\begin{bmatrix}
 t_{11} R_{0,p}(x)  \\
t_{21} R_{1,p}(x)+t_{22} \delta_1(x)R_{0,p}(x) \\
\vdots \\
t_{21} R_{\ell,p}(x)+t_{22} \delta_\ell (x)R_{0,p}(x) \\
\vdots \\
t_{21} R_{p,p}(x)+t_{22} \delta_p(x)R_{0,p}(x) 
\end{bmatrix} + \nonumber \\
 & \ h^2 \mu_2 
\begin{bmatrix}
t_{10}\left\{\beta_1 f'_{\hbox {\tiny $X$}}(x)I(p=0)+\beta_2 f_{\hbox {\tiny $X$}}(x)I(p\le 1)\right\} \\
 t_{21}R_{0,p}(x)f'_{\hbox {\tiny $X$}}(x)\ \\
 t_{21}R_{0,p}(x) f_{\hbox {\tiny $X$}}(x) \\
0 \\
\vdots \\
0 
\end{bmatrix}
+O\left(h^4 \bone_{p+1}\right)\nonumber \\
= &\ f_{\hbox {\tiny $X$}}(x)\left\{t_{21} R_{0,p}(x)\be_1 +t_{21} \bR^*_p(x)+t_{22} R_{0,p}(x)\bDelta^*_0(x) \right\}+\nonumber\\
&\ 0.5 h^2 \mu_2 \left[f''_{\hbox {\tiny $X$}}(x) \left\{t_{21} \bR^*_p(x)+t_{22} R_{0,p}(x)\bDelta^*_0(x) \right\} +\right.\nonumber\\
&\ t_{21}R_{0,p}(x)\left\{f''_{\hbox {\tiny $X$}}(x)\be_1+2f'_{\hbox {\tiny $X$}}(x)\be_2 I(p\ge 1) +2f_{\hbox {\tiny $X$}}(x)\be_3 I(p\ge 2)  \right\}+\nonumber \\
& \left. 2 t_{10} \left\{\beta_1 f'_{\hbox {\tiny $X$}}(x) I(p=0)+\beta_2 f_{\hbox {\tiny $X$}}(x) I(p\le 1) \right\} \be_1\right]+O\left(h^4 \bone_{p+1}\right).
\label{eq:ECpm1}
\end{align}

As for the order of $\textrm{Var}\{\bC^{(1)}_p(x)\}$, by iterated expectations, one has 
\begin{align*}
& \textrm{Var}\{C^{(1)}_{p,\ell}(x)\} \\
= &\ E\left[\textrm{Var}\left\{\left.C^{(1)}_{p,\ell}(x)\right \vert \mathbb{X}\right\}\right]+\textrm{Var}\left[E\left\{\left. C^{(1)}_{p,\ell}(x)\right \vert \mathbb{X}\right\}\right]\\
= &\ E\left[\textrm{Var}\left\{J^{-1}T_{1,\ell}(x)\left \vert \mathbb{X}\right\}\right.\right]+\\
&\ \textrm{Var}\left[ J^{-1}\sum_{j=1}^J \left\{c_j^{-1} \sum_{k=1}^{c_j} (X_{jk}-x)^\ell\right\}  \left\{c_j^{-1} \sum_{k=1}^{c_j} K_h(X_{jk}-x)\right\}\left\{ c_j^{-1} \sum_{j=1}^{c_j}r_p(X_{jk}, x) \right\} \right]\\
= &\  J^{-2}\sum_{j=1}^J E\left[\left\{c_j^{-2}\sum_{j=1}^{c_j}\sigma^2(X_{jk}) \right\} \left\{c_j^{-1} \sum_{k=1}^{c_j} (X_{jk}-x)^\ell\right\}^2  \left\{c_j^{-1} \sum_{k=1}^{c_j} K_h(X_{jk}-x)\right\}^2\right]+ \\
&\ J^{-2}\sum_{j=1}^J \textrm{Var}\left[\left\{c_j^{-1} \sum_{k=1}^{c_j} (X_{jk}-x)^\ell\right\}  \left\{c_j^{-1} \sum_{k=1}^{c_j} K_h(X_{jk}-x)\right\}\left\{ c_j^{-1} \sum_{j=1}^{c_j}r_p(X_{jk}, x)\right\} \right].
\end{align*}
Because the dominating terms in the above expectations and variances are of the same order as $\kappa_{2, 0}(x)=E\{K^2_h(X-x)\}=O(1/h)$ according to (\ref{eq:EXK2}), $\textrm{Var}\{C^{(1)}_{p,\ell}(x)\}=O\{1/(Jh)\}$ for $\ell=0, 1, \ldots, p$. This is also true for $\textrm{Cov}\{C^{(1)}_{p,\ell_1}(x), C^{(1)}_{p,\ell_2}(x)\}$, for $\ell_1 \ne \ell_2 \in \{0, 1, \ldots, p\}$.    

In conclusion,  
\begin{align}
& \bC^{(1)}_p(x) \nonumber \\
= &\ f_{\hbox {\tiny $X$}}(x)\left\{t_{21} R_{0,p}(x)\be_1 +t_{21} \bR^*_p(x)+t_{22} R_{0,p}(x)\bDelta^*_0(x) \right\}+\nonumber\\
&\ 0.5 h^2 \mu_2 \left[f''_{\hbox {\tiny $X$}}(x) \left\{t_{21} \bR^*_p(x)+t_{22} R_{0,p}(x)\bDelta^*_0(x) \right\} +\right.\nonumber\\
&\ t_{21}R_{0,p}(x)\left\{f''_{\hbox {\tiny $X$}}(x)\be_1+2f'_{\hbox {\tiny $X$}}(x)\be_2 I(p\ge 1) +2f_{\hbox {\tiny $X$}}(x)\be_3 I(p\ge 2)  \right\}+\nonumber \\
& \left. 2 t_{10} \left\{\beta_1 f'_{\hbox {\tiny $X$}}(x) I(p=0)+\beta_2 f_{\hbox {\tiny $X$}}(x) I(p\le 1) \right\} \be_1\right]+O\left(h^4 \bone_{p+1}\right)+\nonumber \\
&\ O_{\hbox {\tiny $P$}}\left(\frac{1}{\sqrt{Jh}}\bone_{p+1}\right). \label{eq:C1decom}
\end{align}

Finally, using (\ref{eq:Sinvdecomm1}) and (\ref{eq:C1decom}) yields the asymptotic discrepancy 
$\hat \bbeta_1-\bbeta = \left\{J^{-1} \bS_1(x)\right\}^{-1}\bC_p^{(1)}(x)$. Extracting the first element of $\hat \bbeta_1-\bbeta$ gives 
\begin{align}
&\ \hat m_1(x)-m(x) \nonumber \\
= &\ \be_1^\T \bM_0^{-1}(x) \left\{ \bL_0(x)-h f^{-1}_{\hbox {\tiny $X$}}(x) f'_{\hbox {\tiny $X$}}(x) \bM_1(x) \bM_0^{-1}(x) \bL_0(x) \right. \nonumber \\
& +h^2 f^{-1}_{\hbox {\tiny $X$}}(x) \left(\mu_2 \left\{ \bL_1(x) + \bL_2(x)+0.5 f''_{\hbox {\tiny $X$}}(x)\bL_3(x)\right\} \right.\nonumber \\
& \left.\left.+\left[\frac{\left\{f''_{\hbox {\tiny $X$}}(x)\right\}^2}{f_{\hbox {\tiny $X$}}(x)} \bM_1(x) \bM_0^{-1}(x)\bM_1(x)+0.5f''_{\hbox {\tiny $X$}}(x)\bM_2(x)\right]\bM_0^{-1}\bL_0(x)\right)\right\} \nonumber\\ 
& +O\left(h^4\right)+ O_{\hbox {\tiny $P$}}\left(\frac{1}{\sqrt{Jh}}\right), \label{eq:m1asy}
\end{align}
where 
\begin{align*}
\bM_0(x) =&\ t_{20}\tilde \bmu_0+t_{21}\left\{\tilde \bDelta_0(x)+\bDelta^*_0(x) \bmu_0^{*\T}+\bmu_0^*\bDelta_0^{*\T}(x)\right\} \\
& + t_{22}\bDelta_0^*(x) \bDelta_0^{*\T}(x), \\
\bM_1(x) =&\ t_{20} \tilde \bmu_1+ t_{21}\left\{\bDelta^*_0(x) \bmu_1^{*\T}+\bmu_1^*\bDelta_0^{*\T}(x)\right\}, \\
\bM_2(x) =&\ t_{20}\tilde \bmu_2+ t_{21}\left\{\mu_2\tilde \bDelta_0(x)+\bDelta^*_0(x) \bmu_2^{*\T}+\bmu_2^*\bDelta_0^{*\T}(x)\right\} \\
& + t_{22}\mu_2\bDelta_0^*(x) \bDelta_0^{*\T}(x), \\
\bL_0(x) =&\ t_{21}\left\{R_{0,p}(x) \be_1+\bR_p^*(x)\right\}+t_{22}R_{0,p}(x)\bDelta_0^*(x), \\
\bL_1(x) =&\ t_{10}\left\{ \beta_1 f'_{\hbox {\tiny $X$}}(x)I(p=0)+\beta_2 f_{\hbox {\tiny $X$}}(x)I(p\le 0) \right\}\be_1, \\
\bL_2(x) =&\ t_{21}R_{0,p}(x)\left\{ 0.5 f''_{\hbox {\tiny $X$}}(x)\be_1 + f'_{\hbox {\tiny $X$}}(x)\be_2 I(p\ge 1) +f_{\hbox {\tiny $X$}}(x)\be_3 I(p\ge 2)\right\}, \\
\bL_3(x) =&\ t_{21} \bR^*_p(x)+t_{22}R_{0,p}(x)\bDelta^*(x).
\end{align*}

\subsection{A special case of $\hat m_1(x)$ with $p=0$}
To gain more insight on the effects of pooling and other factors that influence the performance of the average-weighted estimator, we consider a special case for $\hat m_1(x)$ by setting $p=0$ in (\ref{eq:m1asy}) to obtain a local constant estimator of $m(x)$. In particular, setting $p=0$ in (\ref{eq:Sinvdecomm1}) gives 
\begin{align*}
\left\{J^{-1} \bS_1(x)\right\}^{-1}& = f^{-1}_{\hbox {\tiny $X$}}(x)\left\{1+0.5 h^2 \mu_2 \frac{f''_{\hbox {\tiny $X$}}(x)}{f_{\hbox {\tiny $X$}}(x)}(t_{20}+t_{11})\right\}+O(h^4)+O_{\hbox {\tiny $P$}}\left(\frac{1}{\sqrt{Jh}}\right).
\end{align*}
Setting $p=0$ in (\ref{eq:C1decom}) gives 
\begin{align*}
\bC_0^{(1)}(x) = & \left\{f_{\hbox {\tiny $X$}}(x)+0.5 h^2 \mu_2 f''_{\hbox {\tiny $X$}}(x)  \right\}t_{11}R_{0,0}(x)+h^2 \mu_2 t_{10}\left\{\beta_1 f'_{\hbox {\tiny $X$}}(x) +\beta_2 f_{\hbox {\tiny $X$}}(x) \right\}+ \\
&\ O(h^4)+O_{\hbox {\tiny $P$}}\left(\frac{1}{\sqrt{Jh}}\right), 
\end{align*}
where $R_{0,0}(x)=E\{m(X)\}-m(x)$. It follows, for the local constant estimator using the average-weighted estimator,
\begin{align}
& \hat m_1(x)-m(x) \nonumber \\
= &\ E\{m(X)-m(x)\}t_{11} +h^2 \mu_2  t_{10} \left\{\beta_1 \frac{f'_{\hbox {\tiny $X$}}(x)}{f_{\hbox {\tiny $X$}}(x)}+\beta_2\right\}+O(h^4)+O_{\hbox {\tiny $P$}}\left(\frac{1}{\sqrt{Jh}}\right) \nonumber \\
= &\ t_{11}E\{m(X)-m(x)\}+t_{10}\textrm{NW}(x)+O(h^4)+O_{\hbox {\tiny $P$}}\left(\frac{1}{\sqrt{Jh}}\right), \label{eq:biasp0m1}
\end{align}
where $$\textrm{NW}(x)=h^2 \mu_2\left\{\beta_1 \frac{f'_{\hbox {\tiny $X$}}(x)}{f_{\hbox {\tiny $X$}}(x)}+\beta_2\right\}$$ is the dominating bias of the Nadaraya-Watson (NW) estimator. 
Clearly, (\ref{eq:biasp0m1}) reduces to the dominating bias of the NW estimator when $c_j=1$ for all $j=1, \ldots, J$. If $c_j>1$, the result in (\ref{eq:biasp0m1}) suggests that the bias of $\hat m_1(x)$ does not tend  to zero as $J\to \infty$ and $h\to 0$, unless $E\{m(X)-m(x)\}=0$.

\setcounter{equation}{0} 
\setcounter{section}{0} 
\setcounter{subsection}{0} 
\setcounter{subsubsection}{0} 
\def\theequation{B.\arabic{equation}}
\renewcommand\thesubsection{B.\arabic{subsection}}
\renewcommand\thesubsubsection{B.\arabic{subsection}.\arabic{subsubsection}}
\section*{Appendix B: Proof of Theorem~1-(ii)}
The product-weighted estimator $\hat m_2(x)$ results from minimizing the following weighted objective function, 
$$Q_2(\bbeta) = \sum_{j=1}^J \left\{Z_j-\sum_{\ell=0}^p \beta_\ell c_j^{-1}\sum_{k=1}^{c_j}(X_{jk}-x)^\ell \right\}^2 \left\{\prod_{k=1}^{c_j}K_h(X_{jk}-x)\right\}.$$
In the matrix form, $\hat m_2(x)  = \be_1^\T \bS_2^{-1}(x)\bT_2(x)$,
where
\begin{align*}
\bS_2(x) & = \bD_1(x)^\T \bK_2(x) \bD_1(x)=\left[S_{2,\ell_1,\ell_2}(x)\right]_{\ell_1, \ell_2=0, 1, \ldots, p}, \\
\bT_2(x) & = \bD_1(x)^\T \bK_2(x) \bZ = (T_{2,0}(x),\,  T_{2,1}(x), \, \ldots, \, T_{2, p}(x))^\T,\\
\bK_2(x) & =\textrm{diag}\left(\prod_{k=1}^{c_1}K_h(X_{jk}-x), \, \ldots, \, \prod_{k=1}^{c_J} K_h(X_{Jk}-x)\right).
\end{align*}
From above, one can see that entries in $\bS_2(x)$ are, for $\ell_1, \ell_2=0, 1, \ldots, p$,
\begin{align}
& S_{2,\ell_1,\ell_2}(x) \nonumber \\
= &\ \sum_{j=1}^J \left\{c_j^{-1}\sum_{k=1}^{c_j} (X_{jk}-x)^{\ell_1}\right\} \left\{c_j^{-1}\sum_{k=1}^{c_j} (X_{jk}-x)^{\ell_2}\right\} \left\{\prod_{k=1}^{c_j}K_h(X_{jk}-x)\right\};\label{eq:S2entry} 
\end{align}
and entries in $\bT_2(x)$ are, for $\ell=0, 1, \ldots, p$.
\begin{equation}
T_{2,\ell}(x) = \sum_{j=1}^J Z_j \left\{c_j^{-1}\sum_{k=1}^{c_j} (X_{jk}-x)^\ell\right\} \left\{\prod_{k=1}^{c_j}K_h(X_{jk}-x)\right\}. \label{eq:T2entry}
\end{equation}

In what follows, we derive the mean and variance of $J^{-1}\bS_2(x)$ and $\bC^{(2)}_p(x)=J^{-1}\{\bT_2(x)-\bS_2(x)\bbeta\}$ in order to reveal dominating terms of $\hat \bbeta_2-\bbeta$, where $\hat\bbeta_2=\bS_2^{-1}\bT_2(x)$. For notational simplicity in the derivations, we define the following constants, 
\begin{align*}
d(x) & = J^{-1}\sum_{j=1}^J f^{c_j-1}_{\hbox {\tiny $X$}}(x), \\
d_0(x) & = J^{-1}\sum_{j=1}^J c_j^{-1} f^{c_j-1}_{\hbox {\tiny $X$}}(x), \\
d_k(x) & = J^{-1}\sum_{j=1}^J \frac{[\prod_{i=1}^k(c_j-i)]_+}{c_j} f^{c_j-1}_{\hbox {\tiny $X$}}(x), \textrm{ for $k=1, 2$,} \\
w_{20}(x, h, J) & = J^{-2}\sum_{j=1}^J c_j^{-2} f^{c_j-1}_{\hbox {\tiny $X$}}(x)h^{-c_j}, \\
w_{30}(x, h, J) & = J^{-2}\sum_{j=1}^J c_j^{-3} f^{c_j-1}_{\hbox {\tiny $X$}}(x)h^{-c_j},\\
w_{3k}(x, h, J) & = J^{-2}\sum_{j=1}^J \frac{[\prod_{i=1}^k (c_j-i)]_+}{c_j^3} f^{c_j-1}_{\hbox {\tiny $X$}}(x)h^{-c_j}, \textrm{ for $k=1, 2, 3$}. 
\end{align*}
As $J\to \infty$, the first three quantities are of order $O(1)$; and the latter three quantities are of order $O(J^{-1}h^{-c^*})$, where $c^*=\max_{1\le j\le J}c_j$. 

\subsection{Derive $E\{J^{-1}\bS_2(x)\}$}
By (\ref{eq:S2entry}), one has 
\begin{align*}
&\ E\{J^{-1}S_{2,\ell_1, \ell_2}(x)\} \\
 = &\ J^{-1} \sum_{j=1}^J E\left[\left\{c_j^{-1}\sum_{k=1}^{c_j} (X_{jk}-x)^{\ell_1}\right\} \left\{c_j^{-1}\sum_{k=1}^{c_j} (X_{jk}-x)^{\ell_2}\right\} \left\{\prod_{k=1}^{c_j}K_h(X_{jk}-x)\right\}\right]\\
=&\ J^{-1}\sum_{j=1}^J c_j c_j^{-2}E\left\{(X-x)^{\ell_1+\ell_2}K_h(X-x)\right\}\kappa^{c_j-1}_{1,0}(h,x)+\\
&\ J^{-1}\sum_{j=1}^J c_j(c_j-1)c_j^{-2}E\left\{(X-x)^{\ell_1}K_h(X-x)\right\}\times \\
&\ E\left\{(X-x)^{\ell_2}K_h(X-x)\right\}\kappa^{c_j-2}_{1,0}(h,x)\\
=&\ \kappa_{1, \ell_1+\ell_2}(h, x)J^{-1}\sum_{j=1}^J \frac{1}{c_j} \kappa_{1, 0}^{c_j-1}(h,x)+ \\
&\
\kappa_{1, \ell_1}(h,x) \kappa_{1, \ell_2}(h,x) J^{-1}\sum_{j=1}^J \frac{c_j-1}{c_j}\kappa_{1,0}^{c_j-2}(h,x)\\
= &\ h^{\ell_1+\ell_2}J^{-1}\sum_{j=1}^J c_j^{-1}f^{c_j}_{\hbox {\tiny $X$}}(x)\left\{\mu_{\ell_1+\ell_2}+(c_j-1)\mu_{\ell_1}\mu_{\ell_2}\right\}+h^{\ell_1+\ell_2+1}f'_{\hbox {\tiny $X$}}(x)\times\\
&\ J^{-1}\sum_{j=1}^J c_j^{-1}f^{c_j-1}_{\hbox {\tiny $X$}}(x)\left\{\mu_{\ell_1+\ell_2+1}+(c_j-1)(\mu_{\ell_1}\mu_{\ell_2+1}+\mu_{\ell_1+1}\mu_{\ell_2})\right\}+O(h^{\ell_1+\ell_2+2}).
\end{align*}
Written in the matrix form, the above suggests that
\begin{align}
E\left\{J^{-1}\bS_2(x)\right\} = &\ \bH \left[ d_0(x) \left\{f_{\hbox {\tiny $X$}}(x) \tilde\bmu_0+hf'_{\hbox {\tiny $X$}}(x)\tilde \bmu_1\right\}+d_1(x) \left\{f_{\hbox {\tiny $X$}}(x) \bmu^*_0 \bmu^{*\T}_0 \right.\right. \nonumber \\
& \left.\left. +hf'_{\hbox {\tiny $X$}}(x) \left(\bmu^*_0\bmu_1^{*\T}+\bmu^*_1\bmu_0^{*\T}\right)\right\}+O(h^2)\right]\bH.\label{eq:ES2}
\end{align}
Note that, when $c_j=1$ for $j=1, \ldots, J$, (\ref{eq:ES2}) is equal to the familiar counterpart result when individual-level data are available since now $d_0(x)=1$ and $d_1(x)=0$. 

\subsection{The order of $\textrm{Var}\{J^{-1}\bS_2(x)\}$}
\label{s:varS2}
The order of $\textrm{Var}\{J^{-1}S_{2,\ell_1, \ell_2}(x)\}$ is determined by that of 
\begin{equation}
J^{-2} \sum_{j=1}^J E\left[\left\{c_j^{-1}\sum_{k=1}^{c_j} (X_{jk}-x)^{\ell_1}\right\}^2 \left\{c_j^{-1}\sum_{k=1}^{c_j} (X_{jk}-x)^{\ell_2}\right\}^2 \prod_{k=1}^{c_j}K^2_h(X_{jk}-x)\right].
\label{eq:Eprod2}
\end{equation}
According to (\ref{eq:EXK2}) and (\ref{eq:EK2c}), (\ref{eq:Eprod2}) reaches its highest order (i.e., tends to zero the slowest) when $\ell_1=\ell_2=0$, in which case (\ref{eq:Eprod2}) is equal to 
\begin{align*}
 J^{-2} \sum_{j=1}^J E\left\{\prod_{k=1}^{c_j}K^2_h(X_{jk}-x)\right\} = &\ J^{-2}\sum_{j=1}^J \kappa_{2,0}^{c_j} (h, x) \\
= &\ J^{-2}\sum_{j=1}^J \left\{h^{-c_j}\nu_0 f^{c_j}_{\hbox {\tiny $X$}}(x)+O\left(h^{2-c_j}\right)  \right\},\textrm{ by (\ref{eq:EK2c}),}\\
\le &\ J^{-1}h^{-c^*}\nu_0 f_{\hbox {\tiny $X$}}(x) d(x)+O\left(J^{-1}h^{2-c^*}\right).
\end{align*}
Orders of (\ref{eq:Eprod2}) with one or both of $\ell_1$ and $\ell_2$ larger than zero can be similarly derived using (\ref{eq:EXK2}) and (\ref{eq:EK2c}). Putting these orders together reveals that 
\begin{equation}
\textrm{Var}\left\{J^{-1}\bS_2(x)\right\}=O\left(J^{-1}h^{-c^*}\bH^*\right).
\label{eq:VarS2}
\end{equation}

According to (\ref{eq:ES2}) and (\ref{eq:VarS2}), we conclude that 
\begin{align*}
J^{-1}\bS_2(x)  =&\ E\left\{ J^{-1}\bS_2(x)\right\}+O_{\hbox {\tiny $P$}}\left[\textrm{Var}\left\{J^{-1}\bS_2(x)\right\}\right]\\
=&\ \bH \left[ d_0(x) \left\{f_{\hbox {\tiny $X$}}(x) \tilde\bmu_0+hf'_{\hbox {\tiny $X$}}(x) \tilde \bmu_1\right\}+d_1(x) \left\{f_{\hbox {\tiny $X$}}(x) \bmu^*_0 \bmu^{*\T}_0+\right. \right. \\
&\ \left.\left. hf'_{\hbox {\tiny $X$}}(x)\left(\bmu^*_0\bmu_1^{*\T}+\bmu^*_1\bmu_0^{*\T}\right)\right\} +O\left(h^2\right)\right]\bH+O_{\hbox {\tiny $P$}}\left(\bH^{*1/2}/\sqrt{Jh^{c^*}}\right). 
\end{align*}
It follows that 
\begin{equation}
\label{eq:invS2}
\begin{aligned}
	&\ \left\{J^{-1}\bS_2(x)\right\}^{-1}\\
= &\ \bH^{-1}\left[ f^{-1}_{\hbox {\tiny $X$}}(x)\left\{d_0(x) \tilde\bmu_0+d_1(x) \bmu^*_0 \bmu_0^{*\T} \right\}^{-1}-h\frac{f'_{\hbox {\tiny $X$}}(x)}{f^2_{\hbox {\tiny $X$}}(x)}\times \right.\\
&\ \left\{d_0(x) \tilde \bmu_0+d_1(x)  \bmu^*_0 \bmu_0^{*\T} \right\}^{-1}
\left\{d_0(x) \tilde\bmu_1+d_1(x) \left( \bmu^*_0 \bmu_1^{*\T}+ \bmu^*_1 \bmu_0^{*\T}\right) \right\}\times \\
&\ \left.\left\{d_0(x) \tilde\bmu_0+d_1(x) \bmu^*_0 \bmu_0^{*\T} \right\}^{-1}+ O\left(h^2\right)+O_{\hbox {\tiny $P$}}\left(1/\sqrt{Jh^{c^*}}\right)\right]\bH^{-1}.
\end{aligned}
\end{equation}
\vspace{-0.5in}
\subsection{Derive $E\{\bC^{(2)}_p(x)\}$}
\label{s:EC2m2}
View $\bC^{(2)}_p(x)=(C^{(2)}_{p,0}(x), C^{(2)}_{p,1}(x), \ldots, C^{(2)}_{p, p}(x))^\T$. 
By (\ref{eq:S2entry}) and (\ref{eq:T2entry}),
\begin{align}
&\ E\{C^{(2)}_{p,\ell}(x)\} \nonumber\\
= &\ J^{-1}E\left[  \sum_{j=1}^J Z_j \left\{c_j^{-1} \sum_{k=1}^{c_j} (X_{jk}-x)^\ell\right\} \left\{\prod_{k=1}^{c_j} K_h(X_{jk}-x)\right\}-\right.\nonumber\\
& \left. \sum_{\ell_2=0}^p \beta_{\ell_2}\sum_{j=1}^J \left\{c_j^{-1} \sum_{k=1}^{c_j} (X_{jk}-x)^\ell\right\}\left\{c_j^{-1} \sum_{k=1}^{c_j} (X_{jk}-x)^{\ell_2}\right\} \left\{\prod_{k=1}^{c_j} K_h(X_{jk}-x)\right\}\right].
\end{align}
Hence, 
\begin{align}
&\ E\{C^{(2)}_{p,\ell}(x)\} \nonumber\\
= &\ J^{-1}E\left[  \sum_{j=1}^J \left\{c_j^{-1}\sum_{k=1}^{c_j} m(X_{jk}) \right\} \left\{c_j^{-1} \sum_{k=1}^{c_j} (X_{jk}-x)^\ell\right\} \left\{\prod_{k=1}^{c_j} K_h(X_{jk}-x)\right\}-\right.\nonumber\\
& \left. \sum_{j=1}^J \left\{c_j^{-1} \sum_{k=1}^{c_j} (X_{jk}-x)^\ell\right\}  \left\{\prod_{k=1}^{c_j} K_h(X_{jk}-x)\right\}\sum_{\ell_2=0}^p \beta_{\ell_2}\left\{c_j^{-1} \sum_{k=1}^{c_j} (X_{jk}-x)^{\ell_2}\right\}\right]\nonumber\\
= &\ J^{-1}\sum_{j=1}^J E\left[\left\{c_j^{-1} \sum_{k=1}^{c_j} (X_{jk}-x)^\ell\right\}  
\left\{c_j^{-1} \sum_{k=1}^{c_j}r_p(X_{jk}, x)\right\} \left\{\prod_{k=1}^{c_j} K_h(X_{jk}-x)\right\}\right] \nonumber \\
= &\ J^{-1}\sum_{j=1}^J c_j c_j^{-2}E\left\{(X-x)^\ell r_p(X, x)K_h(X-x)\right\}\kappa^{c_j-1}_{1,0}(h,x)+ \nonumber\\
&\ J^{-1}\sum_{j=1}^J c_j (c_j-1)c_j^{-2}E\left\{(X-x)^\ell K_h(X-x)\right\}E\left\{ r_p(X, x)K_h(X-x)\right\}\kappa^{c_j-2}_{1,0}(h,x)\nonumber\\
= &\ E\left\{(X-x)^\ell r_p(X, x)K_h(X-x)\right\} J^{-1}\sum_{j=1}^J c_j^{-1}\kappa^{c_j-1}_{1,0}(h,x)+ \nonumber\\
&\ \kappa_{1,\ell}(h,x)E\left\{ r_p(X, x)K_h(X-x)\right\}J^{-1}\sum_{j=1}^J (c_j-1)c_j^{-1}\kappa^{c_j-2}_{1,0}(h,x) \nonumber\\
= &\ E\left\{(X-x)^\ell r_p(X, x)K_h(X-x)\right\} \left\{ d_0(x) +O\left(h^2\right)\right\}+\nonumber \\
&\ \kappa_{1,\ell}(h,x)E\left\{ r_p(X, x)K_h(X-x)\right\}\left\{ f^{-1}_{\hbox {\tiny $X$}}(x)d_1(x) +O\left(h^2\right)\right\}. \label{eq:ECellm2}
\end{align}

Assuming the $(p+3)$-th derivative of $m(\cdot)$ exists, one has 
\begin{equation}
r_p(X, x)=\beta_{p+1}(X-x)^{p+1}+\beta_{p+2}(X-x)^{p+2}+\beta^*_{p+3}(X-x)^{p+3}, \label{eq:rexpand}
\end{equation} 
where $\beta^*_{p+3}=m^{(p+3)}(x^*)/(p+3)!$, in which $x^*$ lies between $X$ and $x$. Using this expansion of $r_p(X, x)$, one has,
\begin{align}
&\ E\left\{(X-x)^\ell r_p(X, x)K_h(X-x)\right\}\nonumber \\
= &\ h^{\ell+p+1} \mu_{\ell+p+1}\beta_{p+1}f_{\hbox {\tiny $X$}}(x)+h^{\ell+p+2}\mu_{\ell+p+2}\left\{\beta_{p+1}f'_{\hbox {\tiny $X$}}(x)+\beta_{p+2}f_{\hbox {\tiny $X$}}(x) \right\}+\nonumber \\&\ O\left( h^{\ell+p+3}\right).
\label{eq:ErKellm2}
\end{align}
Using (\ref{eq:EXK}) and (\ref{eq:ErKellm2}) in (\ref{eq:ECellm2}) leads to
\begin{align*}
&\ E\left\{C^{(2)}_{p,\ell}(x)\right\}\\
= &\ d_0(x)\left[h^{\ell+p+1} \mu_{\ell+p+1}\beta_{p+1}f_{\hbox {\tiny $X$}}(x)+h^{\ell+p+2}\mu_{\ell+p+2}\left\{\beta_{p+1}f'_{\hbox {\tiny $X$}}(x)+\beta_{p+2}f_{\hbox {\tiny $X$}}(x) \right\} \right]+ \\
&\ d_1(x)f'_{\hbox {\tiny $X$}}(x)\left[ h^{p+1} \mu_{p+1}\beta_{p+1}f_{\hbox {\tiny $X$}}(x)+h^{p+2}\mu_{p+2}\left\{\beta_{p+1}f'_{\hbox {\tiny $X$}}(x)+\beta_{p+2}f_{\hbox {\tiny $X$}}(x) \right\} \right]\times \\
&\ \left\{ h^\ell \mu_{\ell}f_{\hbox {\tiny $X$}}(x) +h^{\ell+1} \mu_{\ell+1}f'_{\hbox {\tiny $X$}}(x) \right\}+O\left(h^{\ell+p+3} \right) \nonumber \\
= &\ h^{\ell+p+1} \beta_{p+1} f_{\hbox {\tiny $X$}}(x)\left\{d_0(x) \mu_{\ell+p+1}+d_1(x) \mu_{\ell}\mu_{p+1} \right\} + \\
&\ h^{\ell+p+2}\left[ \left\{\beta_{p+1} f'_{\hbox {\tiny $X$}}(x)+\beta_{p+2} f_{\hbox {\tiny $X$}}(x)  \right\} \left\{ d_0(x) \mu_{\ell+p+2}+d_1(x) \mu_{\ell}\mu_{p+2}   \right\}+\right. \\
& \left. d_1(x) \mu_{\ell+1}\mu_{p+1}\beta_{p+1} f'_{\hbox {\tiny $X$}}(x)\right]. 
\end{align*}
This is equivalent to, in the matrix form, 
\begin{align}
&\ E\{\bC^{(2)}_p(x)\} \nonumber \\
= &\ h^{p+1} \bH \left(\beta_{p+1} f_{\hbox {\tiny $X$}}(x) \left\{d_0(x)\bmu^*_{p+1}+d_1(x)\mu_{p+1}\bmu^*_0   \right\}  \right. \nonumber \\
&\ + h \left[\left\{\beta_{p+2} f_{\hbox {\tiny $X$}}(x)+\beta_{p+1} f'_{\hbox {\tiny $X$}}(x) \right\}
\left\{d_0(x)\bmu^*_{p+2}+d_1(x)\mu_{p+2}\bmu^*_0 \right\}\right.\nonumber \\
& \left.\left. + d_1(x) \beta_{p+1} f'_{\hbox {\tiny $X$}}(x)\mu_{p+1} \bmu^*_1 \right]+O\left(h^2\bone_{p+1}\right)\right).
\label{eq:ECpm2}
\end{align}

\subsection{The order of $\textrm{Var}\{\bC^{(2)}_p(x)\}$}
By iterated expectations,
\begin{align*}
& \textrm{Var}\{C^{(2)}_{p,\ell}(x)\} \\
= & E\left[\textrm{Var}\left\{C^{(2)}_{p,\ell}(x)\left \vert \mathbb{X}\right\}\right.\right]+\textrm{Var}\left[E\left\{C^{(2)}_{p,\ell}(x)\left \vert \mathbb{X}\right\}\right.\right]\\
= & E\left[\textrm{Var}\left\{J^{-1}T_{2,\ell}(x)\left \vert \mathbb{X}\right\}\right.\right]+\\
& \textrm{Var}\left[ J^{-1}\sum_{j=1}^J \left\{c_j^{-1} \sum_{k=1}^{c_j} (X_{jk}-x)^\ell\right\}  \left\{\prod_{k=1}^{c_j} K_h(X_{jk}-x)\right\}\left\{ c_j^{-1} \sum_{j=1}^{c_j}r_p(X_{jk}, x) \right\} \right]\\
= &  J^{-2}\sum_{j=1}^J E\left[\left\{c_j^{-2}\sum_{j=1}^{c_j}\sigma^2(X_{jk}) \right\} \left\{c_j^{-1} \sum_{k=1}^{c_j} (X_{jk}-x)^\ell\right\}^2  \left\{\prod_{k=1}^{c_j} K_h(X_{jk}-x)\right\}^2\right]+ \\
& J^{-2}\sum_{j=1}^J \textrm{Var}\left[\left\{c_j^{-1} \sum_{k=1}^{c_j} (X_{jk}-x)^\ell\right\}  \left\{ \prod_{k=1}^{c_j} K_h(X_{jk}-x)\right\}\left\{ c_j^{-1} \sum_{j=1}^{c_j}r_p(X_{jk}, x)\right\} \right].
\end{align*}

Following similar derivations as those in Section~\ref{s:varS2}, one can show that $\textrm{Var}\{C^{(2)}_{p,\ell}(x)\}=O(J^{-1}h^{2\ell -c^*})$. Along with (\ref{eq:ECpm2}), now one has 
\begin{align}
&\ \bC^{(2)}_p(x) \nonumber \\
= &\ h^{p+1} \bH \left(\beta_{p+1} f_{\hbox {\tiny $X$}}(x) \left\{d_0(x)\bmu^*_{p+1}+d_1(x)\mu_{p+1}\bmu^*_0   \right\}  \right. \nonumber \\
&\ + h \left[\left\{\beta_{p+2} f_{\hbox {\tiny $X$}}(x)+\beta_{p+1} f'_{\hbox {\tiny $X$}}(x) \right\}
\left\{d_0(x)\bmu^*_{p+2}+d_1(x)\mu_{p+2} \bmu^*_0 \right\}
+ \right. \nonumber \\
& \left. \left. d_1(x) \beta_{p+1} f'_{\hbox {\tiny $X$}}(x)\mu_{p+1} \bmu^*_1 \right]
 +O\left(h^2\right)\right)+O_{\hbox {\tiny $P$}}\left\{\textrm{vecdiag}(\bH)/\sqrt{Jh^{c^*}}\right\}.
\label{eq:Cpm2}
\end{align}

By (\ref{eq:invS2}) and (\ref{eq:Cpm2}), one has 
\begin{align}
&\ \hat \bbeta_2-\bbeta \nonumber \\
= &\ h^{p+1}\bH^{-1}\left[f^{-1}_{\hbox {\tiny $X$}}(x)\left\{d_0(x) \tilde\bmu_0+d_1(x) \bmu^*_0 \bmu_0^{*\T} \right\}^{-1}-h\frac{f'_{\hbox {\tiny $X$}}(x)}{f^2_{\hbox {\tiny $X$}}(x)}\times \right.\nonumber \\
&\ \left\{d_0(x) \tilde\bmu_0+d_1 (x) \bmu^*_0 \bmu_0^{*\T} \right\}^{-1}\left\{d_0(x) \tilde \bmu_1+d_1(x) \left(\bmu^*_0 \bmu_1^{*\T}+ \bmu^*_1 \bmu_0^{*\T}\right) \right\}\times \nonumber \\
& \left.\left\{d_0(x) \tilde \bmu_0+d_1(x)\bmu^*_0 \bmu_0^{*\T} \right\}^{-1}\right]\left(\beta_{p+1} f_{\hbox {\tiny $X$}}(x) \left\{d_0(x) \bmu^*_{p+1}+d_1(x)\mu_{p+1}\bmu^*_0   \right\} +\right.\nonumber\\
& \left. h \left[\left\{\beta_{p+2} f_{\hbox {\tiny $X$}}(x)+\beta_{p+1} f'_{\hbox {\tiny $X$}}(x) \right\} \left\{d_0(x)\bmu^*_{p+2}+d_1(x)\mu_{p+2}\bmu^*_0 \right\}+\right. \right.\nonumber\\
& \left.\left. d_1(x) \beta_{p+1} f'_{\hbox {\tiny $X$}}(x)\mu_{p+1} \bmu^*_1 \right]\right)+\bH^{-1}\bone_{p+1}\left\{o\left(h^{p+2}\right)+O_{\hbox {\tiny $P$}}\left(\frac{1}{\sqrt{Jh^{c^*}}}\right)\right\}. 
\label{eq:biasbetam2}
\end{align}
When $c_j=1$ for $j=1, \ldots, J$, with $d_0(x)=1$ and $d_1(x)=0$, (\ref{eq:biasbetam2}) reduces to the a much simpler expression that matches the counterpart result for non-pooling data. 

\subsection{Special cases of $\hat m_2(x)$ with $p=0, 1$}
Setting $p=0$ in (\ref{eq:biasbetam2}) gives the result regarding the local constant estimator $\hat m_2(x)$, $\hat m_2(x)-m(x)=h^2 \mu_2\left\{{\beta_1f'_{\hbox {\tiny $X$}}(x)}/{f_{\hbox {\tiny $X$}}(x)}+\beta_2\right\}+O(h^4)+O_p\left(\frac{1}{\sqrt{Jh^{c^*}}}\right),$
which coincides with the result for the Nadaraya-Watson estimator in regard to the dominating bias, although the order of the asymptotic variance is inflated whenever $c^*>1$. 

Setting $p=1$ in (\ref{eq:biasbetam2}) and extracting the first entry gives the result regarding the local linear estimator $\hat m_2(x)$, $\hat m_2(x)-m(x)=h^2 \mu_2\beta_2+O(h^4)+O_p\left(\frac{1}{\sqrt{Jh^{c^*}}}\right),$ which suggests the same dominating bias of order $h^2$ that is equal to the dominating bias of the same order associated with the local linear estimator for $m(x)$ based on individual-level data.  

\subsection{Variance of $\hat \bbeta_2$}
By iterative expectations, 
\begin{align*}
&\ \textrm{Var}(\hat \bbeta_2) \\
= &\ \textrm{Var}\left\{\bS_2^{-1}(x)\bT_2(x)\right\} \\
= &\ E\left[\bS_2^{-1}(x)\textrm{Var}\left\{\bT_2(x)|\mathbb{X}\right\}\bS_2^{-\T}(x) \right]+
\textrm{Var}\left[\bS_2^{-1}(x)E\left\{\bT_2(x)|\mathbb{X}\right\} \right] \\
= &\ E\left[J\bS_2^{-1}(x)\textrm{Var}\left\{J^{-1}\bT_2(x)|\mathbb{X}\right\}J\bS_2^{-\T}(x) \right]+
\textrm{Var}\left[J\bS_2^{-1}(x)E\left\{J^{-1}\bT_2(x)|\mathbb{X}\right\} \right].
\end{align*}
On the other hand, we have the dominating term of $J\bS_2^{-1}(x)$ given in (\ref{eq:invS2}) that holds for all $\mathbb{X}$. Hence, after substituting $J\bS_2^{-1}(x)$ with (\ref{eq:invS2}), we will essentially deal with 
\begin{equation}
\textrm{Var}\{J^{-1}\bT_2(x)\}=E\left\{J^{-1}\bT_2(x)J^{-1}\bT^\T_2(x) \right\}-E\left\{J^{-1}\bT_2(x)\right\}E\left\{J^{-1}\bT^\T_2(x) \right\}.
\label{eq:varT2}
\end{equation}
We derive the second term in (\ref{eq:varT2}) first. 

By (\ref{eq:EXK}), (\ref{eq:EKc}), and (\ref{eq:EgXK}), 
\begin{align*}
&\ E\left\{J^{-1}T_{2,\ell}(x)\right\}\\
= &\ J^{-1}\sum_{j=1}^J E\left[\left\{c_j^{-1}\sum_{k=1}^{c_j} m(X_{jk})\right\}\left\{c_j^{-1}\sum_{k=1}^{c_j}(X_{jk}-x)^\ell\right\}\left\{\prod_{k=1}^{c_j}K_h(X_{jk}-x)\right\} \right]\\
= &\ J^{-1}\sum_{j=1}^J c_j c_j^{-2} \kappa^*_{1,\ell}(m)\kappa^{c_j-1}_{1,0}(h,x)+\\
&\ J^{-1}\sum_{j=1}^J c_j(c_j-1) c_j^{-2} \kappa^*_{1,0}(m)\kappa_{1, \ell}(h,x)\kappa^{c_j-2}_{1,0}(h,x) \\
= &\ \left[h^\ell \mu_\ell \beta_0 f_{\hbox {\tiny $X$}}(x)+h^{\ell+1}\mu_{\ell+1}\left\{\beta_0 f'_{\hbox {\tiny $X$}}(x)  +\beta_1f_{\hbox {\tiny $X$}}(x)\right\}+O(h^{\ell+2})\right]\times\\
&\ J^{-1}\sum_{j=1}^J c_j^{-1}\left\{f^{c_j-1}_{\hbox {\tiny $X$}}(x) +(c_j-1)h^2 \mu_2 f^{c_j-2}_{\hbox {\tiny $X$}}(x) f''_{\hbox {\tiny $X$}}(x)/2+O(h^4)\right\}+\\
&\ \left[\beta_0 f_{\hbox {\tiny $X$}}(x)+h^2\mu_2\left\{\beta_0 f''_{\hbox {\tiny $X$}}(x)/2+\beta_1f'_{\hbox {\tiny $X$}}(x) +\beta_2f_{\hbox {\tiny $X$}}(x)\right\}+O\left(h^4\right)\right]\times \\
&\ \left\{ h^\ell \mu_\ell f_{\hbox {\tiny $X$}}(x)+h^{\ell+1} \mu_{\ell+1} f'_{\hbox {\tiny $X$}}(x) +O\left(h^{\ell+2} \right)\right\}\times \\
&\ J^{-1}\sum_{j=1}^J(c_j-1)c_j^{-1}\left\{f^{c_j-2}_{\hbox {\tiny $X$}}(x) +(c_j-2)h^2 \mu_2 f^{c_j-3}_{\hbox {\tiny $X$}}(x) f''_{\hbox {\tiny $X$}}(x)/2+O(h^4)\right\}\\
=&\ h^\ell \mu_\ell \beta_0 f_{\hbox {\tiny $X$}}(x)d(x)+h^{\ell+1}\mu_{\ell+1}\left\{d(x)\beta_0 f'_{\hbox {\tiny $X$}}(x) + d_0(x)\beta_1 f_{\hbox {\tiny $X$}}(x) \right\}+O\left(h^{\ell+2}\right).
\end{align*}
Summarizing the above in matrix form, one has 
\begin{align}
&\ E\left\{J^{-1}\bT_2(x)\right\} \nonumber \\
= &\ \bH \left[\beta_0 f_{\hbox {\tiny $X$}}(x)d(x)\bmu_0^* +h\left\{d(x)\beta_0 f'_{\hbox {\tiny $X$}}(x) + d_0(x)\beta_1 f_{\hbox {\tiny $X$}}(x) \right\}\bmu_1^* +O(h^2)\right]. \label{eq:ET2vec}
\end{align}
It follows that 
\begin{align}
&\ E\left\{J^{-1}\bT_2(x)\right\}E\left\{J^{-1}\bT^\T_2(x)\right\} \nonumber\\
= &\ \beta_0 f_{\hbox {\tiny $X$}}(x)d(x) \bH \left[\beta_0 f_{\hbox {\tiny $X$}}(x)d(x)\bmu^*_0\bmu^{*\T}_0+h \left\{d(x)\beta_0 f'_{\hbox {\tiny $X$}}(x) + d_0(x)\beta_1 f_{\hbox {\tiny $X$}}(x) \right\}\times \right.\nonumber \\
& \left.\left(  \bmu_0^*\bmu_1^{*\T}+\bmu_1^*\bmu_0^{*\T}\right) +O(h^2)\right]\bH.
\label{eq:ET2ET2}
\end{align}

The first term in (\ref{eq:varT2}) relates to $J^{-2}E\{T_{2,\ell_1}(x)T_{2,\ell_2}(x)|\mathbb{X}\}$that we derive next. Because 
\begin{align*}
&\ T_{2, \ell_1}(x)T_{2, \ell_2}(x) \\
=&\ \sum_{j=1}^J Z^2_j \left\{c_j^{-1}\sum_{k=1}^{c_j}(X_{jk}-x)^{\ell_1}\right\}\left\{c_j^{-1}\sum_{k=1}^{c_j}(X_{jk}-x)^{\ell_2}\right\}\prod_{k=1}^{c_j}K^2_h(X_{jk}-x)+\\
&\ \sum_{j_1\ne j_2} Z_{j_1}Z_{j_2} \left\{c_{j_1}^{-1}\sum_{k=1}^{c_{j_1}}(X_{j_1k}-x)^{\ell_1}\right\}\left\{c_{j_2}^{-1}\sum_{k=1}^{c_{j_2}}(X_{j_2k}-x)^{\ell_2}\right\}\times \\
&\ \left\{\prod_{k=1}^{c_{j_1}}K_h(X_{j_1k}-x)\right\}\left\{\prod_{k=1}^{c_{j_2}}K_h(X_{j_2k}-x)\right\},
\end{align*}
hence, 
\begin{align}
&\ J^{-2}E\{T_{2, \ell_1}(x)T_{2, \ell_2}(x)|\mathbb{X}\} \nonumber\\
=&\ J^{-2}\sum_{j=1}^J E(Z^2_j|\tilde \bX_j) \left\{c_j^{-1}\sum_{k=1}^{c_j}(X_{jk}-x)^{\ell_1}\right\}\left\{c_j^{-1}\sum_{k=1}^{c_j}(X_{jk}-x)^{\ell_2}\right\}\prod_{k=1}^{c_j}K^2_h(X_{jk}-x)+ \nonumber \\
&\ J^{-2}\sum_{j_1\ne j_2} E(Z_{j_1}|\tilde \bX_{j_1})E(Z_{j_2}|\tilde \bX_{j_2}) \left\{c_{j_1}^{-1}\sum_{k=1}^{c_{j_1}}(X_{j_1k}-x)^{\ell_1}\right\}\left\{c_{j_2}^{-1}\sum_{k=1}^{c_{j_2}}(X_{j_2k}-x)^{\ell_2}\right\}\times\nonumber \\
&\ \left\{\prod_{k=1}^{c_{j_1}}K_h(X_{j_1k}-x)\right\}\left\{\prod_{k=1}^{c_{j_2}}K_h(X_{j_2k}-x)\right\} \nonumber \\
= &\ J^{-2}\sum_{j=1}^J \left\{c_j^{-2}\sum_{k=1}^{c_j}\sigma^2(X_{jk})\right\} \left\{c_j^{-1}\sum_{k=1}^{c_j}(X_{jk}-x)^{\ell_1}\right\}\left\{c_j^{-1}\sum_{k=1}^{c_j}(X_{jk}-x)^{\ell_2}\right\}\times \nonumber \\
&\ \prod_{k=1}^{c_j}K^2_h(X_{jk}-x)  \label{eq:sig2polyKm2}\\
&\ +J^{-2}\sum_{j=1}^J \left\{c_j^{-1}\sum_{k=1}^{c_j}m(X_{jk})\right\}^2 \left\{c_j^{-1}\sum_{k=1}^{c_j}(X_{jk}-x)^{\ell_1}\right\}\left\{c_j^{-1}\sum_{k=1}^{c_j}(X_{jk}-x)^{\ell_2}\right\}\times \nonumber \\
&\ \prod_{k=1}^{c_j}K^2_h(X_{jk}-x) \label{eq:mean2polyKm2}\\
&\ +J^{-2}\sum_{j_1\ne j_2} \left\{c_{j_1}^{-1}\sum_{k=1}^{c_{j_1}}m(X_{j_1k})\right\}\left\{c_{j_2}^{-1}\sum_{k=1}^{c_{j_2}}m(X_{j_2k})\right\} \left\{c_{j_1}^{-1}\sum_{k=1}^{c_{j_1}}(X_{j_1k}-x)^{\ell_1}\right\}\times\nonumber \\
&\ \left\{c_{j_2}^{-1}\sum_{k=1}^{c_{j_2}}(X_{j_2k}-x)^{\ell_2}\right\}\left\{\prod_{k=1}^{c_{j_1}}K_h(X_{j_1k}-x)\right\}\left\{\prod_{k=1}^{c_{j_2}}K_h(X_{j_2k}-x)\right\}.
\label{eq:mixmeanpolyKm2}
\end{align}

According to our earlier derivation of $E\{J^{-1}T_{2,\ell}(x)\}$, the expectation of (\ref{eq:mixmeanpolyKm2}) is of order $O(J^{-1}h^{\ell_1+\ell_2})$, and thus is dominated by the expectations of (\ref{eq:sig2polyKm2}) and (\ref{eq:mean2polyKm2}), which are of order $O(J^{-1}h^{\ell_1+\ell_2-c^*})$ as we show next. 

By (\ref{eq:EK2c}) and (\ref{eq:EgXK2}), the expectation of the summand of (\ref{eq:sig2polyKm2}) is equal to $c_j^{-4}$ times  
\begin{align*}
&\ E\left[\left\{\sum_{k=1}^{c_j}\sigma^2(X_{jk})\right\} \left\{\sum_{k=1}^{c_j}(X_{jk}-x)^{\ell_1}\right\}\left\{\sum_{k=1}^{c_j}(X_{jk}-x)^{\ell_2}\right\}\prod_{k=1}^{c_j}K^2_h(X_{jk}-x)\right]\\
= &\ c_j \kappa^*_{2, \ell_1+\ell_2}(\sigma^2) \kappa^{c_j-1}_{2,0}(h,x) + c_j(c_j-1)\left\{\kappa^*_{2, 0}(\sigma^2)\kappa_{2,\ell_1+\ell_2}(h,x)+  \right.\\
& \left.\kappa^*_{2, \ell_1}(\sigma^2)\kappa_{2,\ell_2}(h,x)+
    \kappa^*_{2, \ell_2}(\sigma^2)\kappa_{2,\ell_1}(h,x)\right\} \kappa^{c_j-2}_{2,0}(h,x)+\\
&\ c_j(c_j-1)(c_j-2)	\kappa^*_{2, 0}(\sigma^2)\kappa_{2,\ell_1}(h,x)\kappa_{2,\ell_2}(h,x)\kappa^{c_j-3}_{2,0}(h,x)\\
= &\ h^{\ell_1+\ell_2-c_j} \nu_0 \sigma^2(x)c_j f^{c_j}_{\hbox {\tiny $X$}}(x) \{\nu_{\ell_1+\ell_2}
+(c_j-1)(\nu_0\nu_{\ell_1+\ell_2}+2\nu_{\ell_1}\nu_{\ell_2})+\\
&\ [(c_j-1)(c_j-2)]_+ \nu_0 \nu_{\ell_1}\nu_{\ell_2}\}+h^{\ell_1+\ell_2-c_j+1} \nu_0 \sigma(x) c_j 
f^{c_j-1}_{\hbox {\tiny $X$}}(x)\left(\nu_{\ell_1+\ell_2+1}\times \right.\\
&\ \left\{\sigma(x)f'_{\hbox {\tiny $X$}}(x)+2\sigma'(x) f_{\hbox {\tiny $X$}}(x)\right\} 
+(c_j-1)\left[\nu_{\ell_1+\ell_2+1}\sigma(x)f'_{\hbox {\tiny $X$}}(x)+\right.\\
& \left.(\nu_{\ell_1+1}\nu_{\ell_2}+\nu_{\ell_1}\nu_{\ell_2+1})\left\{\sigma(x)f'_{\hbox {\tiny $X$}}(x)+2\sigma'(x) f_{\hbox {\tiny $X$}}(x)\right\}\right]+\\
&\left. [(c_j-1)(c_j-2)]_+\nu_0(\nu_{\ell_1+1}\nu_{\ell_2}+\nu_{\ell_1}\nu_{\ell_2+1})\sigma(x)f'_{\hbox {\tiny $X$}}(x)\right)+O\left(h^{\ell_1+\ell_2+2-c_j}\right). 
\end{align*}
Hence, 
\begin{align*}
&\ J^{-2}\sum_{j=1}^J E\left[\left\{c_j^{-2}\sum_{k=1}^{c_j}\sigma^2(X_{jk})\right\} \left\{c_j^{-1}\sum_{k=1}^{c_j}(X_{jk}-x)^{\ell_1}\right\} \times \right.\\
& \left.\left\{c_j^{-1}\sum_{k=1}^{c_j}(X_{jk}-x)^{\ell_2}\right\}\prod_{k=1}^{c_j}K^2_h(X_{jk}-x)\right]\\
=&\ h^{\ell_1+\ell_2}\nu_0 \sigma^2(x) f_{\hbox {\tiny $X$}}(x)\{w_{30}(x, h, J)\nu_{\ell_1+\ell_2}+
w_{31}(x, h, J)(\nu_0\nu_{\ell_1+\ell_2}+2\nu_{\ell_1}\nu_{\ell_2})+\\
&\ w_{32}(x, h, J)\nu_0\nu_{\ell_1}\nu_{\ell_2}\}+h^{\ell_1+\ell_2+1}\nu_0\sigma^2(x) \left[f'_{\hbox {\tiny $X$}}(x) w_{20}(x, h, J)\nu_{\ell_1+\ell_2+1} + \right.\\
&\ 2\sigma'(x) f_{\hbox {\tiny $X$}}(x)\{w_{30}(x, h, J) \nu_{\ell_1+\ell_2+1}+w_{31}(x, h, J)(\nu_{\ell_1+1}\nu_{\ell_2}+\nu_{\ell_1}\nu_{\ell_2+1})\}+\\
& \left. \sigma(x) f'_{\hbox {\tiny $X$}}(x)\{w_{31}(x, h, J)+\nu_0w_{32}(x, h, J)\}(\nu_{\ell_1+1}\nu_{\ell_2}+\nu_{\ell_1}\nu_{\ell_2+1})\right]+\\
&\ O\left(J^{-1}h^{\ell_1+\ell_2+2-c^*} \right).
\end{align*}
For $\ell_1, \ell_2=0, 1, \ldots, p$, the above expression is the $[\ell_1+1, \ell_2+1]$ entry of the following $(p+1)\times (p+1)$ matrix, 
\begin{align}
&\ \nu_0 \sigma^2(x) \bH\left\{f_{\hbox {\tiny $X$}}(x)\left\{w_{30}(x, h, J) \tilde \bnu_0+w_{31}(x, h, J)\left(\nu_0 \tilde \bnu_0+2\bnu_0^*\bnu_0^{*\T}\right)+  \right. \right. \nonumber \\
& \left.w_{32}(x, h, J)\nu_0 \bnu_0^*\bnu_0^{*\T}\right\}+h \sigma^{-1}(x) \left(\sigma(x)f'_{\hbox {\tiny $X$}}(x)\left[w_{20}(x, h, J)\tilde \bnu_1+  \right.\right. \nonumber \\
& \left.\left\{w_{31}(x, h, J)+\nu_0 w_{32}(x, h, J)\right\}(\bnu_1^* \bnu_0^{*\T}+\bnu_0^*\bnu_1^{*\T})\right]+2\sigma'(x) f_{\hbox {\tiny $X$}}(x) \times \nonumber \\
& \left. \left. \left\{w_{30}(x, h, J)\tilde \bnu_1+  w_{31}(x, h, J)(\bnu^*_1\bnu^{*\T}_0+\bnu_0^*\bnu_1^{*\T})\right\}\right)+O\left(J^{-1}h^{2-c^*}\right)\right\}\bH. \label{eq:varterm1}
\end{align}

The expectation of the summand of (\ref{eq:mean2polyKm2}) is equal to $c_j^{-4}$ times
\begin{align*}
&\ E\left[\left\{\sum_{k=1}^{c_j}m(X_{jk})\right\}^2 \left\{\sum_{k=1}^{c_j}(X_{jk}-x)^{\ell_1}\right\}\left\{\sum_{k=1}^{c_j}(X_{jk}-x)^{\ell_2}\right\}\prod_{k=1}^{c_j}K^2_h(X_{jk}-x)\right]\\
= &\ c_j \kappa^{c_j-1}_{2,0}(h,x)\kappa^*_{2,\ell_1+\ell_2}(m^2)+\\
	&\ c_j(c_j-1)\kappa_{2,0}^{c_j-2}(h,x)\left\{ \kappa^*_{2,0}(m^2)\kappa_{2, \ell_1+\ell_2}(h,x)+
	2 \kappa^*_{2,\ell_1}(m)\kappa^*_{2,\ell_2}(m)+\right.\\
	& \left. 2\kappa^*_{2,0}(m)\kappa^*_{2,\ell_1+\ell_2}(m)
	+\kappa^*_{2,\ell_1}(m^2) \kappa_{2,\ell_2}(h,x)+\kappa^*_{2,\ell_2}(m^2) \kappa_{2,\ell_1}(h,x)\right\}+\\
	&\ [c_j(c_j-1)(c_j-2)]_+\kappa_{2,0}^{c_j-3}(h,x)\left\{\kappa^{*2}_{2,0}(m)\kappa_{2,\ell_1+\ell_2}(h,x)+\right.\\
	& \left.\kappa^*_{2,0}(m^2) \kappa_{2,\ell_1}(h,x)\kappa_{2,\ell_2}(h,x)+ 
	\kappa^*_{2,\ell_1}(m) \kappa^*_{2,0}(m)\kappa_{2,\ell_2}(h,x) \right.\\
	& \left.+\kappa^*_{2,\ell_2}(m) \kappa^*_{2,0}(m)\kappa_{2,\ell_1}(h,x)\right\}+
  [c_j(c_j-1)(c_j-2)(c_j-3)]_+\times\\
	&\ \kappa_{2,0}^{c_j-4}(h,x)\kappa^{*2}_{2,0}(m)\kappa_{2,\ell_1}(h,x)\kappa_{2,\ell_2}(h,x).
\end{align*}
Using (\ref{eq:EXK2}) and (\ref{eq:EgXK2}) in the above gives
\begin{align*}
&\ J^{-2}E\left[\left\{c_j^{-1}\sum_{k=1}^{c_j}m(X_{jk})\right\}^2 \left\{c_j^{-1}\sum_{k=1}^{c_j}(X_{jk}-x)^{\ell_1}\right\}\times \right. \\
& \left.\left\{c_j^{-1}\sum_{k=1}^{c_j}(X_{jk}-x)^{\ell_2}\right\}\prod_{k=1}^{c_j}K^2_h(X_{jk}-x)\right]\\
=&\ h^{\ell_1+\ell_2}w_{30}(x, h, J) \nu_0 \beta_0 \left[ \beta_0 f_{\hbox {\tiny $X$}}(x) \nu_{\ell_1+\ell_2}+h \left\{\beta_0 f'_{\hbox {\tiny $X$}}(x)+2\beta_1 f_{\hbox {\tiny $X$}}(x)  \right\}\nu_{\ell_1+\ell_2+1}\right]+\\
&\ h^{\ell_1+\ell_2} w_{31}(x, h, J) \nu_0 \beta_0 \left(\beta_0 f_{\hbox {\tiny $X$}}(x)(3\nu_0\nu_{\ell_1+\ell_2}+4\nu_{\ell_1}\nu_{\ell_2})+h\left[\nu_0\nu_{\ell_1+\ell_2+1}\times \right. \right.\\
& \left.\left.\left\{2\beta_0 f'_{\hbox {\tiny $X$}}(x)+ \beta_1 f_{\hbox {\tiny $X$}}(x)\right\}+4(\nu_{\ell_1}\nu_{\ell_2+1}+\nu_{\ell_1+1}\nu_{\ell_2})\left\{\beta_0f'_{\hbox {\tiny $X$}}(x)+\beta_1 f_{\hbox {\tiny $X$}}(x)  \right\}\right]\right)+\\
&\ h^{\ell_1+\ell_2}w_{32}(x, h, J)\nu^2_0\beta_0 \left( \beta_0 f_{\hbox {\tiny $X$}}(x)\nu_0 (\nu_{\ell_1+\ell_2}+3\nu_{\ell_1}\nu_{\ell_2})+h\left[ \left\{3\beta_0 f'_{\hbox {\tiny $X$}}(x)+\beta_1 f_{\hbox {\tiny $X$}}(x)  \right\}\times \right.\right.\\
& \left.\left. (\nu_{\ell_1} \nu_{\ell_2+1}+\nu_{\ell_1+1} \nu_{\ell_2})+\beta_0 f'_{\hbox {\tiny $X$}}(x) \nu_0 \nu_{\ell_1+\ell_2+1}\right]\right)+\\
&\ h^{\ell_1+\ell_2} w_{33}(x, h, J)\nu_0^3 \beta_0 \left\{\beta_0 f_{\hbox {\tiny $X$}}(x)\nu_{\ell_1}\nu_{\ell_2}+h\beta_0f'_{\hbox {\tiny $X$}}(x)(\nu_{\ell_1}\nu_{\ell_2+1}+\nu_{\ell_1+1}\nu_{\ell_2})\right\}+\\
&\ O\left(J^{-1}h^{\ell_1+\ell_2+2-c^*} \right)\\
= &\ h^{\ell_1+\ell_2}\nu_0 \beta_0^2 f_{\hbox {\tiny $X$}}(x)\left\{w_{30}(x, h, J) \nu_{\ell_1+\ell_2}+w_{31}(x, h, J)(3 \nu_0\nu_{\ell_1+\ell_2}+4\nu_{\ell_1}\nu_{\ell_2})+ \right.\\
& \left. w_{32}(x, h, J)\nu_0^2 (\nu_{\ell_1+\ell_2}+3\nu_{\ell_1}\nu_{\ell_2})+w_{33}(x, h, J)\nu_0^2 \nu_{\ell_1}\nu_{\ell_2}\right\}+\\
&\ h^{\ell_1+\ell_2+1}\nu_0\beta_0 \left(w_{30}(x, h, J)\left\{\beta_0 f'_{\hbox {\tiny $X$}}(x)+2\beta_1 f_{\hbox {\tiny $X$}}(x)\right\}\nu_{\ell_1+\ell_2+1}+\right. \\
&\ w_{31}(x, h, J)\left[\left\{2\beta_0 f'_{\hbox {\tiny $X$}}(x)+\beta_1 f_{\hbox {\tiny $X$}}(x) \right\}\nu_0\nu_{\ell_1+\ell_2+1}+4 \left\{\beta_0 f'_{\hbox {\tiny $X$}}(x)+\beta_1 f_{\hbox {\tiny $X$}}(x) \right\}\times \right.\\
& \left.(\nu_{\ell_1}\nu_{\ell_2+1}+\nu_{\ell_1+1}\nu_{\ell_2})\right]+w_{32}(x, h, J)\nu_0\left[ 
\left\{3\beta_0 f'_{\hbox {\tiny $X$}}(x)+\beta_1 f_{\hbox {\tiny $X$}}(x)\right\} \times \right.\\
& \left. (\nu_{\ell_1}\nu_{\ell_2+1}+\nu_{\ell_1+1}\nu_{\ell_2})+\beta_0f'_{\hbox {\tiny $X$}}(x) \nu_0 \nu_{\ell_1+\ell_2+1}\right]+\\
& \left. w_{33}(x, h, J)\beta_0 f'_{\hbox {\tiny $X$}}(x)\nu_0^2(\nu_{\ell_1}\nu_{\ell_2+1}+\nu_{\ell_1+1}\nu_{\ell_2})\right)+O\left(J^{-1}h^{\ell_1+\ell_2+2-c^*} \right)
\end{align*}
For $\ell_1, \ell_2=0, 1, \ldots, p$, the above expression is the $[\ell_1+1, \ell_2+1]$ entry of the following $(p+1)\times (p+1)$ matrix, 
\begin{align}
&\ \nu_0 \beta_0 \bH\left[ \beta_0 f_{\hbox {\tiny $X$}}(x)\left\{w_{30}(x, h, J) \tilde \bnu_0+w_{31}(x, h, J)\left(3\nu_0\tilde \bnu_0+4 \bnu_0^*\bnu_0^{*\T}\right)+ \right.\right.\nonumber \\
& \left. w_{32}(x, h, J) \nu^2_0 \left(\tilde \bnu_0+3 \bnu^*_0\bnu^{*\T}_0 \right)+w_{33}(x, h, J)\nu_0^2 \bnu_0^*\bnu_0^{*\T}\right\} +\nonumber \\
&\ h \left(w_{30}(x, h, J)\left\{\beta_0 f'_{\hbox {\tiny $X$}}(x)+2\beta_1 f_{\hbox {\tiny $X$}}(x)  \right\}\tilde \bnu_1 +w_{31}(x, h, J) \left[\nu_0 \tilde \bnu_1\times \right. \right.\nonumber \\
& \left.\left\{2\beta_0 f'_{\hbox {\tiny $X$}}(x)+\beta_1 f_{\hbox {\tiny $X$}}(x)\right\}+
4\left\{\beta_0 f'_{\hbox {\tiny $X$}}(x)+\beta_1 f_{\hbox {\tiny $X$}}(x)\right\}\left(\bnu_0^*\bnu_1^{*\T}+\bnu_1^*\bnu_0^{*\T}  \right)\right]+\nonumber \\
&\ w_{32}(x, h, J)\nu_0\left[\left\{3\beta_0 f'_{\hbox {\tiny $X$}}(x)+\beta_1 f_{\hbox {\tiny $X$}}(x)\right\}\left(\bnu_0^*\bnu_1^{*\T}+\bnu_1^*\bnu_0^{*\T}\right)+\beta_0 f'_{\hbox {\tiny $X$}}(x) \nu_0\tilde \bnu_1 \right]+ \nonumber \\
& \left.\left. w_{33}(x, h, J)\beta_0 f'_{\hbox {\tiny $X$}}(x) \nu_0^2 \left(\bnu_0^*\bnu_1^{*\T}+\bnu_1^*\bnu_0^{*\T}\right)\right)+O\left(J^{-1}h^{2-c^*}\right)\right]\bH.
\label{eq:varterm2}
\end{align}

Summing (\ref{eq:varterm1}) and (\ref{eq:varterm2}) gives the following $(p+1)\times (p+1)$ matrix as the first term in (\ref{eq:varT2}), 
\begin{align}
&\ E\left\{J^{-1}\bT_2(x)J^{-1}\bT^\T_2(x) \right\} \nonumber \\
= &\ E\left[E\left\{J^{-1}\bT_2(x)J^{-1}\bT^\T_2(x)|\mathbb{X} \right\}\right] \nonumber \\
= &\ \nu_0 f_{\hbox {\tiny $X$}}(x) \bH \left[ w_{30}(x, h, J)\left\{\sigma^2(x)+\beta_0^2\right\}\tilde \bnu_0+w_{31}(x, h, J)\left\{ \sigma^2 (x) \left(\nu_0\tilde \bnu_0+2\bnu^*_0\bnu_0^{*\T}\right)\right. \right.\nonumber \\
& \left.+\beta^2_0 \left(3\nu_0 \tilde \bnu_0+4 \bnu_0^*\bnu_0^{*\T} \right)\right\}+w_{32}(x, h, J) \nu_0 \left\{\sigma^2(x) \bnu^*_0\bnu^{*\T}_0+\beta_0^2 \nu_0 \left(\tilde \bnu_0+3 \bnu^*_0\bnu^{*\T}_0\right)  \right\}\nonumber\\
& \left.+w_{33}(x, h, J)\beta_0^2 \nu_0^2 \bnu_0^*\bnu^{*\T}_0\right]\bH+
\nu_0 h \bH\left\{w_{30}(x, h, J)\left[\sigma^2(x) f'_{\hbox {\tiny $X$}}(x)+2\sigma(x)\sigma'(x) f_{\hbox {\tiny $X$}}(x)\right.\right.\nonumber \\
& \left. +\beta_0\left\{ \beta_0 f'_{\hbox {\tiny $X$}}(x)+2\beta_1 f_{\hbox {\tiny $X$}}(x) \right\}\right]\tilde \bnu_1 +w_{31}(x, h, J)\left( \sigma^2(x)f'_{\hbox {\tiny $X$}}(x) \left( \tilde \bnu_1 +\bnu_0^*\bnu_1^{*\T}+ \bnu_1^*\bnu_0^{*\T}\right) \right.\nonumber \\
&\ +2\sigma(x) \sigma'(x) f_{\hbox {\tiny $X$}}(x)\left(\bnu_0^*\bnu_1^{*\T}+ \bnu_1^*\bnu_0^{*\T} \right)+\beta_0 \left[\left\{2\beta_0 f'_{\hbox {\tiny $X$}}(x)+\beta_1 f_{\hbox {\tiny $X$}}(x) \right\}\nu_0\tilde \bnu_1   \right. \nonumber\\
&\left.\left. + 4\left\{\beta_0 f'_{\hbox {\tiny $X$}}(x)+\beta_1 f_{\hbox {\tiny $X$}}(x) \right\}\left(\bnu_0^*\bnu_1^{*\T}+ \bnu_1^*\bnu_0^{*\T}\right) \right]\right)+w_{32}(x, h, J) \nu_0 \left(\sigma^2(x)f'_{\hbox {\tiny $X$}}(x)\times  \right. \nonumber \\
&\left. \left(\bnu_0^*\bnu_1^{*\T}+ \bnu_1^*\bnu_0^{*\T} \right)+\beta_0\left[\left\{3\beta_0 f'_{\hbox {\tiny $X$}}(x)+\beta_1 f_{\hbox {\tiny $X$}}(x) \right\}  \left(\bnu_0^*\bnu_1^{*\T}+ \bnu_1^*\bnu_0^{*\T}\right)+\beta_0 f'_{\hbox {\tiny $X$}}(x) \nu_0\tilde \bnu_1 \right]\right) \nonumber \\
& \left.+ w_{33}(x, h, J)\nu_0^2 \beta_0^2 f'_{\hbox {\tiny $X$}}(x)\left(\bnu_0^*\bnu_1^{*\T}+ \bnu_1^*\bnu_0^{*\T} \right)+O\left(J^{-1}h^{1-c^*} \right)\right\}\bH. \label{eq:ET2T2}
\end{align}

Because (\ref{eq:ET2T2}) is of order $O\{\bH^2/(Jh^{c^*})\}$, whereas (\ref{eq:ET2ET2}) is of order $O(\bH^2)$, the dominating  terms of $\textrm{Var}\{J^{-1}\bT_2(x)\}$ are given by (\ref{eq:ET2T2}).

Using (\ref{eq:invS2}) and (\ref{eq:ET2T2}), we now have 
\begin{align}
& \mbox{Var}\left(\left. \hat \bbeta_2 \right\vert \mathbb{X} \right) \nonumber \\
=&\ \frac{\nu_0}{f_{\hbox {\tiny $X$}}(x)}\bH^{-1}\bD^{-1}_0(x)\left[ \bF(x, h, J)+h\frac{f'_{\hbox {\tiny $X$}}(x)}{f_{\hbox {\tiny $X$}}(x)}\left\{ \bF(x, h, J)\bD^{-1}_0(x)\bD_1(x)-\right. \right. \nonumber \\
& \left. f_{\hbox {\tiny $X$}}(x)\bD_1(x)\bD_0(x)^{-1}\bF(x, h, J) \bD_0(x)+\frac{\bG(x, h, J)}{f'_{\hbox {\tiny $X$}}(x)}\right\}+O(h^2)+\nonumber \\
& \left. O_{\hbox {\tiny $P$}}\left(\frac{1}{Jh^{c^*}} \right)\right]\bD^{-1}_0(x)\bH^{-1}, \label{eq:varbeta2}
\end{align}
where 
\begin{align*}
\bD_0(x) = &\ d_0(x) \tilde \bmu_0+d_1(x) \bmu_0^*\bmu_0^{*\T}, \\
\bD_1(x) = &\ d_0(x) \tilde \bmu_1+d_1(x) \left(\bmu_0^*\bmu_1^{*\T}+\bmu_1^*\bmu_0^{*\T} \right), \\
\bF(x, h, J) = &\ w_{30}(x, h, J)\left\{\sigma^2(x)+\beta_0^2\right\}\tilde \bnu_0+\\
&\ w_{31}(x, h, J)\left\{ \sigma^2 (x) \left(\nu_0\tilde \bnu_0+2\bnu^*_0\bnu_0^{*\T}\right)+
 \beta^2_0 \left(3\nu_0 \tilde \bnu_0+4 \bnu_0^*\bnu_0^{*\T} \right)\right\}+\\
&\ w_{32}(x, h, J) \nu_0 \left\{\sigma^2(x) \bnu^*_0\bnu^{*\T}_0+ \beta_0^2 \nu_0 \left(\tilde \bnu_0+3 \bnu^*_0\bnu^{*\T}_0\right) \right\}+\\
&\ w_{33}(x, h, J)\beta_0^2 \nu_0^2 \bnu_0^*\bnu^{*\T}_0,\\
\bG(x, h, J) = &\ w_{30}(x, h, J)\left[\sigma^2(x) f'_{\hbox {\tiny $X$}}(x)+2\sigma(x)\sigma'(x) f_{\hbox {\tiny $X$}}(x)+ \right. \\
& \left. \beta_0\left\{ \beta_0 f'_{\hbox {\tiny $X$}}(x)+2\beta_1 f_{\hbox {\tiny $X$}}(x) \right\}\right]\tilde \bnu_1 +\\
&\ w_{31}(x, h, J)\left( \sigma^2(x)f'_{\hbox {\tiny $X$}}(x) \left( \tilde \bnu_1 +\bnu_0^*\bnu_1^{*\T}+ \bnu_1^*\bnu_0^{*\T}\right)+2\sigma(x) \sigma'(x) \times \right.\nonumber \\
&\ f_{\hbox {\tiny $X$}}(x)  \left(\bnu_0^*\bnu_1^{*\T}+ \bnu_1^*\bnu_0^{*\T} \right)+\beta_0 \left[\left\{2\beta_0 f'_{\hbox {\tiny $X$}}(x)+\beta_1 f_{\hbox {\tiny $X$}}(x) \right\}\nu_0\tilde \bnu_1+   \right. \\
&\left.\left. 4\left\{\beta_0 f'_{\hbox {\tiny $X$}}(x)+\beta_1 f_{\hbox {\tiny $X$}}(x) \right\}\left(\bnu_0^*\bnu_1^{*\T}+ \bnu_1^*\bnu_0^{*\T}\right) \right]\right)+\\
&\ w_{32}(x, h, J) \nu_0 \left(\sigma^2(x)f'_{\hbox {\tiny $X$}}(x)\left(\bnu_0^*\bnu_1^{*\T}+\bnu_1^*\bnu_0^{*\T} \right)+   \right.  \\
& \left. \beta_0\left[\left\{3\beta_0 f'_{\hbox {\tiny $X$}}(x)+\beta_1 f_{\hbox {\tiny $X$}}(x) \right\}  \left(\bnu_0^*\bnu_1^{*\T}+ \bnu_1^*\bnu_0^{*\T}\right)+\beta_0 f'_{\hbox {\tiny $X$}}(x) \nu_0\tilde \bnu_1 \right]\right)+ \\
&\ w_{33}(x, h, J)\nu_0^2 \beta_0^2 f'_{\hbox {\tiny $X$}}(x)\left(\bnu_0^*\bnu_1^{*\T}+ \bnu_1^*\bnu_0^{*\T} \right).
\end{align*}
One major distinction from the counterpart result when individual-level data are available lies in the fact that the order of the dominating term inside the square brackets in (\ref{eq:varbeta2}) is determined by the order of $\bF(x, h, J)=O\{1/(Jh^{c^*})\}$. Consequently, the asymptotic variance of $\hat \bbeta_2$ is inflated compared to that of $\hat \bbeta_0$, with more inflation when $c^*$ is larger.

\setcounter{equation}{0} 
\setcounter{section}{0} 
\setcounter{subsection}{0} 
\setcounter{subsubsection}{0} 
\def\theequation{C.\arabic{equation}}
\renewcommand\thesubsection{C.\arabic{subsection}}
\renewcommand\thesubsubsection{C.\arabic{subsection}.\arabic{subsubsection}}
\section*{Appendix C: Proof of Theorem~1-(iii)}
\subsection{Bias of the marginal-integration estimator $\hat m_3(x)$}
Consider $\hat\bmbeta_3(x)=\bS_3^{-1}(x)\bT_3(x)$. For $\ell=0, 1, \ldots, p$, the $(\ell+1)$-th element of $\bT_3(x)$ is
\begin{align*}
T_{3,\ell}(x)=&\ \sum_{j=1}^J\sum_{k=1}^{c_j} \{c_jZ_j-(c_j-1)\widehat\mu\}(X_{jk}-x)^\ell K_h(X_{jk}-x)\\
=&\ \sum_{j=1}^J\sum_{k=1}^{c_j} \left\{\sum_{g=1}^{c_j}Y_{jg}-(c_j-1)\widehat\mu\right\}(X_{jk}-x)^\ell K_h(X_{jk}-x)\\
=&\ \sum_{j=1}^J\sum_{k=1}^{c_j}Y_{jk}(X_{jk}-x)^\ell K_h(X_{jk}-x)\\
&\ +\sum_{j=1}^J\sum_{k=1}^{c_j} \left\{\sum_{g=1,g\neq k}^{c_j}Y_{jg}-(c_j-1)\mu\right\}(X_{jk}-x)^\ell K_h(X_{jk}-x)\\
&\ +(\mu-\widehat\mu)\sum_{j=1}^J\sum_{k=1}^{c_j} (c_j-1)(X_{jk}-x)^\ell K_h(X_{jk}-x)\\
\equiv&\ T_{31,\ell}(x)+T_{32,\ell}(x)+(\mu-\widehat\mu)T_{33,l}(x).
\end{align*}
Further, denote $\bT_{31}(x)=(T_{31,0}(x),\dots, T_{31,p}(x))^\T$, $\bT_{32}(x)=(T_{32,0}(x),\dots, T_{32,p}(x))^\T$, and $\bT_{33}(x)=(T_{33,0}(x),\dots, T_{33,p}(x))^\T$. Then
$$
\hat \bmbeta_3(x)=\bS_3^{-1}(x)\bT_{31}(x)+\bS_3^{-1}(x)\bT_{32}(x)+(\mu-\widehat\mu)\bS_3^{-1}(x)\bT_{33}(x).
$$
Note that $\hat \bmbeta_0(x)\equiv \bS_3^{-1}(x)\bT_{31}(x)$ is the local polynomial estimator based on individual-level data, hence, 
\begin{align*}
& \mbox{Bias}\left\{\left.\hat\bmbeta_3(x)\right|\mathbb{X}\right\} \\
= & \mbox{Bias}\left\{\left.\hat \bmbeta_0(x)\right|\mathbb{X}\right\}+E\left\{\left.\bS_3^{-1}(x)\bT_{32}(x)\right|\mathbb{X}\right\}+E\left\{\left.(\mu-\widehat\mu)\bS_3^{-1}(x)\bT_{33}(x)\right|\mathbb{X}\right\},
\end{align*}
in which, by \citet{Fan&Gijbels1996}, 
\begin{align*}
\mbox{Bias}\left\{\hat\bmbeta_0(x)|\mathbb{X}\right\}=&\ h^{p+1}\bH^{-1}\left\{ \beta_{p+1} \tilde \bmmu_0^{-1}\bmmu_{p+1}^*+h\frac{\beta_{p+2}f_{\hbox {\tiny $X$}}(x)+\beta_{p+1}f'_{\hbox {\tiny $X$}}(x)}{f_{\hbox {\tiny $X$}}(x)}\tilde \bmmu_0^{-1}\bmmu_{p+2}^*\right.\\
&\ \left.-h\beta_{p+1}\frac{f'_{\hbox {\tiny $X$}}(x)}{f_{\hbox {\tiny $X$}}(x)}\tilde \bmmu_0^{-1}\tilde \bmmu_1\tilde \bmmu_0^{-1}\bmmu_{p+1}^*+O_{\hbox {\tiny $P$}}(h^2+1/\sqrt{Nh})\right\}.
\end{align*}
Also from \citet{Fan&Gijbels1996},
$$
\bS_3^{-1}(x)=N^{-1}\bH^{-1}\left\{f_X^{-1}(x)\tilde \bmmu_0^{-1}-h \frac{f_X'(x)}{f_X^2(x)}\tilde\bmmu_0^{-1}\tilde\bmmu_1\tilde\bmmu_0^{-1}+O_{\hbox {\tiny $P$}}(h^2)\right\}\bH^{-1},
$$
when $h\rightarrow 0$ and $Nh^3\rightarrow\infty$ as $N\rightarrow\infty$, provided that $f_{\hbox {\tiny $X$}}(x)>0$, $f_{\hbox {\tiny $X$}}(\cdot)$ and $m^{(p+1)}(\cdot)$ (or $f'_{\hbox {\tiny $X$}}(\cdot)$ and $m^{(p+2)}$(x) if $p-\ell$ is even) are continuous in a neighborhood of $x$.

Now we derive $E\{\bS_3^{-1}(x)\bT_{32}(x)|\mathbb{X}\}=\bS_3^{-1}(x)E\{\bT_{32}(x)|\mathbb{X}\}$. Note that
\begin{align*}
E\{T_{32,\ell}(x)|\mathbb{X}\}=\sum_{j=1}^J\sum_{k=1}^{c_j} \left\{\sum_{g=1,g\neq k}^{c_j}m(X_{jg})-(c_j-1)\mu\right\}(X_{jk}-x)^\ell K_h(X_{jk}-x).
\end{align*}
It is easy to see that the expectation of the right-hand side of the last equation is zero. Furthermore, 
\begin{align*}
&\ \mbox{Var}\left[\sum_{j=1}^J\sum_{k=1}^{c_j} \left\{\sum_{g=1,g\neq k}^{c_j}m(X_{jg})-(c_j-1)\mu\right\}(X_{jk}-x)^\ell K_h(X_{jk}-x)\right]\\
= &\ \sum_{j=1}^J E\left[\sum_{k=1}^{c_j}\left\{\sum_{g=1,g\neq k}^{c_j}m(X_{jg})-(c_j-1)\mu\right\} (X_{jk}-x)^\ell K_h(X_{jk}-x)\right]^2\\
=&\ \sum_{j=1}^J \sum_{k=1}^{c_j} E\left[\left\{\sum_{g=1,g\neq k}^{c_j}m(X_{jg})-(c_j-1)\mu\right\}^2 (X_{jk}-x)^{2\ell}K_h^2(X_{jk}-x)\right]\\
&\ +\sum_{j=1}^J \sum_{k_1\neq k_2}^{1,\dots,c_j} E\left[\left\{\sum_{g=1, g\neq k_1, k_2}^{c_j}m(X_{jg})+m(X_{jk_1})-(c_j-1)\mu\right\} (X_{jk_1}-x)^{\ell}K_h(X_{jk_1}-x)\right.\\
&\ \quad\quad\quad\quad\times \left. \left\{\sum_{g=1,g\neq k_1,k_2}^{c_j}m(X_{jg})+m(X_{jk_2})-(c_j-1)\mu\right\} (X_{jk_2}-x)^{\ell}K_h(X_{jk_2}-x)\right],
\end{align*}
which is
\begin{align*}
&\ \sum_{j=1}^J \sum_{k=1}^{c_j} E\left[\left\{\sum_{g=1,g\neq k}^{c_j}m(X_{jg})-(c_j-1)\mu\right\}^2 (X_{jk}-x)^{2\ell}K_h^2(X_{jk}-x)\right]\\
&\ +\sum_{j=1}^J \sum_{k_1\neq k_2}^{1,\dots,c_j} (c_j-2)\tilde\sigma^2E\left[ (X_{jk_1}-x)^{\ell}K_h(X_{jk_1}-x_0)(X_{jk_2}-x)^{\ell}K_h(X_{jk_2}-x)\right]\\
&\ + \sum_{j=1}^J \sum_{k_1\neq k_2}^{1,\dots,c_j}  E[\{m(X_{jk_1})-\mu\}\{m(X_{jk_2})-\mu\} (X_{jk_1}-x)^{\ell}K_h(X_{jk_1}-x)(X_{jk_2}-x)^{\ell}K_h(X_{jk_2}-x)]\\
=&\ \sum_{j=1}^J c_j (c_j-1)h^{2\ell-1} \tilde\sigma^2 f_{\hbox {\tiny $X$}}(x) \nu_{2\ell}\{1+o(1)\}\\
&\ + \sum_{j=1}^J c_j(c_j-1) (c_j-2)O(h^{2\ell})
 + \sum_{j=1}^J c_j(c_j-1) O(h^{2\ell}),
\end{align*}
in which $\tilde\sigma^2=E[\{m(X)-\mu\}^2]=\mbox{Var}\{m(X)\}$.
It follows that
\begin{align*}
E[T_{32,\ell}(x)|\mathbb{X}]=&\ 0+O_{\hbox {\tiny $P$}}\left(\sqrt{\sum_{j=1}^J c_j (c_j-1)h^{2\ell-1} \tilde\sigma^2 f_{\hbox {\tiny $X$}}(x)\nu_{2\ell}}\right)\\
=&\ Nh^\ell \tilde\sigma \sqrt{\sum_{j=1}^Jc_j(c_j-1)/\sum_{j=1}^Jc_j} \times O_{\hbox {\tiny $P$}}\left(1/\sqrt{Nh}\right).
\end{align*}
Thus
\begin{align*}
& E\left\{\bS_3^{-1}(x)\bT_{32}(x)|\mathbb{X}\right\}\\
=&\ \bH^{-1}\left\{f_{\hbox {\tiny $X$}}^{-1}(x)\tilde\bmmu_0^{-1}-h \frac{f_{\hbox {\tiny $X$}}'(x)}{f_{\hbox {\tiny $X$}}^2(x)}\tilde\bmmu_0^{-1}\tilde\bmmu_1\tilde\bmmu_0^{-1}+O_{\hbox {\tiny $P$}}(h^2)\right\}\bm{1}\times O_{\hbox {\tiny $P$}}(1/\sqrt{Nh})\\
=&\ \bH^{-1}\sqrt{\frac{\sum_{j=1}^Jc_j(c_j-1)}{\sum_{j=1}^Jc_j}} \times O_{\hbox {\tiny $P$}}(1/\sqrt{Nh}).
\end{align*}

To calculate
$E\{(\mu-\widehat\mu)\bS_3^{-1}(x)\bT_{33}(x)|\mathbb{X}\}$, where
$$
T_{33,l}(x)=\sum_{j=1}^J\sum_{k=1}^{c_j} (c_j-1)(X_{jk}-x)^\ell K_h(X_{jk}-x),
$$
we have
\begin{align*}
E\left\{(\mu-\widehat\mu)\bS_3^{-1}(x)\bT_{33}(x)|\mathbb{X}\right\}=E\left(\mu-\widehat\mu|\mathbb{X}\right)\bS_3^{-1}(x)\bT_{33}(x).
\end{align*}
It is easy to see that
$$
E\left(\mu-\widehat\mu|\mathbb{X}\right)=\mu-N^{-1}\sum_{j=1}^J\sum_{k=1}^{c_j}m(X_{ij})=O_{\hbox {\tiny $P$}}(1/\sqrt{N}),
$$
and
$$
\bT_{33}(x)=N\bH \{O(1)+O_{\hbox {\tiny $P$}}(1/\sqrt{Nh})\}.
$$
Thus
$$
E\left\{(\mu-\widehat\mu)\bS_3^{-1}(x)\bT_{33}(x)\right\}=\bH^{-1}O_{\hbox {\tiny $P$}}(1/\sqrt{N})=o_{\hbox {\tiny $P$}}\left[E\left\{\bS_3^{-1}(x)\bT_{32}(x)|\mathbb{X}\right\}\right]
$$
if $h\rightarrow 0$.
Thus, we have the bias term given by
\begin{align*}
\hat\bmbeta_3(x)-\bbeta=&\ h^{p+1}\bH^{-1}\left\{ \beta_{p+1} \tilde\bmmu_0^{-1}\bmmu_{p+1}^*+h\frac{\beta_{p+2}f_{\hbox {\tiny $X$}}(x)+\beta_{p+1}f'_X(x)}{f_{\hbox {\tiny $X$}}(x)}\tilde\bmmu_0^{-1}\bmmu_{p+2}^*\right.\\
&\ \quad\quad\quad\quad\left.-h\beta_{p+1}\frac{f'_{\hbox {\tiny $X$}}(x)}{f_{\hbox {\tiny $X$}}(x)}\tilde\bmmu_0^{-1}\tilde\bmmu_1\tilde\bmmu_0^{-1}\bmmu_{p+1}^*+O_{\hbox {\tiny $P$}}(h^2+1/\sqrt{Nh})\right\}\\
&\ +\bH^{-1}\sqrt{\frac{\sum_{j=1}^Jc_j(c_j-1)}{\sum_{j=1}^Jc_j}} \bm{1}\times O_{\hbox {\tiny $P$}}(1/\sqrt{Nh})\\
=&\ \hat\bmbeta_0(x)-\bbeta+\bH^{-1}\sqrt{\frac{\sum_{j=1}^Jc_j(c_j-1)}{\sum_{j=1}^Jc_j}} \bm{1}\times O_{\hbox {\tiny $P$}}(1/\sqrt{Nh}),
\end{align*}
where $\bm{1}$ is a vector of size $1$ of an appropriate length.
The effect of pooling in terms of bias is reflected by the second term in the last line, which disappears if $c_j=1$ for all $j$'s (i.e., no pooling).

When $p=0$, the marginal-integration estimator returns a local constant estimator with dominating bias given by 
\begin{align*}
& \mbox{Bias}\left\{\hat m_3(x)|\mathbb{X}\right\} \\
=&\ \left\{\frac{1}{2}m^{''}(x)+\frac{f_{\hbox {\tiny $X$}}'(x)}{f_X(x)}m'(x)\right\}\mu_2h^2+\sqrt{\frac{\sum_{j=1}^Jc_j(c_j-1)}{\sum_{j=1}^Jc_j}} \times O_{\hbox {\tiny $P$}}(1/\sqrt{Nh})\\
&\ +o_{\hbox {\tiny $P$}}(h^2+1/\sqrt{Nh}).
\end{align*}

Setting $p=1$, we have the local linear marginal-integration estimator with dominating bias given by 
\begin{align*}
\mbox{Bias}\left\{\hat m_3(x)|\mathbb{X}\right\}=&\ \frac{1}{2}m''(x)\mu_2h^2+\sqrt{\frac{\sum_{j=1}^Jc_j(c_j-1)}{\sum_{j=1}^Jc_j}} \times O_{\hbox {\tiny $P$}}(1/\sqrt{Nh})\\
& +o_{\hbox {\tiny $P$}}(h^2+1/\sqrt{Nh}).
\end{align*}

\subsection{Variance of $\hat m_3(x)$}
Again, we focus on $\hat\bmbeta_3(x)=\bS_3^{-1}(x)\bT_3(x)$, where
\begin{align*}
T_{3,\ell}(x)=&\ \sum_{j=1}^J\sum_{k=1}^{c_j} \{c_jZ_j-(c_j-1)\widehat\mu\}(X_{jk}-x)^\ell K_h(X_{jk}-x)\\
=&\ \sum_{j=1}^Jc_jZ_j \sum_{k=1}^{c_j}(X_{jk}-x)^\ell K_h(X_{jk}-x)\\
&\ -\widehat\mu\sum_{j=1}^J\sum_{k=1}^{c_j} (c_j-1)(X_{jk}-x)^\ell K_h(X_{jk}-x)\\
\equiv&\ T_{31,\ell}^*(x)+T_{32,\ell}^*(x).
\end{align*}
First, we have
\begin{align*}
\mbox{Cov}\left\{T_{31,\ell_1}^*(x), T_{31,\ell_2}^*(x)|\mathbb{X}\right\}=&\ \sum_{j=1}^J \sum_{k=1}^{c_j}\sigma^2(X_{jk})\sum_{k=1}^{c_j}(X_{jk}-x)^{\ell_1}K_h(X_{jk}-x)\\
&\ \times \sum_{k=1}^{c_j}(X_{jk}-x)^{\ell_2}K_h(X_{jk}-x).
\end{align*}
Straightforward calculation presents, by (\ref{eq:EXK2}) and (\ref{eq:EgXK2}),
\begin{align*}
&\ E\left\{\mbox{Cov}[T_{31,\ell_1}^*(x), T_{31,\ell_2}^*(x)|\mathbb{X}]\right\}\\ 
=&\ \sum_{j=1}^JE\left\{\sum_{k=1}^{c_j}\sigma^2(X_{jk})\sum_{k=1}^{c_j}(X_{jk}-x)^{\ell_1+\ell_2}K_h^2(X_{jk}-x)\right\}+\mbox{Lower-order terms}\\
= & \ \sum_{j=1}^J \left[ c_j \kappa^*_{2,\ell_1+\ell_2}(\sigma^2, h, x)+c_j(c_j-1)E\left\{\sigma^2(X) \right\}\kappa_{2,\ell_1+\ell_2}(h,x)\right]+\mbox{Lower-order terms} \\
=&\ Nh^{\ell_1+\ell_2-1}f_{\hbox {\tiny $X$}}(x)\nu_{\ell_1+\ell_2}\left\{\sigma^2(x)+\bar\sigma^2N^{-1}\sum_{j=1}^J\sum_{k=1}^{c_j}(c_j-1)\right\}+O\left(h^{\ell_1+\ell_2}  \right), 
\end{align*}
where $\bar \sigma^2=E\{\sigma^2(X)\}$.
Moreover,
\begin{align*}
&\ \mbox{Var}\left[\mbox{Cov}\left\{T_{31,\ell_1}^*(x), T_{31,\ell_2}^*(x)|\mathbb{X}\right\}\right]\\
=&\ \sum_{j=1}^J \mbox{Var}\left[\sum_{k=1}^{c_j}\sigma^2(X_{jk})\sum_{k=1}^{c_j}(X_{jk}-x)^{\ell_1}K_h(X_{jk}-x)\sum_{k=1}^{c_j}(X_{jk}-x)^{\ell_2}K_h(X_{jk}-x)\right]\\
=&\ O(Nh^{2\ell_1+2\ell_2-3}).
\end{align*}
Thus
$$
\mbox{Cov}\left\{\bT_{31}^*(x)|\mathbb{X}\right\}= h^{-1}N\bH \left\{\sigma^2(x)+\bar\sigma^2N^{-1}\sum_{j=1}^J\sum_{k=1}^{c_j}(c_j-1)\right\}f_{\hbox {\tiny $X$}}(x)\tilde \bnu_0\{1+o_p(1)\}\bH.
$$
Then
\begin{align*}
& \mbox{Cov}[\bS_3^{-1}(x)\bT_{31}^*(x)|\mathbb{X}]\\
=&\  N^{-1}\bH^{-1}\left\{f_{\hbox {\tiny $X$}}^{-1}(x)\tilde\bmu_0^{-1}+O_{\hbox {\tiny $P$}}(h)\right\}\bH^{-1} \\
&\ \times h^{-1}N\bH \left\{\sigma^2(x)+\bar\sigma^2N^{-1}\sum_{j=1}^J\sum_{k=1}^{c_j}(c_j-1)\right\}f_{\hbox {\tiny $X$}}(x)\tilde \bnu_0\{1+o_{\hbox {\tiny $P$}}(1)\}\bH\\
&\ \times N^{-1}\bH^{-1}\left\{f_{\hbox {\tiny $X$}}^{-1}(x)\tilde \bmu_0^{-1}+O_{\hbox {\tiny $P$}}(h)\right\}\bH^{-1}\\
=&\ \frac{\sigma^2(x)+\bar\sigma^2N^{-1}\sum_{j=1}^J\sum_{k=1}^{c_j}(c_j-1)}{hf_{\hbox {\tiny $X$}}(x)}N^{-1}\bH^{-1} \tilde \bmu_0^{-1}\tilde \bnu_0 \tilde \bmu_0^{-1}\bH^{-1}\{1+o_{\hbox {\tiny $P$}}(1)\},
\end{align*}
in which $\sigma^2(x)\{hf_X(x)\}^{-1} N^{-1}\bH^{-1} \tilde \bmu_0^{-1}\tilde \bnu_0\tilde \bmu_0^{-1}\bH^{-1}\{1+o_p(1)\}$ is the same as that in \citet{Fan&Gijbels1996} and $\bar\sigma^2N^{-1}\sum_{j=1}^J\sum_{k=1}^{c_j}(c_j-1)$ accounts for the effect of pooling and disappears when $c_j=1$ for all $j$'s.

Now we calculate $\mbox{Cov}\{\bS_3^{-1}(x)\bT_{32}^*(x)|\mathbb{X}\}$.  
\begin{align*}
\mbox{Cov}\left\{T_{32,\ell_1}^*(x),T_{32,\ell_2}^*(x)|\mathbb{X}\right\}=&\ \sum_{j=1}^J\sum_{k=1}^{c_j} (c_j-1)(X_{jk}-x)^{\ell_1}K_h(X_{jk}-x)\mbox{Var}(\widehat\mu|\mathbb{X})\\
&\ \times \sum_{j=1}^J\sum_{k=1}^{c_j} (c_j-1)(X_{jk}-x)^{\ell_2}K_h(X_{jk}-x),
\end{align*}
in which
\begin{align*}
\mbox{Var}(\widehat\mu|\mathbb{X})=\frac{\sum_{j=1}^J\sum_{k=1}^{c_j}\sigma^2(X_{ij})}{N^2}=N^{-1}\{\bar\sigma^2+O_{\hbox {\tiny $P$}}(1/\sqrt{N})\}
\end{align*}
and
\begin{align*}
&\ \sum_{j=1}^J\sum_{k=1}^{c_j} (c_j-1)(X_{jk}-x)^\ell K_h(X_{jk}-x)\\
=&\ f_X(x)h^\ell \mu_\ell\sum_{j=1}^J\sum_{k=1}^{c_j}(c_j-1)\{1+o(1)\}\\
&\ +h^\ell O_{\hbox {\tiny $P$}}\left(\sqrt{h^{-1}\sum_{j=1}^J\sum_{k=1}^{c_j}(c_j-1)^2}\right).
\end{align*}
Hence
\begin{align*}
& \mbox{Cov}\left\{T_{32,\ell_1}^*(x),T_{32,\ell_2}^*(x)|\mathbb{X}\right\}\\
=&\ Nh^{\ell_1}\mu_{\ell_1}f_{\hbox {\tiny $X$}}(x)\frac{\sum_{j=1}^J\sum_{k=1}^{c_j}(c_j-1)}{N}\{1+o(1)+O_{\hbox {\tiny $P$}}(1/\sqrt{Nh})\}\\
&\ \times Nh^{\ell_2}\mu_{\ell_2}f_{\hbox {\tiny $X$}}(x)\frac{\sum_{j=1}^J\sum_{k=1}^{c_j}(c_j-1)}{N}\{1+o(1)+O_{\hbox {\tiny $P$}}(1/\sqrt{Nh})\}\\
&\ \times N^{-1}\{\bar\sigma^2+O_{\hbox {\tiny $P$}}(1/\sqrt{N})\};
\end{align*}
i.e., 
\begin{align*}
\mbox{Cov}[\bT_{33}^*(x)|\mathbb{X}]=&\  \bar\sigma^2f_X^2(x)\left(\frac{\sum_{j=1}^J\sum_{k=1}^{c_j}(c_j-1)}{N}\right)^2 N\bH\tilde\bmmu_0\tilde\bmmu_0^\t \bH\{1+o_{\hbox {\tiny $P$}}(1)\}\\
=&\ \mbox{Cov}[\bT_{31}^*(x)|\mathbb{X}]\times o_{\hbox {\tiny $P$}}(1),
\end{align*}
which becomes negligible.

The last term is $\mbox{Cov}\{\bS_3^{-1}(x)\bT_{31}^*(x),\bS_3^{-1}(x)\bT_{32}^*(x)|\mathbb{X}\}$, in which $\mbox{Cov}\{T_{31, \ell_1}^*(x), T_{32,\ell_2}^*(x)|\mathbb{X}\}$ equals to
\begin{align*}
&\ \mbox{Cov}\left\{\sum_{j=1}^Jc_jZ_j \sum_{k=1}^{c_j}(X_{jk}-x)^{\ell_1}K_h(X_{jk}-x), \widehat\mu\sum_{j=1}^J\sum_{k=1}^{c_j} (c_j-1)(X_{jk}-x)^{\ell_2}K_h(X_{jk}-x)|\mathbb{X}\right\}\\
=&\ \sum_{j=1}^J \mbox{Cov}\left(c_jZ_j,\widehat\mu|\mathbb{X}\right)\sum_{k=1}^{c_j}(X_{jk}-x)^{\ell_1}K_h(X_{jk}-x)\times \sum_{j=1}^J\sum_{k=1}^{c_j} (c_j-1)(X_{jk}-x)^{\ell_2}K_h(X_{jk}-x)\\
=&\ N^{-1} \sum_{j=1}^J \left\{\sum_{k=1}^{c_j}\sigma^2(X_{jk})\sum_{k=1}^{c_j}(X_{jk}-x)^{\ell_1}K_h(X_{jk}-x)\right\}\times \sum_{j=1}^J\sum_{k=1}^{c_j} (c_j-1)(X_{jk}-x)^{\ell_2}K_h(X_{jk}-x),
\end{align*}
in which
\begin{align*}
\sum_{j=1}^J \left\{\sum_{k=1}^{c_j}\sigma^2(X_{jk})\sum_{k=1}^{c_j}(X_{jk}-x)^{\ell_1}K_h(X_{jk}-x)\right\}=&\ O(Nh^{\ell_1})+O_{\hbox {\tiny $P$}}(\sqrt{Nh^{2\ell_1-1}})\\
=&\ Nh^{\ell_1}\{O(1)+O_p(1/\sqrt{Nh})\},\\
\sum_{j=1}^J\sum_{k=1}^{c_j} (c_j-1)(X_{jk}-x)^\ell K_h(X_{jk}-x)
=&\ O(Nh^{\ell_2})+O_{\hbox {\tiny $P$}}(\sqrt{Nh^{\ell_2-1}})\\
=&\ Nh^{\ell_2}\{O(1)+O_{\hbox {\tiny $P$}}(1/\sqrt{Nh})\}.
\end{align*}
Thus
$$
\mbox{Cov}\left\{T_{31,\ell_1}^*(x), T_{32,\ell_2}^*(x)|\mathbb{X}\right\}=Nh^{\ell_1+\ell_2}\{O(1)+O_{\hbox {\tiny $P$}}(1/\sqrt{Nh})\},
$$
i.e.,
$$
\mbox{Cov}\left\{\bT_{31}^*(x),\bT_{32}^*(x)|\mathbb{X}\right\}=\mbox{Cov}\left\{\bT_{31}^*(x)|\mathbb{X}\right\}\times o_{\hbox {\tiny $P$}}(1),
$$
which also becomes negligible.

Finally, we have
\begin{align*}
\mbox{Var}\left\{\left.\hat\bbeta_3(x)\right|\mathbb{X}\right\}=\frac{\sigma^2(x)+\bar\sigma^2N^{-1}\sum_{j=1}^J\sum_{k=1}^{c_j}(c_j-1)}{hf_X(x)}N^{-1}\bH^{-1} \tilde \bmu_0^{-1}\tilde \bnu_0\tilde \bmu_0^{-1}\bH^{-1}\{1+o_{\hbox {\tiny $P$}}(1)\}.
\end{align*}

\subsubsection{Variance of $\hat m_3(x)$ when $p=0, 1$}
With $p=0$ or 1, we have 
\begin{align*}
&\ \mbox{Var}\left\{\hat m^{(3)}(x)|\mathbb{X}\right\}\\
=&\ \nu_0\frac{\sigma^2(x)+\bar\sigma^2N^{-1}\sum_{j=1}^J\sum_{k=1}^{c_j}(c_j-1)}{f_X(x)Nh}\{1+o_{\hbox {\tiny $P$}}(1)\}.
\end{align*}

\setcounter{equation}{0} 
\setcounter{section}{0} 
\setcounter{subsection}{0} 
\setcounter{subsubsection}{0} 
\def\theequation{D.\arabic{equation}}
\renewcommand\thesubsection{D.\arabic{subsection}}
\renewcommand\thesubsubsection{D.\arabic{subsection}.\arabic{subsubsection}}
\section*{Appendix D: Proof of Theorem~2}
It is assumed that $c_j=c$, for $j=1, \ldots, J$, in this appendix. Suppose that covariate data are sorted before forming pools, resulting in ordered covariate data, $X_{(11)}\le X_{(12)} \le \ldots \le X_{(1c)} \le X_{(21)}\le \ldots \le X_{(2c)}\le \ldots \le X_{(Jc)}$, so that $\tilde \bX_{(j)}=(X_{(j1)}, \ldots, X_{(jc)})^\t$ are covariate data in the $j$th pool, and $Z_{(j)}$ is the corresponding pooled response, for $j=1, \ldots, J$. 

Throughout this section, we consider the kernel function $K(\cdot)$ satisfying that $K(|t|)=0$ if $|t|>1$. Kernels that satisfy this condition include the Epanechnikov, quartic, triweight, and tricube kernel. Along with the condition that $h\rightarrow 0$ as $N\rightarrow\infty$, this condition on the kernel shares the same spirit as Condition (T5) in \citet{delaigle2012nonparametric}.

\subsection{Bias and variance of $\hat m_1(x)$}
Using pooled data from homogeneous pooling, the weighted least squares objective function $Q_1(\bbeta)$ defined in the main article is  
$$Q_1(\bbeta)=
\sum_{j=1}^{J}\left\{Z_{(j)}-\sum_{\ell=0}^p \beta_{\ell} c^{-1} \sum_{k=1}^{c}\left(X_{(j k)}-x\right)^{\ell}\right\}^2\left\{c^{-1} \sum_{k=1}^{c} K_{h}\left(X_{(jk)}-x\right)\right\}.
$$
Then the averaged-weighted local polynomial estimator of order $p$ for $m(x)$ is $\hat m_1(x)= \be_1^\T\bS_1^{-1}(x)\bT_1(x)$, where entries in 
$\bS_1(x)=\ [S_{1,\ell_1, \ell_2}(x)]_{\ell_1,\ell_2=0,1,\dots, p}$ and $\bT_1(x)=\ (T_{1,0}(x), T_{1,1}(x),\dots, T_{1,p}(x))^\T$ are 
\begin{align*}
S_{1,\ell_1,\ell_2}(x)=&\ J^{-1}\sum_{j=1}^{J}\left\{c^{-1} \sum_{k=1}^{c}\left(X_{(jk)}-x\right)^{\ell_1}\right\}\left\{c^{-1} \sum_{k=1}^{c}\left(X_{(jk)}-x\right)^{\ell_2}\right\}\times\\
&\left\{c^{-1} \sum_{k=1}^{c} K_{h}\left(X_{(jk)}-x\right)\right\}, \mbox{ for } \ell_1, \ell_2=0,1,\dots, p,\\
T_{1,\ell}(x)=&\ J^{-1}\sum_{j=1}^JZ_{(j)}\left\{c^{-1} \sum_{k=1}^{c}\left(X_{(jk)}-x\right)^{\ell}\right\}\left\{c^{-1} \sum_{k=1}^{c} K_{h}\left(X_{(jk)}-x\right)\right\}, \\
& \mbox{ for } \ell=0,1,\dots, p.
\end{align*}

We focus on deriving the bias of $\hat\bmbeta_1(x)=\bS_1^{-1}(x)\bT_1(x)$ conditioning on the collection of all covariate data $\mathbb{X}$ first,
\begin{align*}
\mbox{Bias}[\hat\bmbeta_1(x)|\mathbb{X}]=\blS_1^{-1}(x)E[\blT_1(x)-\blS_1(x)\bmbeta(x)|\mathbb{X}].
\end{align*}
Because, for $\ell=0,1,\dots, p$, the $\ell$th component of $\bT_1(x)-\bS_1(x)\bbeta(x)$ is 
\begin{align*}
T_{1,\ell}(x)-\bS_{1}(x)[\ell,\,  ]\bmbeta(x)=&\ J^{-1}\sum_{j=1}^J\left\{Z_{(j)}-\sum_{\ell'=0}^p \beta_{\ell'} c^{-1} \sum_{k=1}^{c}\left(X_{(j k)}-x\right)^{\ell'}\right\}\\
&\ \times \left\{c^{-1} \sum_{k=1}^{c}\left(X_{(jk)}-x\right)^{\ell}\right\}\left\{c^{-1} \sum_{k=1}^{c} K_{h}\left(X_{(jk)}-x\right)\right\}, 
\end{align*}
one has 
\begin{align}
 & E\{T_{1,\ell}(x)-\bS_{1}(x)[\ell,\, ]\bmbeta(x)|\mathbb{X}\} \nonumber \\
=&\ N^{-1} \sum_{j=1}^J \left\{\sum_{k=1}^{c} K_{h}\left(X_{(jk)}-x\right)\right\}\left\{c^{-1} \sum_{k=1}^{c}\left(X_{(jk)}-x\right)^{\ell}\right\} \nonumber \\
& \times \left[ c^{-1}\sum_{k=1}^{c}\left\{m(X_{(jk)})-\sum_{\ell'=0}^p \beta_{\ell'} \left(X_{(j k)}-x\right)^{\ell'}\right\}\right]\label{e:T1l}.
\end{align}

Here, we only consider $x$ being in the interior set of $\cI$ that is a compact, nondegenerate interval. Besides continuity, we also assume $f_X(\cdot)$ bounded away from zero on an open interval $\cJ$ such that $\cI\subset\cJ$, which is precisely Condition (T1) in \citet{delaigle2012nonparametric}. For any such $x$, we consider the cumulative distribution function (cdf) of $X$ evaluated at $x$, $F_X(x)$. Because $x$ is an interior point of $\cI$, and $f_X(x)$ is bounded away from zero in $\cJ$, there exist $a$ and $b$ such that $a<x<b$ and $0<F_X(a)<F_X(x)<F_X(b)<1$.

We argue that when $N$ is large, for any $j$, if one of $X_{(j1)}, \dots, X_{(jc)}$ is in $[x-h, x+h]$, then all covariate data in pool $j$ fall in $[a,b]$, where we note that $h \rightarrow 0$ as $N\rightarrow\infty$. To signify the dependence on $N$ of this part of the discussions, we write these covariate data as $X_{(j1:N)}, \dots, X_{(jc:N)}$ and the bandwidth as $h_N$. 
Now suppose the opposite is true. Then there exists a sequence of sample sizes $c< N_1<N_2<\cdots< N_m<\cdots<\infty$, where $\lim_{m\rightarrow\infty}N_m=\infty$, such that for each $m$, $-h_{N_m}<X_{(j1:N_m)}-x<h_{N_m}$ and $X_{(jc:N_m)}>b+\epsilon$ for some $j\in\{1, \dots, J\}$, where $\epsilon$ is a very small constant such that $F_X(b+\epsilon)>F_X(b)$. We have that $F_{N_k}(X_{(j1:N_m)})=((j-1)c+1)/N_m$ and $F_{N_k}(X_{(jc:N_m)})=\{(j-1)c+k\}/N_m$, where $F_N$ is the empirical function of $F_X$ when the sample size is $N$. Because $h_{N_m}$ goes to zero as $m$ goes to infinity, $\lim_{m\rightarrow\infty}X_{(j1:N_m)}=x$. Further more, because $F_{N_k}(\cdot)$ converges to $F_X(\cdot)$ from the classic uniform convergence of an empirical process, and that $F_X$ is continuous, we conclude that $\lim_{m\rightarrow\infty}\{(j-1)c+1\}/N_m=F_X(x)<F_X(b)$. Consequently, $\lim_{m\rightarrow\infty}\{(j-1)c+c\}/N_m=F_X(x)<F_X(b)$; i.e., $\lim_{m\rightarrow\infty}F_{N_k}(X_{jc:N_m})=F_X(x)<F_X(b)$. However, $F_{N_k}(X_{jc:N_m})>F_{N_k}(b+\epsilon)$ implies that  $\lim_{m\rightarrow\infty}F_{N_k}(X_{jc:N_m})\geq F_X(b+\epsilon)> F_X(b)$, which provides a contradiction. 

Define a partition of the index set $\{1, \ldots, J\}=\cJ_1\cup \cJ_2$, where $\cJ_1=\{j\in \{1, \ldots, J\}: \, \textrm{at least one of $X_{(j1)}, \dots, X_{(jc)}$ is in $[x-h,x+h]$}\}$, and $\cJ_2=\{j\in \{1, \ldots, J\}: \, \textrm{none of $X_{(j1)}, \dots, X_{(jc)}$ are in $[x-h,x+h]$}\}$. Following this partition, \eqref{e:T1l} can be re-expressed as
$$
E\left\{\left.T_{1,\ell}(x)-\bS_{1,\ell,\cdot}(x)\bmbeta(x)\right|\mathbb{X}\right\}=\cL_1+\cL_2,
$$
where, for $a=1,2$, $\cL_a=$
$$
N^{-1}\sum_{j\in\cJ_a}\sum_{k=1}^{c} K_{h}\left(X_{(jk)}-x\right)\left\{c^{-1} \sum_{k=1}^{c}\left(X_{(jk)}-x\right)^{\ell}\right\}c^{-1}\sum_{k=1}^{c}\left\{m(X_{(jk)})-\sum_{\ell=0}^p \beta_{\ell} \left(X_{(j k)}-x\right)^{\ell}\right\}.
$$
When $j\in\cJ_2$, all the $K_{h}\left(X_{(jk)}-x\right)$'s are zero. Hence
\begin{align*}
\cL_2=0=N^{-1}\sum_{j\in\cJ_2}\sum_{k=1}^{c} K_{h}\left(X_{(jk)}-x\right)\left(X_{(jk)}-x\right)^{\ell} \left\{m(X_{(jk)})-\sum_{\ell=0}^p \beta_{\ell} \left(X_{(j k)}-x\right)^{\ell}\right\}.
\end{align*}
When $j\in\cJ_1$, we known that when $N$ is large, all the $X_{(j1)}, \dots, X_{(jc)}$ will be in $[a,b]$. Using the classical result of an empirical quantile process, we know that
$$
\sup_{u\in[F_X(a),F_X(b)]}\sqrt{N}|F_N^{-1}(u)-F_X^{-1}(u)|=O_{\hbox {\tiny $P$}}(1).
$$
where we denote by $F_N^{-1}$ the empirical quantile function using the random sample $X_{jk}$'s and $F^{-1}$ the quantile function of the distribution of $X$. Then we have 
$X_{(jk)}=F_N^{-1}(\{(j-1)c+k\}/N)$,
and
\begin{align}
& \sup_{j\in\cJ_1} \sqrt{N}|X_{(jc)}-X_{(j1)}|\nonumber \\
\leq &\ \sup_{j\in\cJ_1} \sqrt{N}\left|F_N^{-1}\left(\frac{\sum_{m=1}^{j-1}c_m+c}{N}\right)-F_X^{-1}\left(\frac{\sum_{m=1}^{j-1}c_m+c}{N}\right)\right|\nonumber\\
&\ +\sup_{j\in\cJ_1} \sqrt{N}\left|F_N^{-1}\left(\frac{\sum_{m=1}^{j-1}c_m+1}{N}\right)-F_X^{-1}\left(\frac{\sum_{m=1}^{j-1}c_m+1}{N}\right)\right|\nonumber\\
&\ +\sup_{j\in\cJ_1} \sqrt{N}\left|F_X^{-1}\left(\frac{\sum_{m=1}^{j-1}c_m+c}{N}\right)-F_X^{-1}\left(\frac{\sum_{m=1}^{j-1}c_m+1}{N}\right)\right|\nonumber\\
\leq &\ 2\sup_{u\in[F_X(a),F_X(b)]} \sqrt{N}|F_N^{-1}(u)-F_X^{-1}(u)|\nonumber\\
&\ +\sup_{j\in\cJ_1} \sqrt{N}\left|F_X^{-1}\left(\frac{\sum_{m=1}^{j-1}c_m+c}{N}\right)-F_X^{-1}\left(\frac{\sum_{m=1}^{j-1}c_m+1}{N}\right)\right|.\label{e:dis}
\end{align}
Hence, when $j\in\cJ_1$, $\sup_j|X_{(jc)}-X_{(j1)}|=O_{\hbox {\tiny $P$}}(1/\sqrt{N})$.

Then, for $j\in\cJ_1$, we have $|X_{(jk)}-x|\leq h$. Consequently,
\begin{align*}
c^{-1} \sum_{k=1}^{c}\left(X_{(jk)}-x\right)^{\ell}=&\ \left(X_{(jk)}-x\right)^{\ell}+c^{-1}\sum_{k_1\neq k}^c \{(X_{(jk_1)}-x)^\ell-(X_{(jk)}-x)^\ell\}\\
=&\ \left(X_{(jk)}-x\right)^{\ell}+O_p(h^{\ell-1}/\sqrt{N})\{1-I(\ell=0)\}.
\end{align*}
Assuming that $m(\cdot)$ is $(p+2)$-th order continuously differentiable,
\begin{align*}
&\ c^{-1}\sum_{k_1=1}^{c}\left\{m(X_{(jk_1)})-\sum_{\ell=0}^p \beta_{\ell} \left(X_{(j k_1)}-x\right)^{\ell}\right\}-\left\{m(X_{(jk)})-\sum_{\ell=0}^p \beta_{\ell} \left(X_{(j k)}-x\right)^{\ell}\right\}\\
=&\ c^{-1}\sum_{k_1\neq k}\left[\left\{m(X_{(jk_1)})-\sum_{\ell=0}^p \beta_{\ell} \left(X_{(j k_1)}-x\right)^{\ell}\right\}-\left\{m(X_{(jk)})-\sum_{\ell=0}^p \beta_{\ell} \left(X_{(j k)}-x\right)^{\ell}\right\}\right]\\
=&\ c^{-1}\sum_{k_1\neq k}\left\{m^{(p+1)}(X_{(jk_1)}^*)(X_{(jk_1)}-x)^{p+1}-m^{(p+1)}(X_{(jk)}^*)(X_{(jk)}-x)^{p+1}\right\}/(p+1)!\\
=&\ O_{\hbox {\tiny $P$}}(h^p/\sqrt{N}),
\end{align*}
where $X_{(jk_1)}^*$ lies between $X_{(jk_1)}$ and $x$, and $X_{(jk)}^*$ lies between $X_{(jk)}$ and $x$. 
 
Consequently, recalling the definition of $\cJ_2$, we can write $E\{T_{1,\ell}(x_0)-\blS_{1,\ell,\cdot}(x)\bmbeta(x)|\mathbb{X}\}=\cL_1$ as follows, 
\begin{align*}
&\ N^{-1}\sum_{j=1}^J\sum_{k=1}^{c} K_{h}\left(X_{(jk)}-x\right)\left(X_{(jk)}-x\right)^{\ell}\left\{m(X_{(jk)})-\sum_{\ell=0}^p \beta_{\ell} \left(X_{(j k)}-x\right)^{\ell}\right\}\\
&\ +O_{\hbox {\tiny $P$}}(h^{\ell-1}/\sqrt{N})\{1-I(l=0)\}N^{-1}\sum_{j=1}^J\sum_{k=1}^{c} K_{h}\left(X_{(jk)}-x\right)\left\{m(X_{(jk)})-\sum_{\ell=0}^p \beta_{\ell} \left(X_{(j k)}-x\right)^{\ell}\right\}\\
&\ +O_{\hbox {\tiny $P$}}(h^p/\sqrt{N})N^{-1}\sum_{j=1}^J\sum_{k=1}^{c} K_{h}\left(X_{(jk)}-x\right)\left(X_{(jk)}-x\right)^{\ell}\\
&\ +O_{\hbox {\tiny $P$}}(h^{\ell-1}/\sqrt{N})\{1-I(\ell=0)\}O_{\hbox {\tiny $P$}}(h^p/\sqrt{N}) N^{-1}\sum_{j=1}^J\sum_{k=1}^{c} K_{h}\left(X_{jk}-x\right)\\
=&\ N^{-1}\sum_{j=1}^J\sum_{k=1}^{c} K_{h}\left(X_{(jk)}-x\right)\left(X_{(jk)}-x_0\right)^{\ell}\left\{m(X_{(jk)})-\sum_{\ell=0}^p \beta_{\ell} \left(X_{(j k)}-x\right)^{\ell}\right\}\{1+o_{\hbox {\tiny $P$}}(1)\}
\end{align*}
provided that $Nh^4\rightarrow\infty$.
Similarly, we have
\begin{align*}
S_{1,\ell_1,\ell_2}(x)=&\ N^{-1}\sum_{j=1}^J\sum_{k=1}^{c} K_{h}\left(X_{jk}-x\right)\left(X_{jk}-x\right)^{\ell_1+\ell_2}\{1+o_{\hbox {\tiny $P$}}(1)\}.
\end{align*}
Hence, $\mbox{Bias}\{\hat\bmbeta_1(x)|\mathbb{X}\}=\mbox{Bias}\{\hat\bmbeta_0(x)|\mathbb{X}\}\{1+o_{\hbox {\tiny $P$}}(1)\}$.

Now we consider the variance,
$$
\mbox{Cov}\{\bS_1(x)^{-1}\bT_1(x)|\mathbb{X}\}=\bS_1(x)^{-1}\mbox{Cov}\{\bT_1(x)|\mathbb{X}\}\bS_1(x)^{-1}.
$$
We see that
\begin{align*}
\mbox{Cov}\left\{T_{1,\ell_1}(x), T_{1,\ell_2}(x)|\mathbb{X}\right\}=&\ J^{-2}\sum_{j=1}^Jc^{-2}\sum_{k=1}^{c}\sigma^2(X_{(jk)})\left\{c^{-1} \sum_{k=1}^{c}\left(X_{(jk)}-x\right)^{\ell_1}\right\}\\
&\ \times \left\{c^{-1} \sum_{k=1}^{c}\left(X_{(jk)}-x\right)^{\ell_2}\right\}\left\{c^{-1} \sum_{k=1}^{c} K_{h}\left(X_{(jk)}-x\right)\right\}^2.
\end{align*}
Similarly, we define $\cJ_1$ and $\cJ_2$. For $j\in\cJ_2$,
\begin{align*}
&\ J^{-2}\sum_{j\in\cJ_2}c^{-2}\sum_{k=1}^{c}\sigma^2(X_{(jk)})\left\{c^{-1} \sum_{k=1}^{c}\left(X_{(jk)}-x\right)^{l_1}\right\}\\
&\ \times \left\{c^{-1} \sum_{k=1}^{c}\left(X_{(jk)}-x\right)^{l_2}\right\}\left\{c^{-1} \sum_{k=1}^{c} K_{h}\left(X_{(jk)}-x\right)\right\}^2\\
=&\ (Nh)^{-1}N^{-1}\sum_{j\in\cJ_2}\sum_{k=1}^c\sigma^2(X_{jk})\left(X_{jk}-x\right)^{\ell_1+\ell_2}
K^\ddagger_{h}\left(X_{jk}-x\right),
\end{align*}
where $K^\ddagger(t)=K^2(t)$.
For $j\in\cJ_1$, assuming smoothness of $\sigma^2(\cdot)$, we have
\begin{align*}
c^{-1}\sum_{k=1}^{c}\sigma^2(X_{(jk)})=&\ \sigma^2(X_{jk})+O_{\hbox {\tiny $P$}}(1/\sqrt{N})\\
c^{-1} \sum_{k=1}^{c}\left(X_{(jk)}-x\right)^{\ell}=&\ \left(X_{jk}-x\right)^{\ell}+O_{\hbox {\tiny $P$}}(h^{l-1}/\sqrt{N})\{1-I(\ell=0)\}.
\end{align*}
Further, provided that $K'(t)$ is bounded,
\begin{align*}
&\ \left\{c^{-1} \sum_{k=1}^{c} K_{h}\left(X_{(jk)}-x\right)\right\}^2= c^{-2}\sum_{k=1}^{c} \left\{K_h\left(X_{(jk)}-x\right)\right\}^2\\
&\ +c^{-2}\sum_{k_1\neq k_2}K_h\left(X_{(jk_1)}-x\right)K_h\left(X_{(jk_2)}-x\right)\\
=&\ c^{-2}\sum_{k=1}^{c} \left\{K_h\left(X_{(jk)}-x\right)\right\}^2+c^{-2}(c-1)\sum_{k_1=1}^{c}K_h\left(X_{(jk_1)}-x\right)\{K_h\left(X_{(jk_1)}-x\right)+O_{\hbox {\tiny $P$}}(1/\sqrt{Nh^4})\}\\
=&\ c^{-1}h^{-1}\sum_{k=1}^{c} K^\ddagger_h\left(X_{(jk)}-x\right)+\frac{c-1}{c^2}h^{-1}\sum_{k=1}^{c}K_h(X_{(jk)}-x)O_{\hbox {\tiny $P$}}(1/\sqrt{Nh^2}).
\end{align*}
Then
\begin{align*}
&\ J^{-2}\sum_{j\in\cJ_1}c^{-2}\sum_{k=1}^{c}\sigma^2(X_{(jk)})\left\{c^{-1} \sum_{k=1}^{c}\left(X_{(jk)}-x\right)^{l_1}\right\}\\
&\ \times \left\{c^{-1} \sum_{k=1}^{c}\left(X_{(jk)}-x\right)^{l_2}\right\}\left\{c^{-1} \sum_{k=1}^{c} K_{h}\left(X_{(jk)}-x\right)\right\}^2\\
=&\ (Nh)^{-1}N^{-1}\sum_{j\in\cJ_1}\left\{\sigma^2(X_{(jk)})+O_{\hbox {\tiny $P$}}(1/\sqrt{N})\right\}\left\{\left(X_{(jk)}-x\right)^{\ell_1}+O_{\hbox {\tiny $P$}}(h^{\ell_1-1}/\sqrt{N})\{1-I(\ell_1=0)\}\right\}\\
&\ \times \left\{\left(X_{(jk)}-x\right)^{\ell_2}+O_{\hbox {\tiny $P$}}(h^{\ell_2-1}/\sqrt{N})\{1-I(\ell_2=0)\}\right\}\\
&\ \times\left\{ \sum_{k=1}^{c} K^\ddagger_h\left(X_{(jk)}-x\right)+(c-1)\sum_{k=1}^{c}K_h(X_{(jk)}-x)O_{\hbox {\tiny $P$}}(1/\sqrt{Nh^2})\right\}\\
=&\ (Nh)^{-1}N^{-1}\sum_{j=1}^J\sum_{k=1}^{c}K^\ddagger_h\left(X_{(jk)}-x\right)\sigma^2(X_{(jk)})\left(X_{(jk)}-x_0\right)^{\ell_1+\ell_2}\{1+o_{\hbox {\tiny $P$}}(1)\}.
\end{align*}
provided that $Nh^4\rightarrow\infty$.

Thus, we conclude that
\begin{align*}
&\ (Nh)\mbox{Cov}\left\{T_{1,\ell_1}(x), T_{1,\ell_2}(x)|\mathbb{X}\right\}\\
=&\ N^{-1}\sum_{j=1}^J\sum_{k=1}^{c}K^\ddagger_h\left(X_{jk}-x\right)\sigma^2(X_{jk})\left(X_{jk}-x\right)^{\ell_1+\ell_2}\{1+o_{\hbox {\tiny $P$}}(1)\}.
\end{align*}
Hence, $\mbox{Var}\{\hat\bmbeta_1(x)|\mathbb{X}\}=\mbox{Var}\{\hat\bmbeta_0(x)|\mathbb{X}\}\{1+o_{\hbox {\tiny $P$}}(1)\}.$ Note that, homogeneous pooling uses $J$ tests while individual uses $cJ$ tests. 

\subsection{Bias and variance of $\hat m_2(x)$}
Under homogeneous pooling, the weighted least squares objective function $Q_2(\bbeta)$ is 
$$
Q_2(\bbeta)=\sum_{j=1}^{J}\left\{Z_{(j)}-\sum_{\ell=0}^p \beta_{\ell} c^{-1} \sum_{k=1}^{c}\left(X_{(j k)}-x\right)^{\ell}\right\}^2\left\{\prod_{k=1}^c K_{h}\left(X_{(jk)}-x\right)\right\}. 
$$
Then the product-weighted $p$-th order local polynomial estimator for $m(x)$, is
$$
\hat m_2(x)=\be_1^\t\bS_2^{-1}(x)\bT_2(x),
$$
where
\begin{align*}
\bS_2(x)=&\ [S_{1,\ell_1,\ell_2}(x)]_{\ell_1, \ell_2=0,1,\dots, p},\\
\bT_2(x)=&\ (T_{1,0}(x), T_{1,1}(x),\dots, T_{1,p}(x))^\t, \mbox{ in which}\\
S_{2,\ell_1,\ell_2}(x)=&\ J^{-1}\sum_{j=1}^{J}\left\{c^{-1} \sum_{k=1}^{c}\left(X_{(jk)}-x\right)^{\ell_1}\right\}\left\{c^{-1} \sum_{k=1}^{c}\left(X_{(jk)}-x\right)^{\ell_2}\right\}\left\{h^{c-1}\prod_{k=1}^c K_{h}\left(X_{(jk)}-x\right)\right\},\\
& \mbox{ for } \ell_1, \ell_2=0,1,\dots, p,\\
T_{2,\ell}(x)=&\ J^{-1}\sum_{j=1}^JZ_{(j)}\left\{c^{-1} \sum_{k=1}^{c}\left(X_{(jk)}-x\right)^{\ell}\right\}\left\{h^{c-1}\prod_{k=1}^c K_{h}\left(X_{(jk)}-x\right)\right\}, \mbox{ for } \ell=0,1,\dots, p.
\end{align*}

Study the bias term first,
\begin{align*}
\mbox{Bias}\left\{\left.\hat\bmbeta_2(x)\right|\mathbb{X}\right\}=\bS_2^{-1}(x)E\left\{\left.\bT_2(x)-\bS_2(x)\bmbeta(x)\right|\mathbb{X}\right\}
\end{align*}
Rewrite the $\ell$th component of $\bT_2(x)-\bS_2(x)\beta(x)$, where $\ell=0,1,\dots, p$,
\begin{align*}
T_{2,\ell}(x)-\bS_{2,l,\cdot}(x)\bmbeta(x)=&\ J^{-1}\sum_{j=1}^J\left\{Z_{(j)}-\sum_{\ell=0}^p \beta_{\ell} c^{-1} \sum_{k=1}^{c}\left(X_{(j k)}-x\right)^{\ell}\right\}\\
&\ \times \left\{c^{-1} \sum_{k=1}^{c}\left(X_{(jk)}-x\right)^{\ell}\right\}\left\{h^{c-1}\prod_{k=1}^cK_{h}\left(X_{(jk)}-x\right)\right\}.
\end{align*}
Then 
\begin{align}
E\left\{\left.T_{1,\ell}(x)-\bS_{1,\ell,\cdot}(x)\bmbeta(x)\right|\mathbb{X}\right\}=&\ J^{-1} \sum_{j=1}^J \left\{h^{c-1}\prod_{k=1}^cK_{h}\left(X_{(jk)}-x\right)\right\}\nonumber\\
&\ \times \left\{c^{-1} \sum_{k=1}^{c}\left(X_{(jk)}-x\right)^{\ell}\right\}\nonumber\\
&\ \times c^{-1}\sum_{k=1}^{c}\left\{m(X_{(jk)})-\sum_{\ell=0}^p \beta_{\ell} \left(X_{(j k)}-x\right)^{\ell}\right\}\label{e:T1l}
\end{align}
Similarly, we break the summation $\sum_{j=1}^J$ in \eqref{e:T1l} into two parts as $\sum_{j\in\cJ_1}$ and $\sum_{j\in\cJ_2}$ where for $j\in\cJ_1$, at least one of $X_{(j1)}, \dots, X_{(jc)}$ is in $[x-h,x+h]$; and for $j\in\cJ_2$, none of $X_{(j1)}, \dots, X_{(jc)}$ are in $[x-h,x+h]$. We rewrite \eqref{e:T1l} as
$$
E\left\{\left.T_{1,\ell}(x)-\bS_{1,\ell,\cdot}(x)\bmbeta(x)\right|\mathbb{X}\right\}=\cL_1+\cL_2.
$$
When $j\in\cJ_2$, all the $K_{h}\left(X_{(jk)}-x\right)$'s are zero. Hence
\begin{align*}
\cL_2=0=N^{-1}\sum_{j\in\cJ_2}\sum_{k=1}^{c} K^{\dagger}_{h}\left(X_{(jk)}-x\right)\left(X_{(jk)}-x\right)^{\ell} \left\{m(X_{(jk)})-\sum_{\ell=0}^p \beta_{\ell} \left(X_{(j k)}-x\right)^{\ell}\right\},
\end{align*}
where $K^\dagger(t)=K^c(t)$.
When $j\in\cJ_1$, we have $|X_{(jk)}-x|\leq h$. Consequently,
\begin{align*}
c^{-1} \sum_{k=1}^{c}\left(X_{(jk)}-x\right)^{\ell}=&\ \left(X_{(jk)}-x\right)^{\ell}+c^{-1}\sum_{k_1\neq k}^c \{(X_{(jk_1)}-x)^\ell-(X_{(jk)}-x)^\ell\}\\
=&\ \left(X_{(jk)}-x\right)^{\ell}+O_{\hbox {\tiny $P$}}(h^{\ell-1}/\sqrt{N})\{1-I(\ell=0)\}.
\end{align*}
By the smoothness of $m(\cdot)$,
\begin{align*}
&\ c^{-1}\sum_{k=1}^{c}\left\{m(X_{(jk)})-\sum_{\ell=0}^p \beta_{\ell} \left(X_{(j k)}-x\right)^{\ell}\right\}-\left\{m(X_{(jk)})-\sum_{\ell=0}^p \beta_{\ell} \left(X_{(j k)}-x\right)^{\ell}\right\}\\
=&\ c^{-1}\sum_{k_1\neq k}\left[\left\{m(X_{(jk_1)})-\sum_{\ell=0}^p \beta_{\ell} \left(X_{(j k_1)}-x\right)^{\ell}\right\}-\left\{m(X_{(jk)})-\sum_{\ell=0}^p \beta_{\ell} \left(X_{(j k)}-x\right)^{\ell}\right\}\right]\\
=&\ c^{-1}\sum_{k_1\neq k}\left\{m^{(p+1)}(X_{(jk_1)}^*)(X_{(jk_1)}-x)^{p+1}-m^{(p+1)}(X_{(jk)}^*)(X_{(jk)}-x)^{p+1}\right\}/(p+1)!\\
=&\ O_{\hbox {\tiny $P$}}(h^p/\sqrt{N}).
\end{align*}
And, when $K'(t)$ is bounded and $Nh^2\rightarrow\infty$,
\begin{align*}
h^{c-1}\prod_{k=1}^cK_{h}\left(X_{(jk)}-x\right)=&\ h^{-1}\prod_{k=1}^c K\left(\frac{X_{(jk)-x}}{h}\right)\\
=&\ h^{-1}\prod_{k=1}^c \left\{K\left(\frac{X_{(j1)-x}}{h}\right)+O_p(1/\sqrt{Nh^2})\right\}\\
=&\ h^{-1}K^c \left(\frac{X_{(j1)-x}}{h}\right)\{1+o_{\hbox {\tiny $P$}}(1)\}.
\end{align*}
Hence, we have
\begin{align*}
h^{c-1}\prod_{k=1}^cK_{h}\left(X_{(jk)}-x\right)=&\ c^{-1}\sum_{k=1}^c K^\dagger_h \left(\frac{X_{(jk)-x}}{h}\right)\{1+o_{\hbox {\tiny $P$}}(1)\}.
\end{align*}
Consequently, recalling the definition of $\cJ_2$, we can write $E[T_{1,l}(x)-\bS_{1,l,\cdot}(x)\bmbeta(x)|\mathbb{X}]=\cL_1$ as 
\begin{align*}
N^{-1}\sum_{j=1}^J\sum_{k=1}^{c} K^\dagger_{h}\left(X_{(jk)}-x\right)\left(X_{(jk)}-x\right)^{\ell}\left\{m(X_{(jk)})-\sum_{\ell=0}^p \beta_{\ell} \left(X_{(j k)}-x\right)^{\ell}\right\}\{1+o_{\hbox {\tiny $P$}}(1)\}
\end{align*}
provided that $Nh^2\rightarrow\infty$.
Similarly, we have
\begin{align*}
S_{1,\ell_1,\ell_2}(x)=&\ N^{-1}\sum_{j=1}^J\sum_{k=1}^{c} K^\dagger_{h}\left(X_{jk}-x\right)\left(X_{jk}-x\right)^{\ell_1+\ell_2}\{1+o_{\hbox {\tiny $P$}}(1)\}.
\end{align*}
Hence, $\mbox{Bias}\{\hat\bmbeta_1(x)|\mathbb{X}\}=\mbox{Bias}[\hat\bmbeta_0(x)|\mathbb{X}]\{1+o_{\hbox {\tiny $P$}}(1)\}$ when using $K^\dagger(t)$ as the kernel function.

Similarly, we can conclude that
$\mbox{Var}\{\hat\bmbeta_2(x)|\mathbb{X}\}=\mbox{Var}\{\hat\bmbeta_0(x)|\mathbb{X}\}\{1+o_{\hbox {\tiny $P$}}(1)\}$, again where the kernel function is $K^\dagger(t)$ instead of $K(t)$. Note that, homogeneous pooling uses $J$ tests while individual uses $cJ$ tests.

\setcounter{equation}{0} 
\setcounter{section}{0} 
\setcounter{subsection}{0} 
\setcounter{subsubsection}{0} 
\def\theequation{E.\arabic{equation}}
\renewcommand\thesubsection{E.\arabic{subsection}}
\renewcommand\thesubsubsection{E.\arabic{subsection}.\arabic{subsubsection}}
\section*{Appendix E: Additional simulation results}
\begin{landscape}
\begin{figure} 
	\centering
	\setlength{\linewidth}{1.4\textwidth}
	\includegraphics[width=\linewidth]{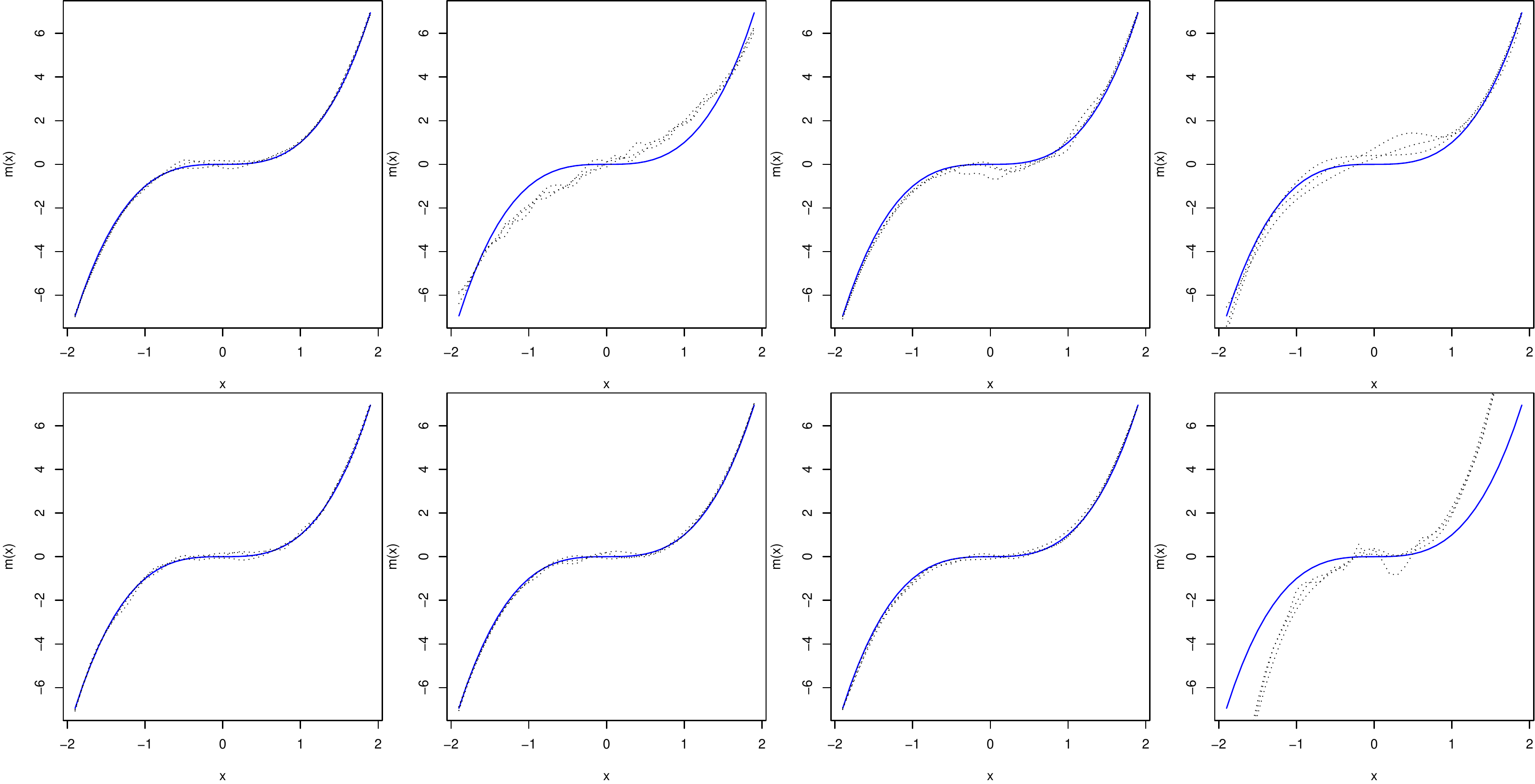} 
	\caption{\linespread{1.3}\selectfont{}Four estimates for $m(x)$ under (D1): the local linear estimate based on individual-level data (the first column), $\hat m_0(x)$, the average-weighted estimate (the second column), $\hat m_1(x)$, the product-weighted estimate (the third column), $\hat m_2(x)$, and the marginal-integration estimate (the fourth column), $\hat m_3(x)$. The latter three estimates are based on random pooled data in the upper panels, and are based on homogeneous pooled data in the lower panels. Within each panel, the blue curve is the true function $m(x)$, and the three dotted lines are three realizations of the estimator depicted in that panel whose ISE's are equal to the three quartiles among the 500 realizations of ISE's associated with the estimator.}
	\label{f:D1}
\end{figure}
\end{landscape}

\begin{landscape}
\begin{figure} 
	\centering
	\setlength{\linewidth}{1.4\textwidth}
	\includegraphics[width=\linewidth]{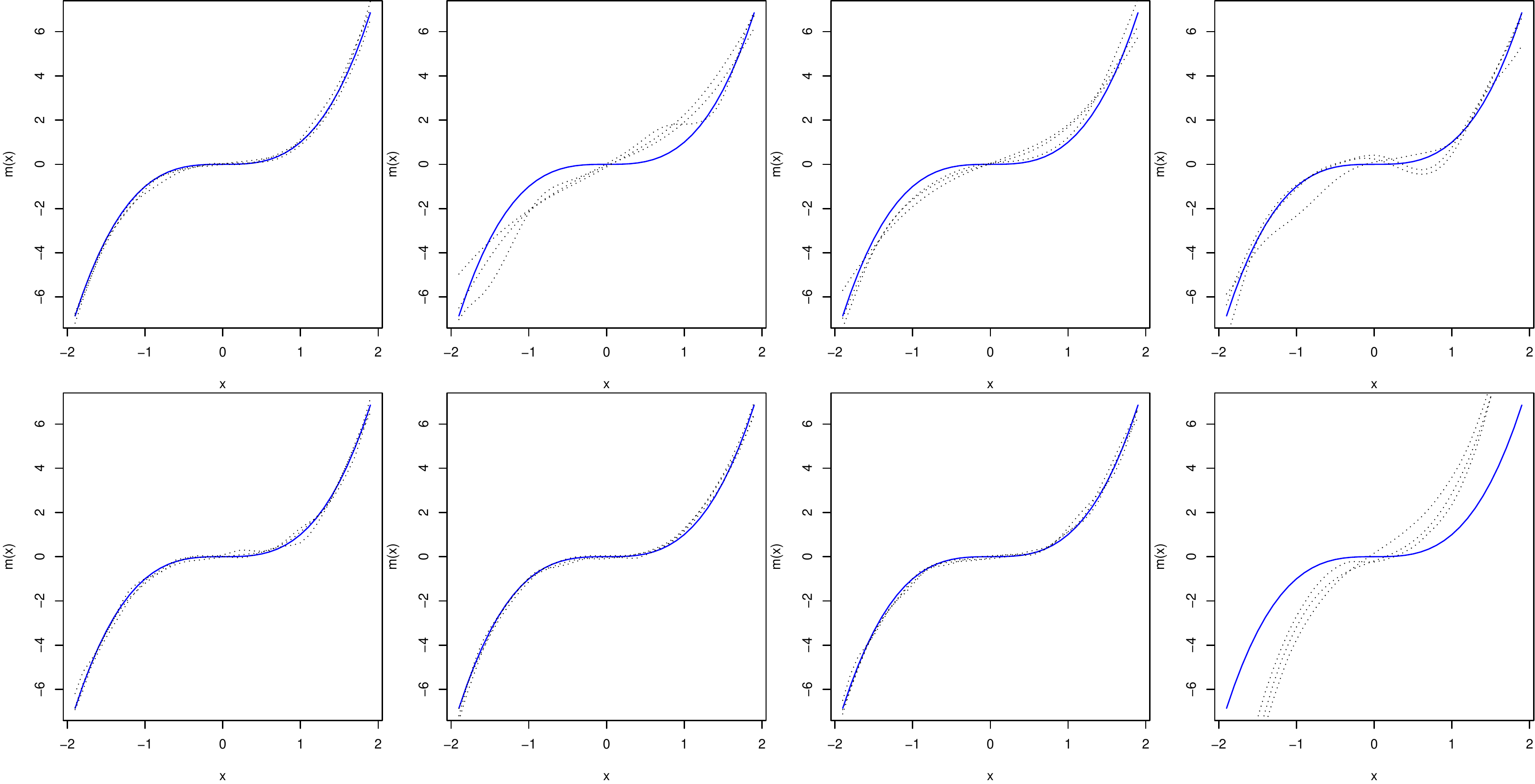} 
	\caption{\linespread{1.3}\selectfont{}Four estimates for $m(x)$ under (D3): the local linear estimate based on individual-level data (the first column), $\hat m_0(x)$, the average-weighted estimate (the second column), $\hat m_1(x)$, the product-weighted estimate (the third column), $\hat m_2(x)$, and the marginal-integration estimate (the fourth column), $\hat m_3(x)$. The latter three estimates are based on random pooled data in the upper panels, and are based on homogeneous pooled data in the lower panels. Within each panel, the blue curve is the true function $m(x)$, and the three dotted lines are three realizations of the estimator depicted in that panel whose ISE's are equal to the three quartiles among the 500 realizations of ISE's associated with the estimator.}
	\label{f:D3}
\end{figure}
\end{landscape}

\begin{landscape}
\begin{figure} 
	\centering
	\setlength{\linewidth}{1.4\textwidth}
	\includegraphics[width=\linewidth]{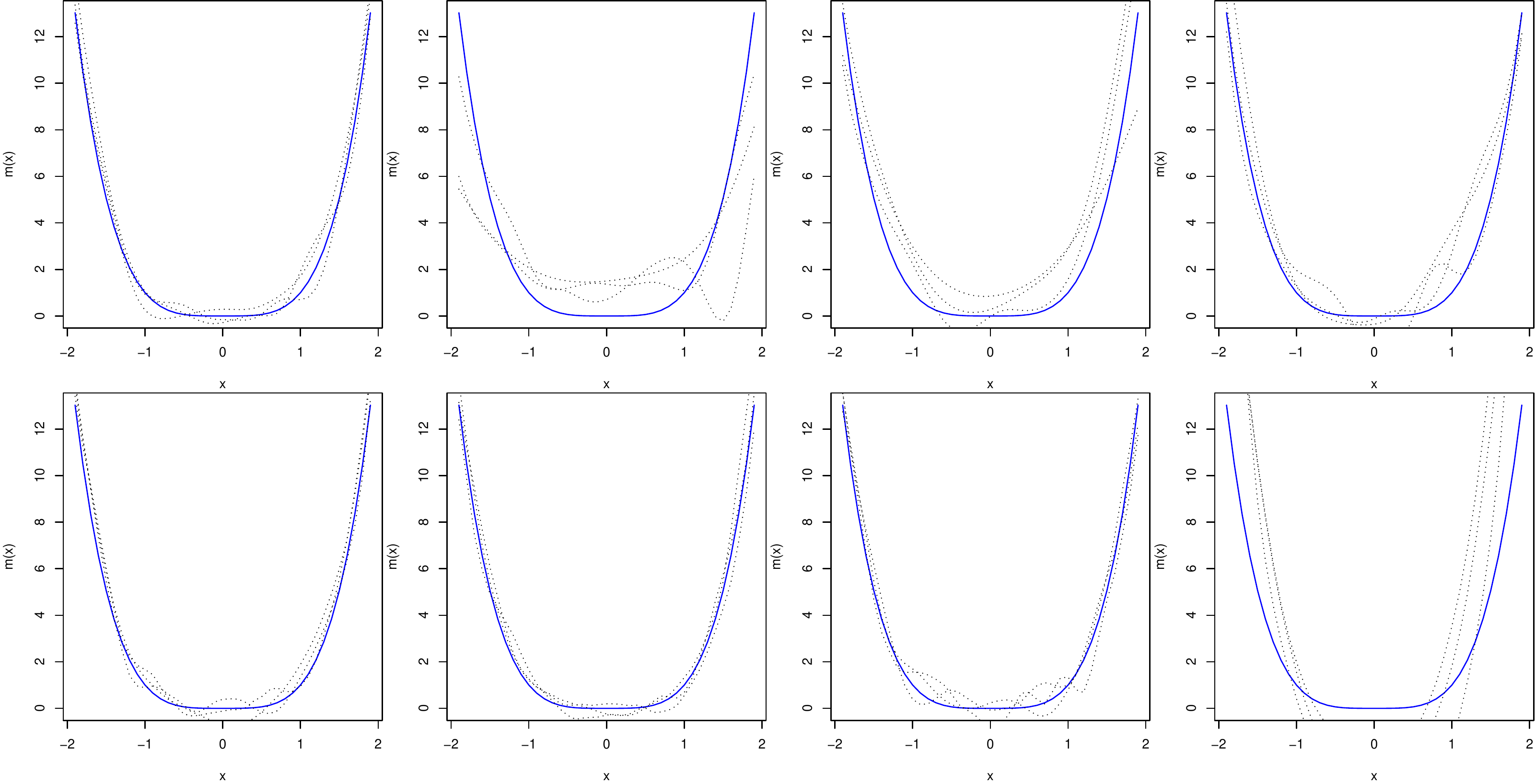} 
	\caption{\linespread{1.3}\selectfont{}Four estimates for $m(x)$ under (D4): the local linear estimate based on individual-level data (the first column), $\hat m_0(x)$, the average-weighted estimate (the second column), $\hat m_1(x)$, the product-weighted estimate (the third column), $\hat m_2(x)$, and the marginal-integration estimate (the fourth column), $\hat m_3(x)$. The latter three estimates are based on random pooled data in the upper panels, and are based on homogeneous pooled data in the lower panels. Within each panel, the blue curve is the true function $m(x)$, and the three dotted lines are three realizations of the estimator depicted in that panel whose ISE's are equal to the three quartiles among the 500 realizations of ISE's associated with the estimator.}
	\label{f:D4}
\end{figure}
\end{landscape}

\begin{landscape}
\begin{figure} 
	\centering
	\setlength{\linewidth}{1.4\textwidth}
	\includegraphics[width=\linewidth]{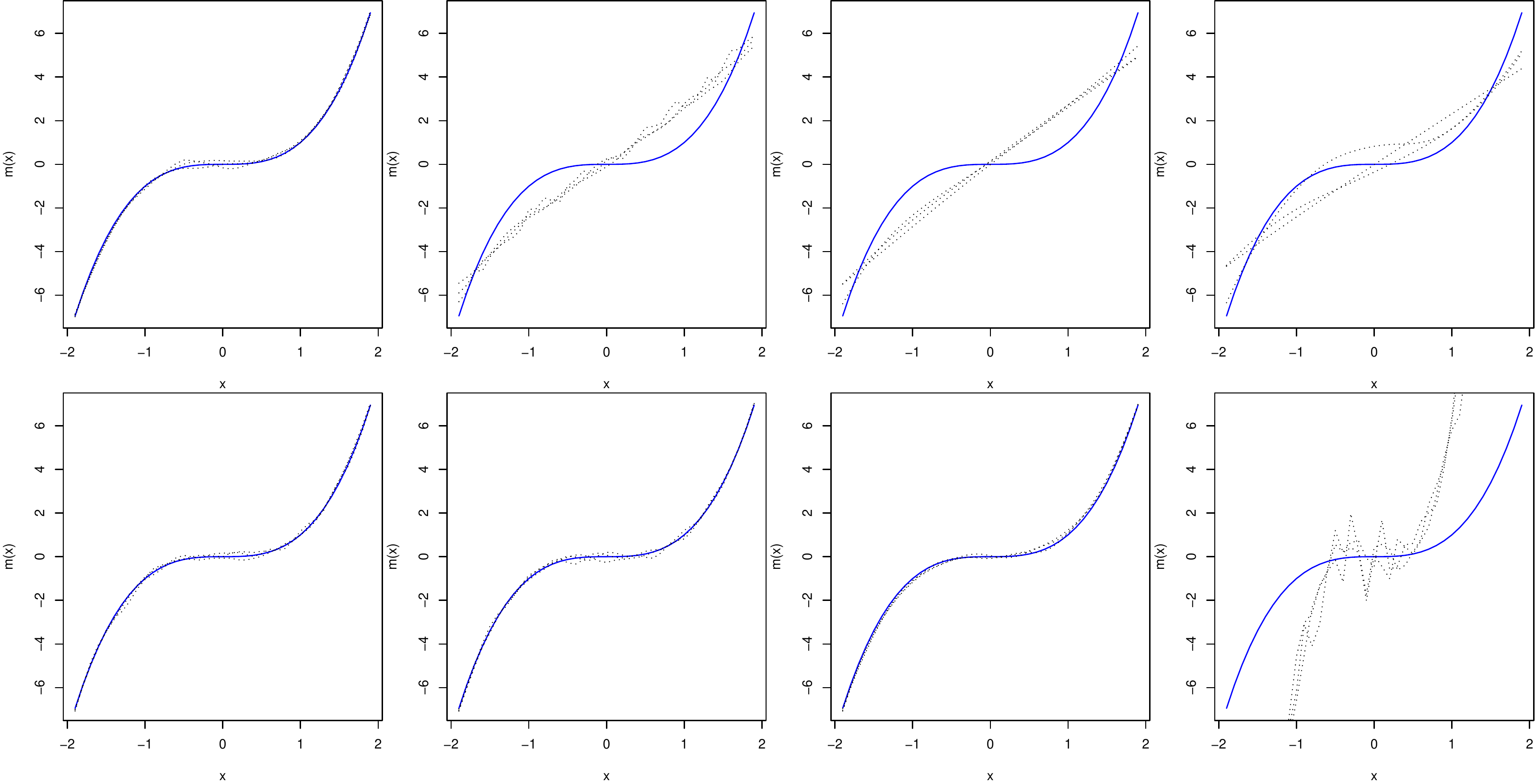} 
	\caption{\linespread{1.3}\selectfont{}Four estimates for $m(x)$ under (D1): the local linear estimate based on individual-level data (the first column), $\hat m_0(x)$, the average-weighted estimate (the second column), $\hat m_1(x)$, the product-weighted estimate (the third column), $\hat m_2(x)$, and the marginal-integration estimate (the fourth column), $\hat m_3(x)$. The latter three estimates are based on random pooling data in the upper panels, and based on homogeneous pooling data in the lower panels. Within each panel, the blue curve is the true function $m(x)$, and the three dotted lines are three realizations of the estimator depicted in that panel whose ISE's are equal to the three quartiles among the 500 realizations of ISE's associated with the estimator.}
	\label{f:D1c6}
\end{figure}
\end{landscape}

\begin{landscape}
\begin{figure} 
	\centering
	\setlength{\linewidth}{1.4\textwidth}
	\includegraphics[width=\linewidth]{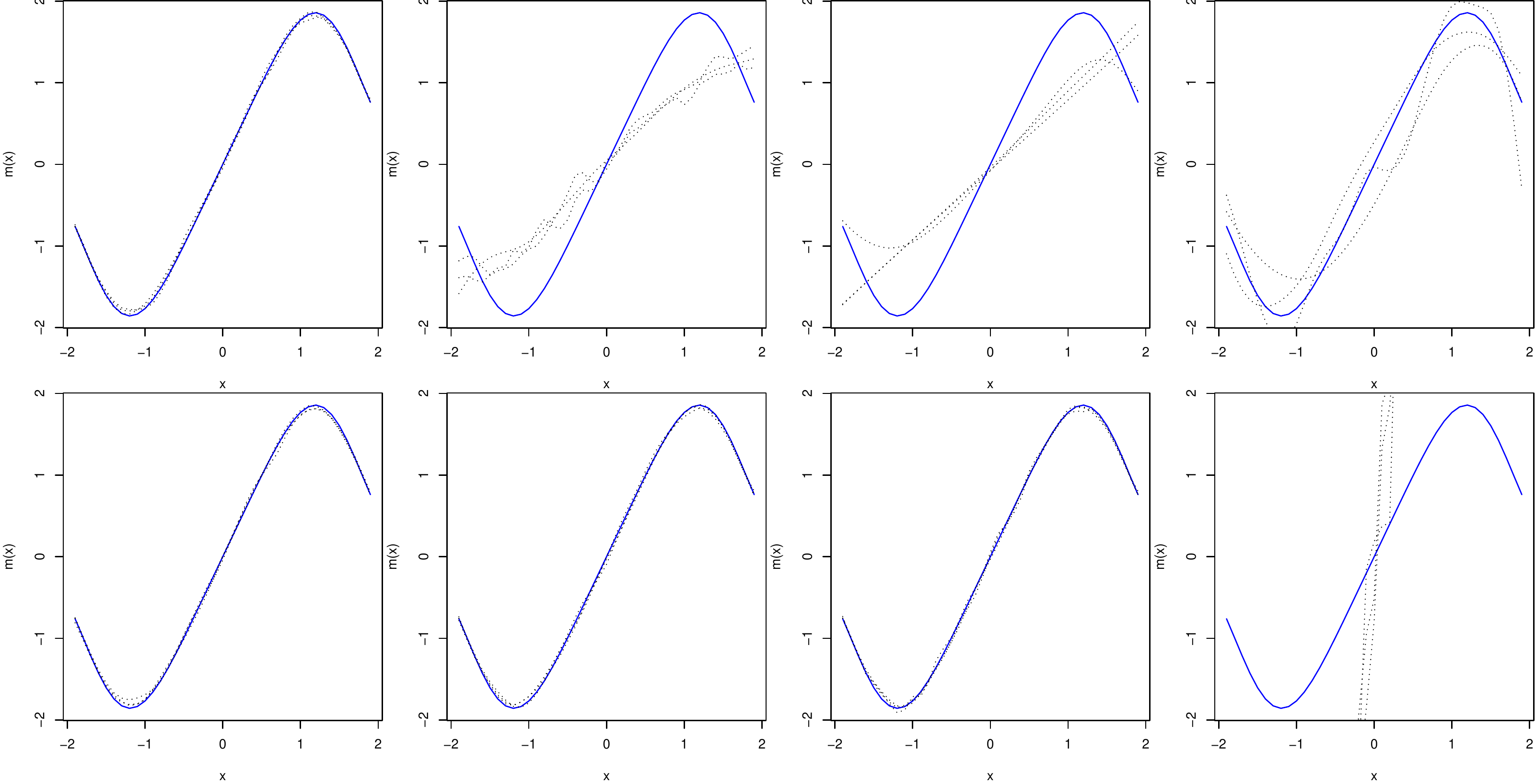} 
	\caption{\linespread{1.3}\selectfont{}Four estimates for $m(x)$ under (D2): the local linear estimate based on individual-level data (the first column), $\hat m_0(x)$, the average-weighted estimate (the second column), $\hat m_1(x)$, the product-weighted estimate (the third column), $\hat m_2(x)$, and the marginal-integration estimate (the fourth column), $\hat m_3(x)$. The latter three estimates are based on random pooling data in the upper panels, and based on homogeneous pooling data in the lower panels. Within each panel, the blue curve is the true function $m(x)$, and the three dotted lines are three realizations of the estimator depicted in that panel whose ISE's are equal to the three quartiles among the 500 realizations of ISE's associated with the estimator.}
	\label{f:D2c6}
\end{figure}
\end{landscape}

\begin{landscape}
\begin{figure} 
	\centering
	\setlength{\linewidth}{1.4\textwidth}
	\includegraphics[width=\linewidth]{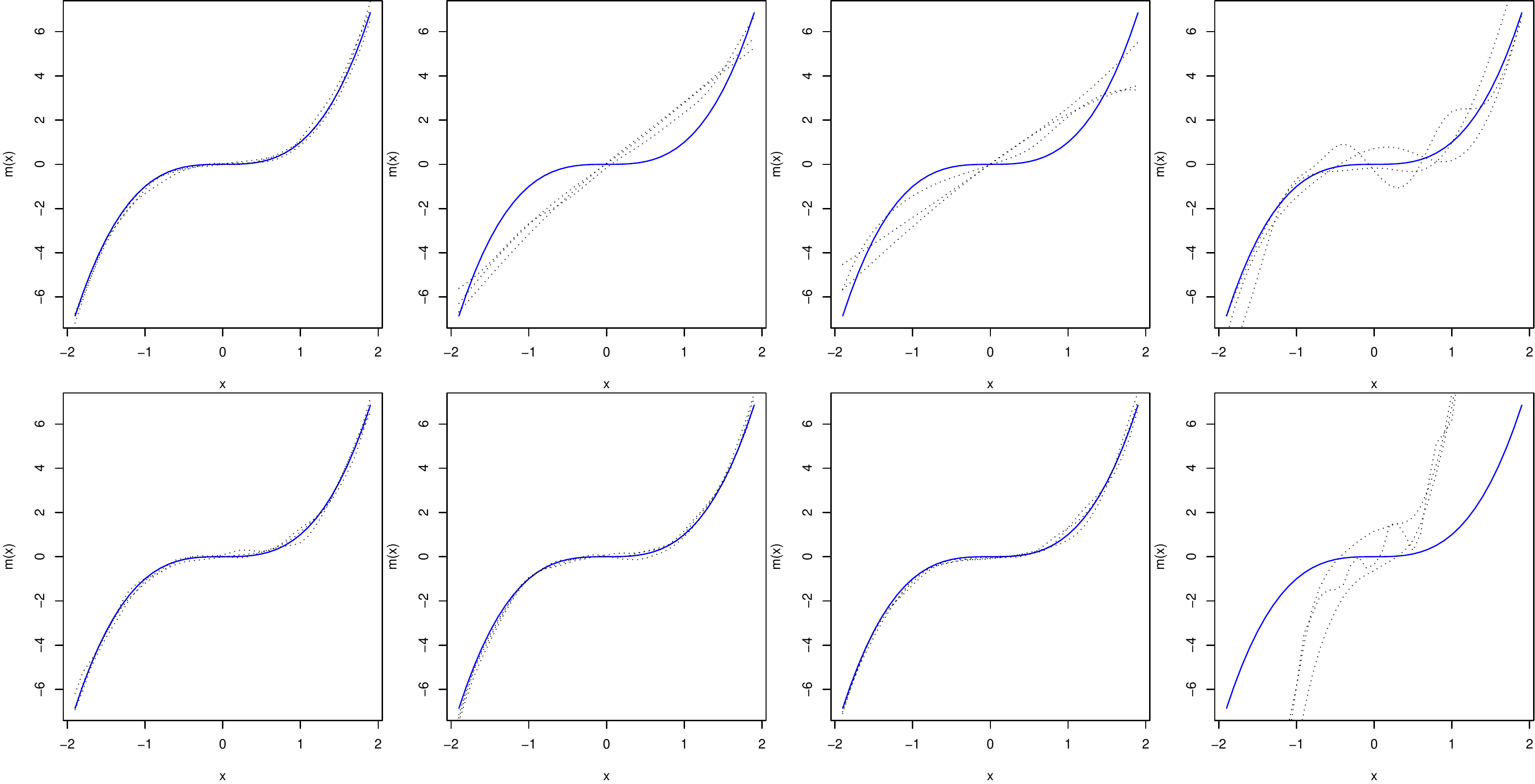} 
	\caption{\linespread{1.3}\selectfont{}Four estimates for $m(x)$ under (D3): the local linear estimate based on individual-level data (the first column), $\hat m_0(x)$, the average-weighted estimate (the second column), $\hat m_1(x)$, the product-weighted estimate (the third column), $\hat m_2(x)$, and the marginal-integration estimate (the fourth column), $\hat m_3(x)$. The latter three estimates are based on random pooling data in the upper panels, and based on homogeneous pooling data in the lower panels. Within each panel, the blue curve is the true function $m(x)$, and the three dotted lines are three realizations of the estimator depicted in that panel whose ISE's are equal to the three quartiles among the 500 realizations of ISE's associated with the estimator.}
	\label{f:D3c6}
\end{figure}
\end{landscape}

\begin{landscape}
\begin{figure} 
	\centering
	\setlength{\linewidth}{1.4\textwidth}
	\includegraphics[width=\linewidth]{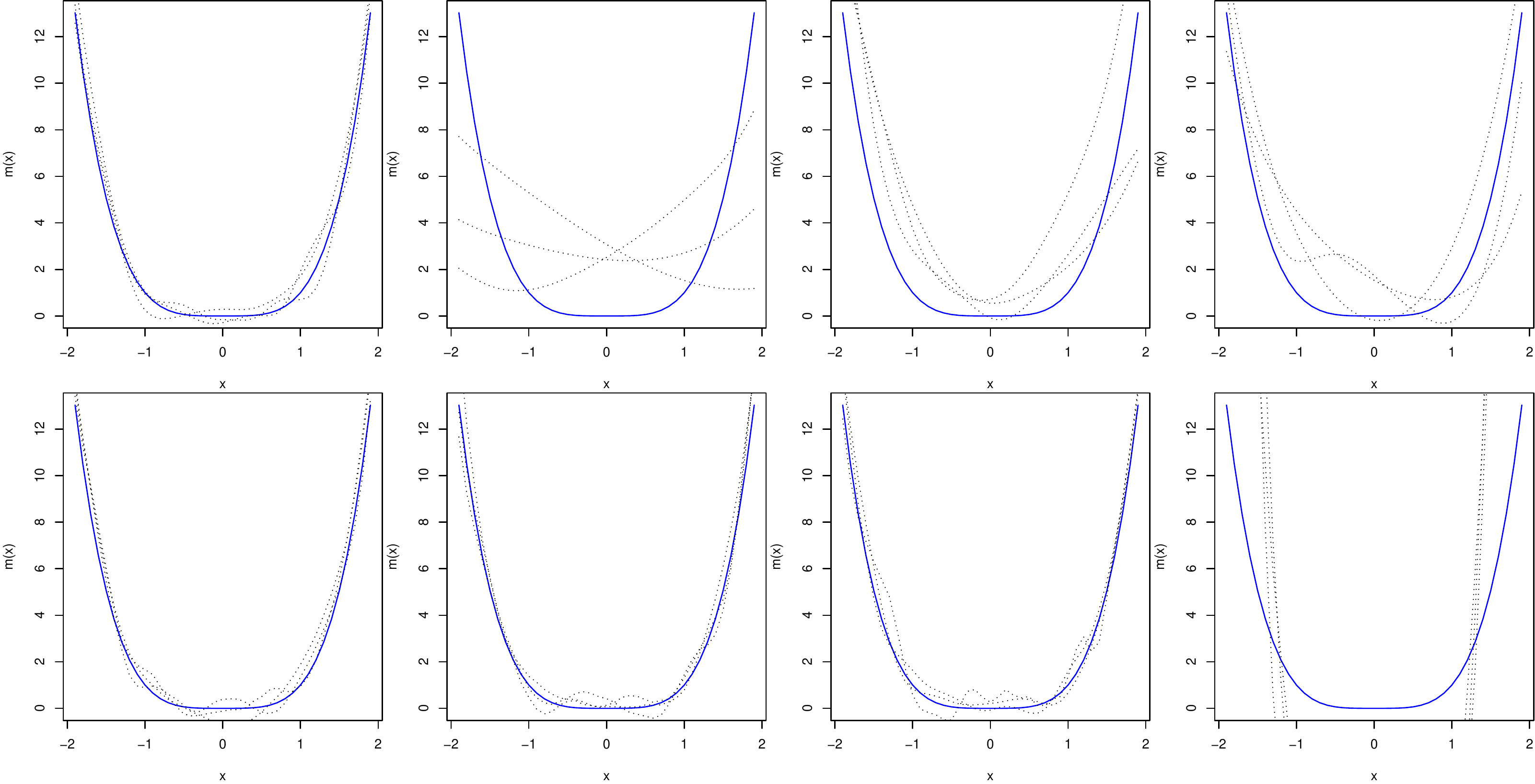} 
	\caption{\linespread{1.3}\selectfont{}Four estimates for $m(x)$ under (D4): the local linear estimate based on individual-level data (the first column), $\hat m_0(x)$, the average-weighted estimate (the second column), $\hat m_1(x)$, the product-weighted estimate (the third column), $\hat m_2(x)$, and the marginal-integration estimate (the fourth column), $\hat m_3(x)$. The latter three estimates are based on random pooling data in the upper panels, and based on homogeneous pooling data in the lower panels. Within each panel, the blue curve is the true function $m(x)$, and the three dotted lines are three realizations of the estimator depicted in that panel whose ISE's are equal to the three quartiles among the 500 realizations of ISE's associated with the estimator.}
	\label{f:D4c6}
\end{figure}
\end{landscape}

Using data generated from models specified in (D1), (D3), and (D4) in Section 6 in the main article, Figures \ref{f:D1}, \ref{f:D3}, and \ref{f:D4} depict the three proposed estimators based on pooled data when $c=2$ and the benchmark estimate, $\hat m_0(x)$, based on individual-level data. Figures \ref{f:D1c6}--\ref{f:D4c6} provide parallel results when $c=6$ under (D1), (D2), (D3), and (D4). 

\noindent {\large\bf Acknowledgments}

Wang's work was supported by the National Institutes of Health (NIH) under the Award Number R03 AI135614.
\par

\bibliographystyle{apalike}
\bibliography{ref4pooly}

\begin{thebibliography}{}

\bibitem[Bilder and Tebbs, 2009]{bilder2009bias}
Bilder, C.~R. and Tebbs, J.~M. (2009).
\newblock Bias, efficiency, and agreement for group-testing regression models.
\newblock {\em Journal of Statistical Computation and Simulation},
  79(1):67--80.

\bibitem[Delaigle et~al., 2009]{delaigle2009design}
Delaigle, A., Fan, J., and Carroll, R. (2009).
\newblock A design-adaptive local polynomial estimator for the
  errors-in-variables problem.
\newblock {\em Journal of the American Statistical Association},
  104(485):348--359.

\bibitem[Delaigle and Hall, 2012]{delaigle2012nonparametric}
Delaigle, A. and Hall, P. (2012).
\newblock Nonparametric regression with homogeneous group testing data.
\newblock {\em The Annals of Statistics}, 40(1):131--158.

\bibitem[Dhawan and Richmond, 2002]{dhawan2002role}
Dhawan, P. and Richmond, A. (2002).
\newblock Role of cxcl1 in tumorigenesis of melanoma.
\newblock {\em Journal of Leukocyte Biology}, 72(1):9--18.

\bibitem[Fan and Gijbels, 1996]{Fan&Gijbels1996}
Fan, J. and Gijbels, I. (1996).
\newblock {\em Local Polynomial Modelling and Its Applications: Monographs on
  Statistics and Applied Probability 66}, volume~66.
\newblock Chapman \& Hall/CRC.

\bibitem[Genuis et~al., 2013]{genuis2013gastrointestinal}
Genuis, S.~J., Curtis, L., and Birkholz, D. (2013).
\newblock Gastrointestinal elimination of perfluorinated compounds using
  cholestyramine and chlorella pyrenoidosa.
\newblock {\em ISRN Toxicology}, 2013.

\bibitem[K{\"a}rrman et~al., 2006]{karrman2006levels}
K{\"a}rrman, A., Mueller, J.~F., Van~Bavel, B., Harden, F., Toms, L.-M.~L., and
  Lindstr{\"o}m, G. (2006).
\newblock Levels of 12 perfluorinated chemicals in pooled australian serum,
  collected 2002- 2003, in relation to age, gender, and region.
\newblock {\em Environmental Science \& Technology}, 40(12):3742--3748.

\bibitem[Kendziorski et~al., 2003]{kendziorski2003efficiency}
Kendziorski, C., Zhang, Y., Lan, H., and Attie, A. (2003).
\newblock The efficiency of pooling mrna in microarray experiments.
\newblock {\em Biostatistics}, 4(3):465--477.

\bibitem[Lin and Wang, 2018]{LinWang2018}
Lin, J. and Wang, D. (2018).
\newblock Single-index regression for pooled biomarker data.
\newblock {\em Journal of Nonparametric Statistics}, 30(4):813--833.

\bibitem[Linton and Whang, 2002]{linton2002nonparametric}
Linton, O. and Whang, Y.-J. (2002).
\newblock Nonparametric estimation with aggregated data.
\newblock {\em Econometric Theory}, 18(2):420--468.

\bibitem[Liu et~al., 2017]{liu2017general}
Liu, Y., McMahan, C., and Gallagher, C. (2017).
\newblock A general framework for the regression analysis of pooled biomarker
  assessments.
\newblock {\em Statistics in Medicine}, 36(15):2363--2377.

\bibitem[Ma et~al., 2011]{ma2011cost}
Ma, C.-X., Vexler, A., Schisterman, E.~F., and Tian, L. (2011).
\newblock Cost-efficient designs based on linearly associated biomarkers.
\newblock {\em Journal of Applied Statistics}, 38(12):2739--2750.

\bibitem[Malinovsky et~al., 2012]{malinovsky2012pooling}
Malinovsky, Y., Albert, P.~S., and Schisterman, E.~F. (2012).
\newblock Pooling designs for outcomes under a gaussian random effects model.
\newblock {\em Biometrics}, 68(1):45--52.

\bibitem[McMahan et~al., 2016]{mcmahan2016estimating}
McMahan, C.~S., McLain, A.~C., Gallagher, C.~M., and Schisterman, E.~F. (2016).
\newblock Estimating covariate-adjusted measures of diagnostic accuracy based
  on pooled biomarker assessments.
\newblock {\em Biometrical Journal}, 58(4):944--961.

\bibitem[Mitchell et~al., 2014]{mitchell2014regression}
Mitchell, E.~M., Lyles, R.~H., Manatunga, A.~K., Danaher, M., Perkins, N.~J.,
  and Schisterman, E.~F. (2014).
\newblock Regression for skewed biomarker outcomes subject to pooling.
\newblock {\em Biometrics}, 70(1):202--211.

\bibitem[Mitchell et~al., 2015]{mitchell2015semiparametric}
Mitchell, E.~M., Lyles, R.~H., Manatunga, A.~K., and Schisterman, E.~F. (2015).
\newblock Semiparametric regression models for a right-skewed outcome subject
  to pooling.
\newblock {\em American Journal of Epidemiology}, 181(7):541--548.

\bibitem[Shih et~al., 2004]{shih2004effects}
Shih, J.~H., Michalowska, A.~M., Dobbin, K., Ye, Y., Qiu, T.~H., and Green,
  J.~E. (2004).
\newblock Effects of pooling mrna in microarray class comparisons.
\newblock {\em Bioinformatics}, 20(18):3318--3325.

\bibitem[Tsou et~al., 2007]{tsou2007critical}
Tsou, C.-L., Peters, W., Si, Y., Slaymaker, S., Aslanian, A.~M., Weisberg,
  S.~P., Mack, M., and Charo, I.~F. (2007).
\newblock Critical roles for ccr2 and mcp-3 in monocyte mobilization from bone
  marrow and recruitment to inflammatory sites.
\newblock {\em The Journal of Clinical Investigation}, 117(4):902--909.

\bibitem[Vansteelandt et~al., 2000]{vansteelandt2000regression}
Vansteelandt, S., Goetghebeur, E., and Verstraeten, T. (2000).
\newblock Regression models for disease prevalence with diagnostic tests on
  pools of serum samples.
\newblock {\em Biometrics}, 56(4):1126--1133.

\end{thebibliography}

\end{document}